\documentclass[twocolumn,twocolappendix,times]{aastex631}

\usepackage{color} 
\usepackage[flushleft]{threeparttable}
\usepackage{amsmath}
\usepackage{float}
\usepackage{amssymb}
\usepackage{graphicx}
\usepackage{natbib}
\usepackage{multirow}

\graphicspath{{./}{figures/}}

\definecolor{emerald}{rgb}{0.1,0.5,0.3}

\definecolor{purple}{HTML}{DA70D6}

\definecolor{red}{HTML}{DC143C}

\definecolor{brown}{HTML}{B8860B}

\shorttitle{Measuring the ISM content of nearby, luminous QSOs}
\shortauthors{}

\begin{document}

\title{Measuring the ISM Content of Nearby, Luminous, Type 1 and Type 2 QSOs through CO and [\ion{C}{2}]}

\correspondingauthor{Yuanze Luo}
\email{yluo37@jhu.edu}

\author[0000-0002-0696-6952]{Yuanze Luo}
\affiliation{William H. Miller III Department of Physics and Astronomy, Johns Hopkins University, Baltimore, MD 21218, USA}

\author[0000-0003-4030-3455]{A. O. Petric}
\affiliation{William H. Miller III Department of Physics and Astronomy, Johns Hopkins University, Baltimore, MD 21218, USA}
\affiliation{Space Telescope Science Institute, 3700 San Martin Dr, Baltimore, MD 21218, USA}

\author[0000-0001-9122-9668]{R.M.J. Janssen}
\affiliation{Jet Propulsion Laboratory, California Institute of Technology, Pasadena, CA 91109, USA}

\author[0000-0002-3698-7076]{D. Fadda}
\affiliation{Space Telescope Science Institute, 3700 San Martin Dr, Baltimore, MD 21218, USA}

\author[0000-0002-8763-1555]{N. Flagey}
\affiliation{Space Telescope Science Institute, 3700 San Martin Dr, Baltimore, MD 21218, USA}

\author[0000-0002-4721-3922]{A. Omont}
\affiliation{Sorbonne Université, UPMC Université Paris 6 and CNRS, UMR 7095, Institut d’Astrophysique de Paris, France}

\author[0000-0001-7836-3425]{A. M. Jacob}
\affiliation{William H. Miller III Department of Physics and Astronomy, Johns Hopkins University, Baltimore, MD 21218, USA}

\author[0000-0001-7883-8434]{K. Rowlands}
\affiliation{William H. Miller III Department of Physics and Astronomy, Johns Hopkins University, Baltimore, MD 21218, USA}
\affiliation{AURA for ESA, Space Telescope Science Institute,
3700 San Martin Drive, Baltimore, MD 21218, USA}

\author[0000-0002-4261-2326]{K. Alatalo}
\affiliation{William H. Miller III Department of Physics and Astronomy, Johns Hopkins University, Baltimore, MD 21218, USA}
\affiliation{Space Telescope Science Institute, 3700 San Martin Dr, Baltimore, MD 21218, USA}

\author[0000-0003-3429-3836]{N. Billot}
\affiliation{Université de Genève, Department of Astronomy, Chemin Pegasi 51/b, 1290 Versoix, Switzerland}

\author[0000-0001-6670-6370]{T. Heckman}
\affiliation{William H. Miller III Department of Physics and Astronomy, Johns Hopkins University, Baltimore, MD 21218, USA}
\affiliation{School of Earth and Space Exploration, Arizona State University, Tempe, AZ 85287-1404, USA}

\author[0000-0003-2901-6842]{B. Husemann}
\affiliation{European Organisation for the Exploitation of Meteorological Satellites, Eumetsat Allee 1, 64295 Darmstadt, Germany}

\author[0000-0002-2603-2639]{D. Kakkad}
\affiliation{Space Telescope Science Institute, 3700 San Martin Dr, Baltimore, MD 21218, USA}

\author[0000-0002-3032-1783]{M. Lacy}
\affiliation{National Radio Astronomy Observatory, Charlottesville, VA 22903, USA}

\author{J. Marshall}
\affiliation{Department of Astronomy and Astrophysics, University of California, Santa Cruz, CA 95064, USA}

\author[0000-0002-1261-6641]{R. Minchin}
\affiliation{National Radio Astronomy Observatory, P.O. Box O, Socorro, NM 87801, USA}

\author{R. Minsley}
\affiliation{Steward Observatory, University of Arizona, Tucson, AZ 85721, also Department of Planetary Sciences, USA}

\author[0000-0001-5783-6544]{N. Nesvadba}
\affiliation{Universit\'e C\^ote d'Azur, Observatoire de la C\^ote d'Azur,
  CNRS, Laboratoire Lagrange, Bd de l'Observatoire, CS 34229, 06304
  Nice cedex 4, France}

\author[0000-0003-3191-9039]{J. A. Otter}
\affiliation{William H. Miller III Department of Physics and Astronomy, Johns Hopkins University, Baltimore, MD 21218, USA}

\author[0000-0002-9471-8499]{P. Patil}
\affiliation{William H. Miller III Department of Physics and Astronomy, Johns Hopkins University, Baltimore, MD 21218, USA}

\author[0000-0001-6746-9936]{T. Urrutia}
\affiliation{Leibniz-Institut für Astrophysik, Potsdam (AIP), An der Sternwarte 16, 14482 Potsdam, Germany}


\begin{abstract}
We present observations of CO(1--0) and CO(2--1) lines from the Institut de radioastronomie millim\'etrique (IRAM) 30m telescope toward 20 nearby, optically luminous type 2 quasars (QSO2s) and observations of [\ion{C}{2}] 158$\mu$m line from the Stratospheric Observatory For Infrared Astronomy (SOFIA) for 5 QSO2s in the CO sample and 5 type 1 quasars (QSO1s). In the traditional evolutionary scenario explaining different types of QSOs, obscured QSO2s emerge from gas-rich mergers observed as luminous infrared galaxies (LIRGs) and then turn into unobscured QSO1s as the black holes clear out the obscuring material in a blow-out phase. We test the validity of this theoretical prediction by comparing the gas fractions and star formation efficiencies among LIRGs and QSOs. We find that CO luminosity, CO-derived gas masses and gas fractions in QSO1s are consistent with those estimated for QSO2s, while LIRGs exhibit a closer resemblance to QSO2s in terms of CO-derived gas masses and gas fractions. Comparisons between [\ion{C}{2}] luminosity and star formation tracers such as the CO and infrared luminosity imply additional sources of [\ion{C}{2}] emission in QSO1s likely tracing neutral atomic or ionized gas with the caveat of a small sample size. All three types of galaxies have statistically indistinguishable distributions of star formation efficiency. Our results are consistent with part of the evolutionary scenario where nearby QSO2s could emerge from LIRGs, but they may not be the precursors of nearby QSO1s.

\end{abstract}


\section{Introduction} \label{sec:intro}
Accreting massive black holes at galaxy centers are called active galactic nuclei (AGN), the brightest of which are known as quasars (QSOs), which vastly outshine their host galaxies. AGN and QSOs are classified into two types based on their spectral features: broad-line type 1 QSOs (QSO1s, unobscured black hole), and narrow-line type 2 QSOs (QSO2s, obscured black hole). There have been two leading theories for the relationship between QSO1s and QSO2s. The orientation theory \citep[e.g.,][]{Antonucci_1993} postulates that QSO1s and QSO2s both contain dusty tori, but are viewed at different orientations. 

The other theory postulates an evolutionary scenario, where mergers give rise to QSO2s which then evolve into QSO1s \citep[e.g.,][]{Hopkins_2008,Alexander_2012}. This theory is based on the fact that some quasar activity can be triggered by gas-rich mergers \citep[e.g.,][]{Sanders_1988b,Sanders_1988,Sanders_1996,Urrutia_2008,petric2011,Glikman_2015,elli2019}. At higher redshift ($z \gtrsim 1$), studies have discovered that the farthest, most luminous, and dustiest of QSOs\footnote{Hereafter QSOs refer to their host galaxies.} bear the marks of gravitational interactions from tidal tails to complex nuclear structures\footnote{Although see \citet{Villforth_2017} for a counter-argument.} \citep[e.g.,][]{Urrutia_2008,Glikman_2012,Glikman_2015,Lacy_2018,Kakkad_2023}. \citet{Urrutia_2008} used the \textit{Hubble Space Telescope} and ground-based adaptive optics to image obscured AGN to separate the emission from the nucleus and the host, and found that the hosts of obscured AGN tend to be major mergers. The observations in \citet{Glikman_2012,Glikman_2015} are particularly telling: at the peak epoch for galaxy and black hole growth, the most luminous QSOs are also the most dust reddened, and their host galaxies are major mergers. These observations along with simulations \citep[e.g.,][]{Hopkins_2006} prompt the evolutionary link between mergers, QSO2s, and QSO1s: the merger and associated starburst stir up the gas and dust near the black hole, triggering quasar activity that gives rise to QSO2s; the black hole emerges as QSO1s only when its winds and UV emission clear the obscuring dust and gas from the nuclear region, e.g., in a ``blow-out phase" (see Fig.1 in \citealt{Hopkins_2006}). This evolutionary scenario predicts a decrease in the gas fraction from mergers to QSO2s and then to QSO1s. If AGN produce negative feedback on the star formation (SF) of the host galaxies, the star-formation efficiency (SFE) is also expected to decrease from mergers to QSO1s. These changes should be observable on a galaxy-wide scale as simulations show that the processes that shape the evolution of galaxies and their central SMBH are multi-scale and multiphase \citep[e.g.,][]{Harrison_2024}.

At $z\leqslant0.5$ some QSO2s are consistent with being differently oriented QSO1s expected from traditional, orientation-based unification models \citep[e.g.,][]{Zakamska_2005}, while other QSO2s appear to be intrinsically different. Relative to QSO1s, these QSO2s which appear to be intrinsically different seem to have higher star formation rates\footnote{After accounting for extinction and AGN contribution via photoionization calculation and multiwavelength data (e.g., PAH features and FIR luminosity). See references in the text for more details.} \citep[SFR; e.g.,][]{Kim_2006,Lacy_2007,Zakamska_2008} and more chaotic ionized gas velocity fields \citep[e.g.,][]{Greene_2011}. These properties are qualitatively consistent with these QSO2s being the evolutionary precursors to QSO1s, formed during the turbulent aftermath of gas-rich mergers.  

As many as $\sim$90\% of nearby luminous infrared galaxies (LIRGs) are gas-rich mergers \citep[e.g.,][]{stier2013, larson2016, petric2018}. Gas-rich mergers in general and LIRGs, in particular, have been proposed as precursors of QSOs \citep{Sanders_1988b,Sanders_1988,Sanders_1996,hopkins2008}. A progression from gas-rich mergers (e.g., LIRGs) to QSO2s would imply higher gas fractions in LIRGs relative to QSO2s. If feedback from the AGN effectively impacts the molecular gas and hence the host galaxy's ability to make new stars, we would expect QSO1s to have lower gas fractions and SFE than QSO2s. Therefore, comparing the interstellar medium (ISM) in LIRGs, QSO2s, and QSO1s can provide valuable insights into the evolutionary link between them. As the second most abundant molecule in space, CO via observations of its $J$ = 1--0 line in emission has traditionally been used as a molecular gas tracer and in turn used to infer the total H$_2$ gas mass in both the Milky Way and external galaxies \citep[e.g.,][]{Bolatto_2013}. However, there is increasing evidence for the existence of molecular gas not traced by CO \citep{Grenier_2005}, the ``CO-dark" H$_2$. The properties of the CO-dark H$_2$ gas have been widely studied via both observations of the [\ion{C}{2}] emission \citep{Pineda_2013,Madden_2013,Hall_2020, Fadda2021, Liszt2023} and simulations \citep[e.g.,][]{Li2018}.

The [\ion{C}{2}] 158$\mu$m fine structure line of singly ionized carbon is one of the major coolants of neutral gas in the ISM and photodissociation regions \citep[PDRs;][]{Wolfire_2003,Narayanan2016}. The [\ion{C}{2}] emission primarily arises as a result of FUV radiation from O and B stars ionizing carbon in the molecular clouds of PDRs. While the abundance of carbon is predominantly locked into CO in dense regions of molecular clouds that are well-shielded from FUV radiation, the [\ion{C}{2}] emission arises from multiple layers of PDRs but most likely originates from the outer, more diffuse gas layers with overall lower molecular fractions. Additionally, high radiation fields associated with AGN can lead to photoionization, heating, and compression of the gas. Radiation pressure can increase the density of ionized gas, which in turn may enhance the [\ion{C}{2}] emission. Significant gas turbulence from jets or galaxy mergers could also impact the distribution and thus the density of the gas, affecting both local and global [\ion{C}{2}] emission \citep[e.g.,][]{SmirnovaPinchukova_2019}. Thus [\ion{C}{2}] provides a complementary probe of the ISM to CO which primarily traces denser gas that is physically related to star formation \citep{Narayanan2016,garcia2023}.

In this paper, we present observations of the CO(1--0) and (2--1) lines using the Institut de radioastronomie millim\'etrique (IRAM) 30m telescope for 20 nearby, optically luminous QSO2s that have existing FIR observations from the \textit{Herschel Space Observatory} \citep{Pilbratt_2010}. We also present [\ion{C}{2}] 158$\mu$m observations from the Stratospheric Observatory For Infrared Astronomy \citep[SOFIA;][]{Young_2012}, of unprecedented sensitivity, for 5 QSO2s from the CO sample as well as for 5 QSO1s from the Palomar–Green (PG) bright quasar survey \citep{Schmidt_1983,Boroson_1992}. Combined with observations of nearby LIRGs (e.g., the Great Observatories All-sky Survey, \citealt{Armus_2009}) from the literature, our data permits a meaningful comparison in the CO and [\ion{C}{2}] observations of local QSOs and LIRGs. The SOFIA data will assess the utility of the [\ion{C}{2}] cooling line as a SF tracer compared to other tracers such as the CO and FIR luminosity. In addition, the ability of [\ion{C}{2}] and CO to trace gas in different phases offers a unique opportunity to uncover the similarities and differences in the ISM compositions within our samples. Comparison of [\ion{C}{2}] between QSOs and LIRGs with varying degrees of AGN prominence could also shed light on the extent to which high radiation fields and gas turbulence could affect FIR emission lines. Given the scarcity of CO and [\ion{C}{2}] observations of nearby QSOs for which we have optical spectroscopy, our data form the basis for a benchmark sample of nearby QSOs to compare with QSOs at higher redshifts.

AGN are believed to dramatically affect the gas within their hosts and play an important role in suppressing SF \citep[e.g.,][]{Kakkad_2017}. Yet direct observational evidence for AGN quenching remains somewhat elusive. Multi-wavelength studies of AGN are necessary to gain a more comprehensive understanding of the effects of AGN on their host environments. The galaxies in our sample are chosen specifically to have data across multiple wavelengths from the mid-infrared (MIR) to the far-infrared (FIR) regimes. The combination of our data with archival data sets will help compare the physical properties of the ISM (e.g., molecular gas mass M$_{\rm{H_2}}$ and molecular gas fraction M$_{\rm{H_2}}$/M$_{*}$) and SFE between QSO1s, QSO2s, and LIRGs, with the aim of gleaning a more global understanding of how the ISM conditions differ in different types of QSOs and shed light on the evolutionary link between QSOs and LIRGs. We note that the evolutionary scenario described above is framed on a global, galaxy-wide scale. ISM conditions at smaller spatial scales could be more complex and often vary case by case. In this study, we aim to compare the ISM across different galaxy populations and evaluate the proposed evolutionary scenario from a global perspective. Additionally, spatially resolved data will offer deeper insights into the origins of the observed emissions, as elaborated in the discussion sections.

This paper is organized as follows. We present our sample selection and observation details in \S2. The results of our observations (detection of molecular gas and derived physical quantities) are presented in \S3. We compare observed and derived ISM properties across different samples and explore the implications of our study in \S4. We summarize our conclusions in \S5 and provide ancillary data gathered from the literature in \S6. Throughout this paper we assume the cosmological parameters $H_0 = 70 \  \mathrm{km\ s^{-1}\ Mpc^{-1}}$, $\Omega_m = 0.3$, and $\Omega_{\Lambda} = 0.7$.

\begin{figure*}
\plotone{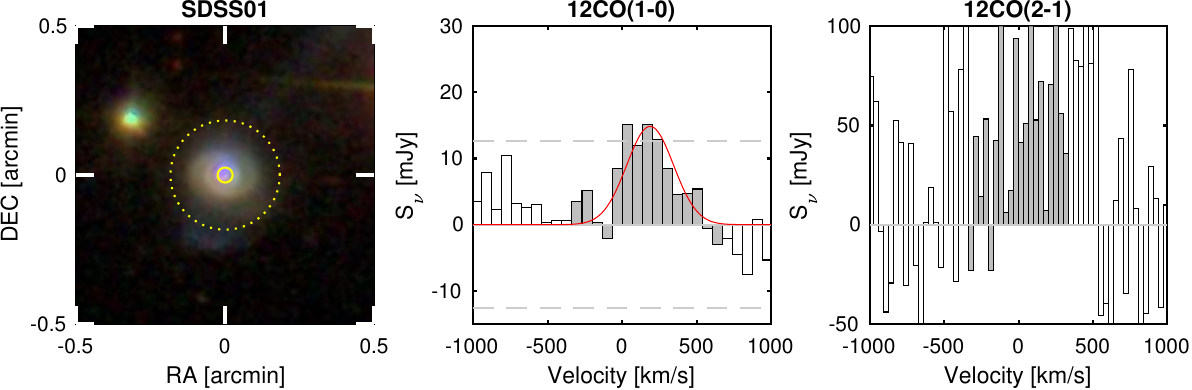}
\caption{\textbf{Left}: $60\arcsec \times 60\arcsec$ SDSS ($gri$) cutouts of the QSO2s. The solid and dotted yellow circles show the extent of the $3\arcsec$ SDSS spectroscopy fiber and the $22\arcsec$ IRAM beam at 3 mm, respectively. \textbf{Middle}: CO(1--0) spectra in units of mJy. \textbf{Right}: CO(2--1) spectra in units of mJy. For both spectra, the x-axis indicates the velocity offset of the line from the systemic velocity of the galaxy as determined by the optical spectrum. Spectra are shown at 25 MHz resolution and were taken from the WILMA backend. The dark grey shaded area indicates the CO line region used for the determination of $I_{\mathrm{CO}}$ and the red line is the Gaussian fit. Horizontal grey lines indicate the baseline (solid) and the $3\sigma$ scatter around the baseline (dashed). The complete figure set (20 images) is available in the online journal. \label{fig:COspectra}}
\end{figure*}

\section{Sample and Data} \label{sec:data}

\citet{Shangguan_2019} selected 87 QSO2s from the Sloan Digital Sky Survey \citep[SDSS;][]{Reyes_2008} that match the PG QSO1s in both redshift and [\ion{O}{3}] $\lambda$5007 luminosity (Fig.1 in \citealt{Shangguan_2019}, see also \citealt{Petric_2015,Zakamska_2016}). Both samples of QSO2s and QSO1s satisfy the historical criterion of $M_B<$-23 mag for quasars \citep{Schmidt_1983}, representing the extremely luminous end of accreting black holes. These closely matched QSO2s and QSO1s make an excellent sample for studying the differences in the ISM properties between the two types of QSOs. To obtain information on the ISM, we select the 20 most nearby QSO2s from \citet{Shangguan_2019} to observe with IRAM for CO emission. We also select the 5 nearest QSO2s from our CO sample and the 5 nearest PG QSO1s observable within the SOFIA atmospheric window for [\ion{C}{2}] observations. A complete list of our targets and observations are tabulated in Table \ref{tab:CO10observations}, \ref{tab:CO21observations}, and \ref{tab:CIIobservations}. Details about our observations and data reduction are presented in the following sections.

\subsection{IRAM CO Observations and Data Reduction}
We observe our sample of QSO2s with the IRAM 30m telescope, in Pico Veleta, Spain, during two observation periods: 14--20 August 2012 and 8--10 January 2013 (PI: Petric). The Eight MIxer Receiver (EMIR) instrument was used in wobbler switching mode with a reference position offset of 120\arcsec\ at a frequency of 1 Hz or 0.8 Hz. We take advantage of the 32 GHz wide available bandwidth to simultaneously observe the CO(1--0) and CO(2--1) using the E090 and E230 EMIR bands with the Wideband Line Multiple Autocorrelator (WILMA) backend. 

The CO(1--0) line is anticipated at 106.153 $\leq \nu_{\rm CO(1-0)} \leq$ 111.158 GHz and the CO(2--1) line 212.301 $\leq \nu_{\rm CO(2-1)} \leq$ 222.312 GHz at our targets' optical redshifts. Local oscillator (LO) frequencies were chosen to place the CO(1--0) of multiple targets in the Upper Inner sideband of EMIR. This reduces the overheads to 1 or 2 LO tunings per 12 hours of observation. The anticipated CO line was kept at least 400 MHz from the band edge to allow for base-lining. We take the data in eight-minute scans comprised of twelve 30-second long sub-scans. Observations are interspersed with instrument calibrations every 2--3 scans and pointing calibrations every 1--2 hours on sources from the IRAM Flux Monitoring Program.  Telescope refocusing is carried out approximately every 2--4 hours and following sunrise and sunset preferentially on Mars, Venus, Saturn, Jupiter, or the QSO 0316+413 (3C 84). If these were not available or very distant, bright sources from the IRAM Flux Monitoring Program were used for refocusing. SDSS10 is observed with position switching mode because we experienced problems with the wobbler switching modes.

We use the standard data reduction software, the Multichannel Imaging and Calibration Software for Receiver Arrays (\texttt{MIRA}), to calibrate each science scan using the instrument calibration scan taken closest in time. The data are then reduced using the Continuum and Line Analysis Single-dish Software (\texttt{CLASS}\footnote{Part of the GILDAS software package \url{https://www.iram.fr/IRAMFR/GILDAS/}.}). 
A first-order polynomial baseline was subtracted from each individual scan. Baselines are calculated using a linear fit after exclusion of the anticipated CO-line frequencies $\frac{\nu_{\mathrm{CO}}}{1+z} \pm 400$ MHz. All individual scans are coadded using weights based on their RMS baseline fluctuations $\sigma\left(T_A^*\right)$, although using weighting based on observation time $t_{\mathrm{obs}}$ gives negligible changes to the results $(\Delta \sigma\left(T_A^*\right) < 3$\%).

Following \citet{French_2015}, line intensities are estimated by summing over the line profile at velocities $v_0 \pm 3\sigma(v)$, where $v_0$ and $\sigma(v)$ are the mean and standard deviation of a Gaussian profile fit to the CO line intensity $I_{\rm{CO}}$ at a spectral resolution of 25 MHz. While not all lineshapes are Gaussian, this method estimates an appropriate integration interval systematically. We calculate the error in the integrated CO line intensity as
\begin{equation}
\sigma_I^2 = (\Delta v)^2 \sigma^2 N_l \left(1+\frac{N_b}{N_l} \right),
\end{equation}
where $\Delta v = 71$ km/s is the channel velocity width, $\sigma$ is the channel RMS noise, $N_l$ the number of channels used to integrate over the CO-line, and $N_b$ the number of channels used to fit the baseline (typically 32). We also take into account an estimated flux calibration error of 10\%. For non-detections, we calculate upper limits on $I_{\rm{CO}}$ as $<3\sigma_I$.

From the line intensity ($I_{\rm{CO}}$ in $\mathrm{K \ km/s \ pc^2}$)\footnote{$I_{\rm{CO}}$ and $L^{\prime}_{\rm{CO}}$ are computed assuming the main-beam temperature $T_{\rm mb} = 1.33T_{A}^{*}$ on \url{https://www.iram.fr/IRAMFR/ARN/jan95/node46.html}.} we calculate the CO line luminosity $L^{\prime}_{\rm{CO}}$ (in $\mathrm{K \ km/s \ pc^2}$) following \citep{Solomon_1997}
\begin{equation}
L^{\prime}_{\rm CO} = 23.5 \Omega_{s*b} D_L^2 I_{\rm CO} (1+z)^{-3},
\end{equation}
where $z$ is the redshift from \citet{Reyes_2008} and $D_L$ is the corresponding luminosity distance in Mpc. $\Omega_{s*b}$ is the solid angle of the source convolved with the beam:
\begin{equation}
 \Omega_{s*b} = \frac{\pi(\theta_s^2+\theta_b^2)}{4\ln(2)},
\end{equation}
where $\theta_s$ and $\theta_b$ are the half-power beam widths of the source and beam, respectively. Given the size of our sources with respect to $\theta_b$ (see Fig.\ref{fig:COspectra}), we assume that the beam is much larger than the source size, i.e., $ \Omega_{s*b} \approx  \Omega_{b}$. The final spectra of our observed sources alongside image cutouts from SDSS are shown in Fig.\ref{fig:COspectra}. The derived CO luminosities are summarized in Tables \ref{tab:CO10observations} and \ref{tab:CO21observations}.

\begin{figure*}
\centering
\includegraphics[width=0.46\textwidth]{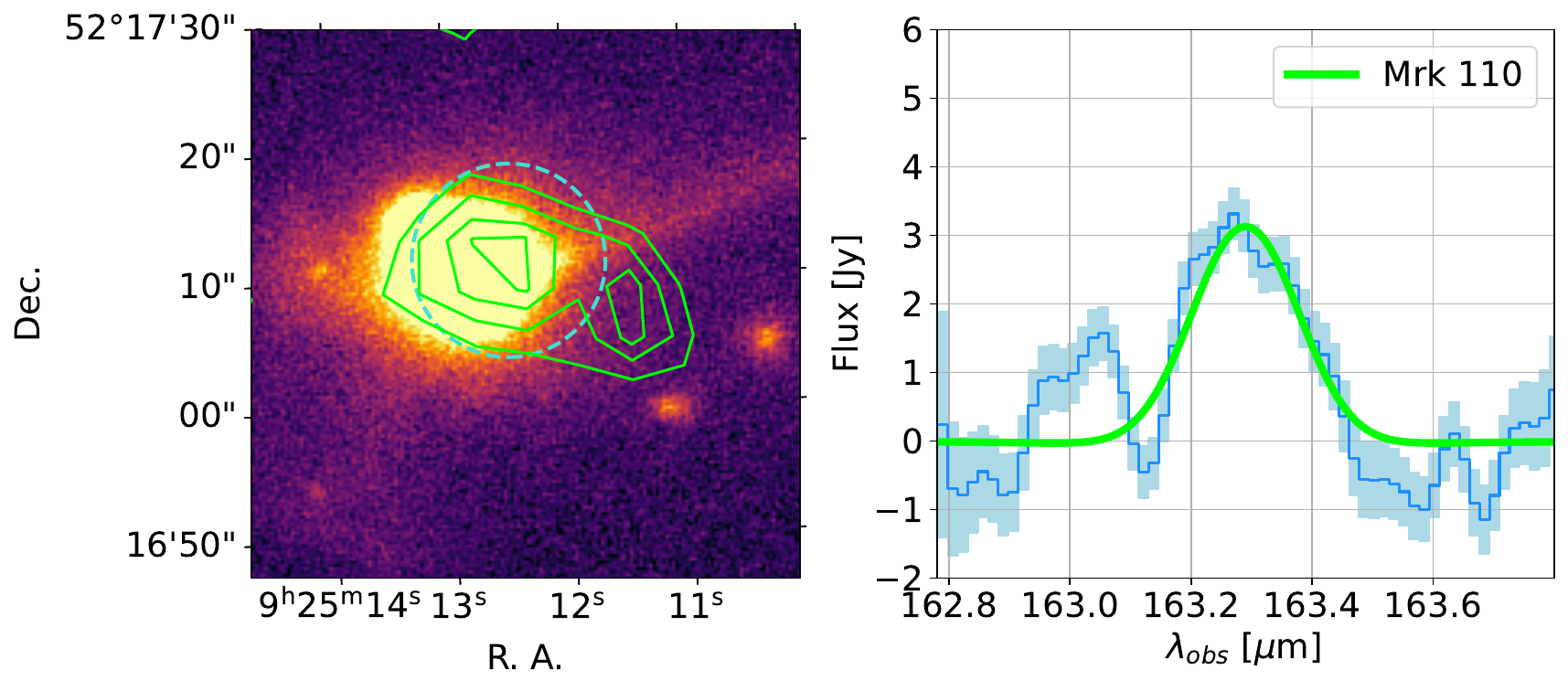}
\includegraphics[width=0.46\textwidth]{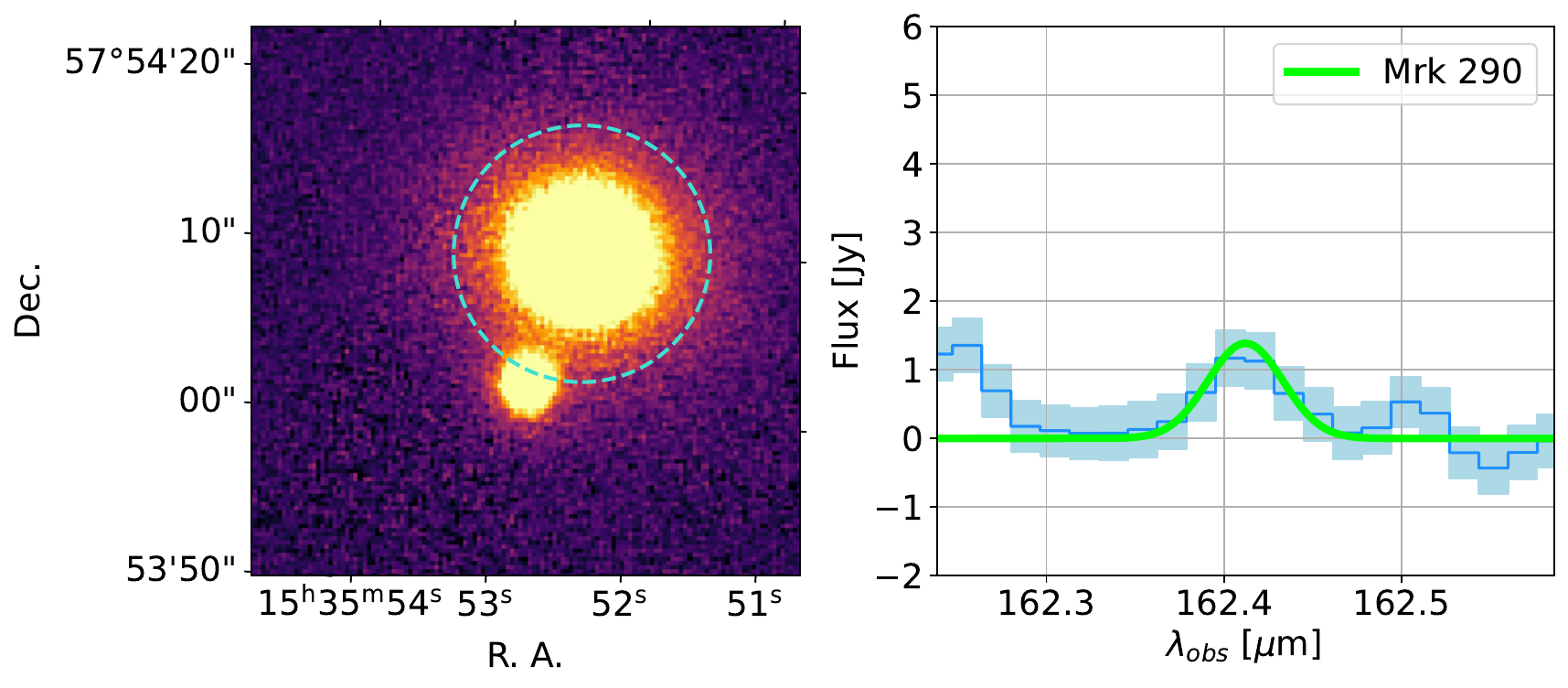}
\includegraphics[width=0.46\textwidth]{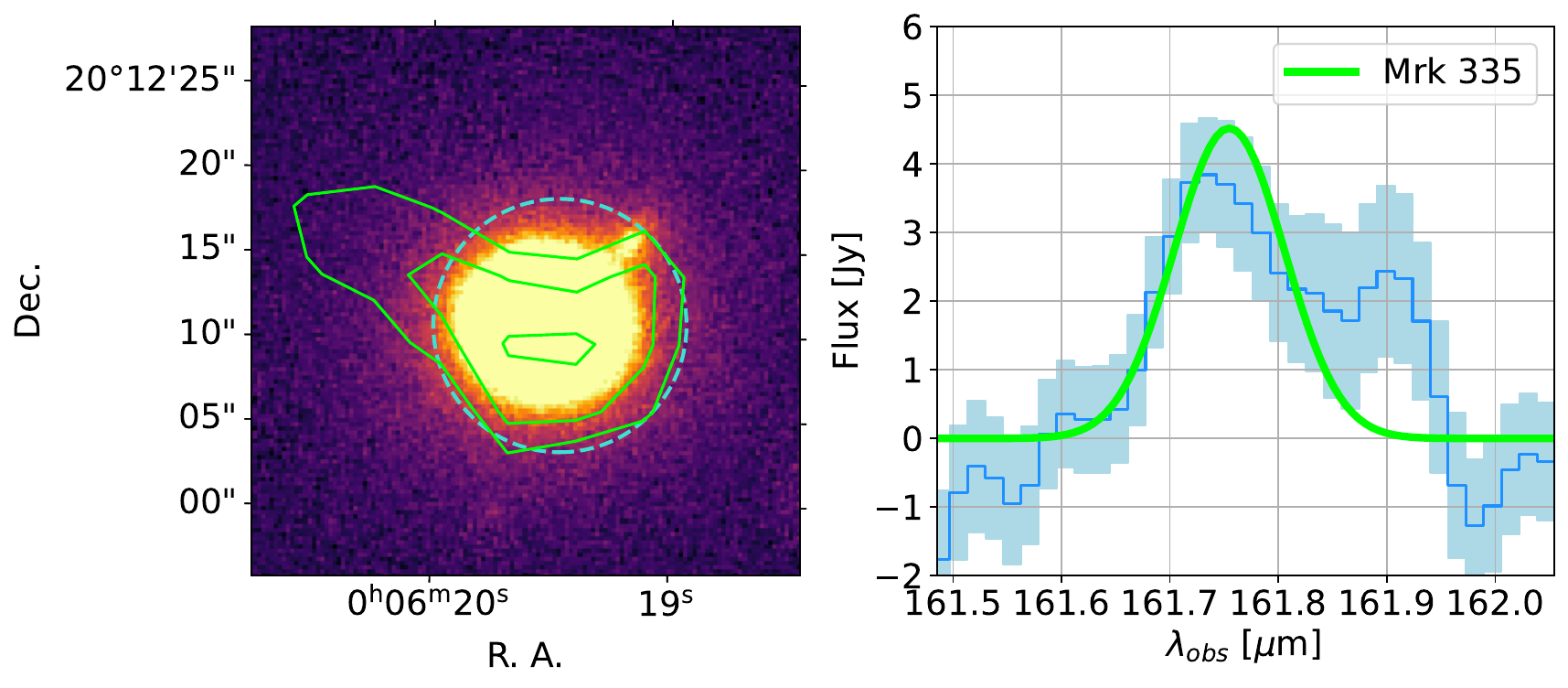}
\includegraphics[width=0.46\textwidth]{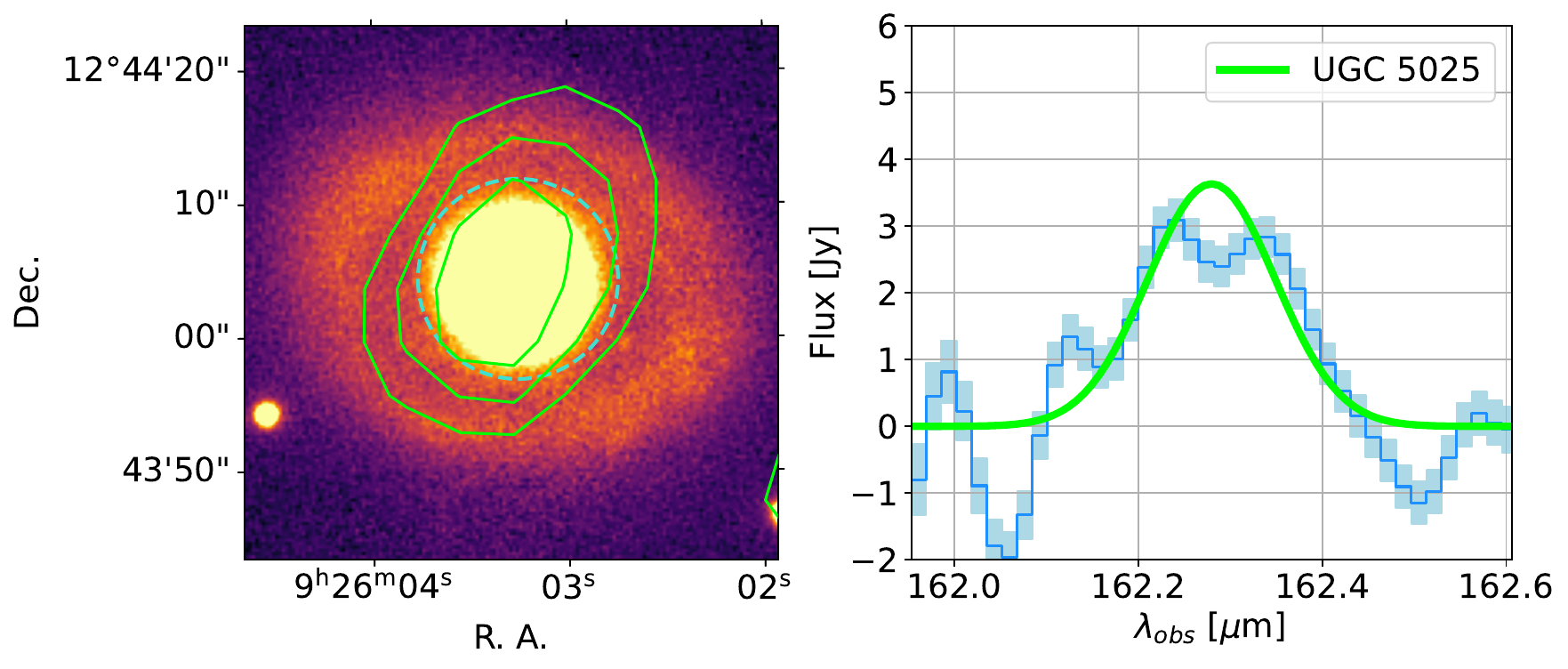}
\includegraphics[width=0.46\textwidth]{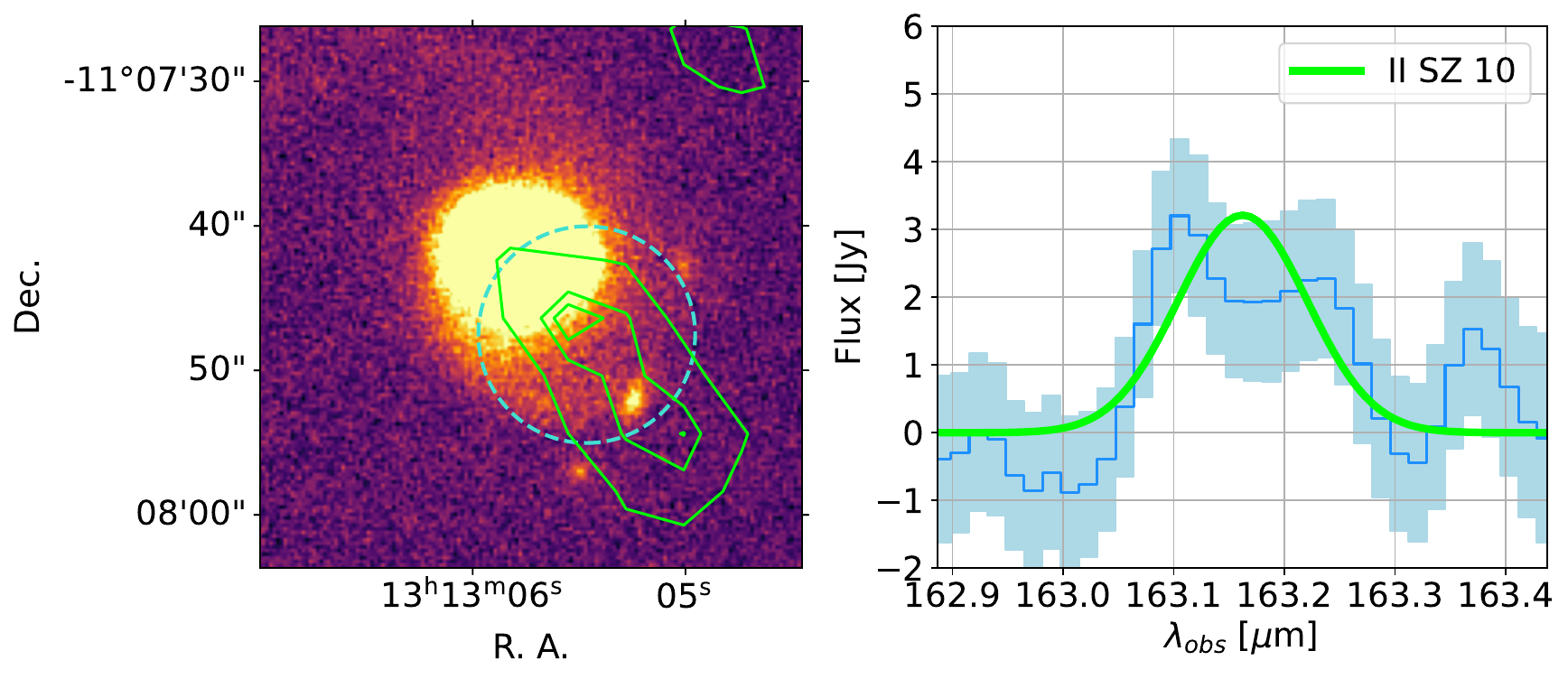}
\includegraphics[width=0.46\textwidth]{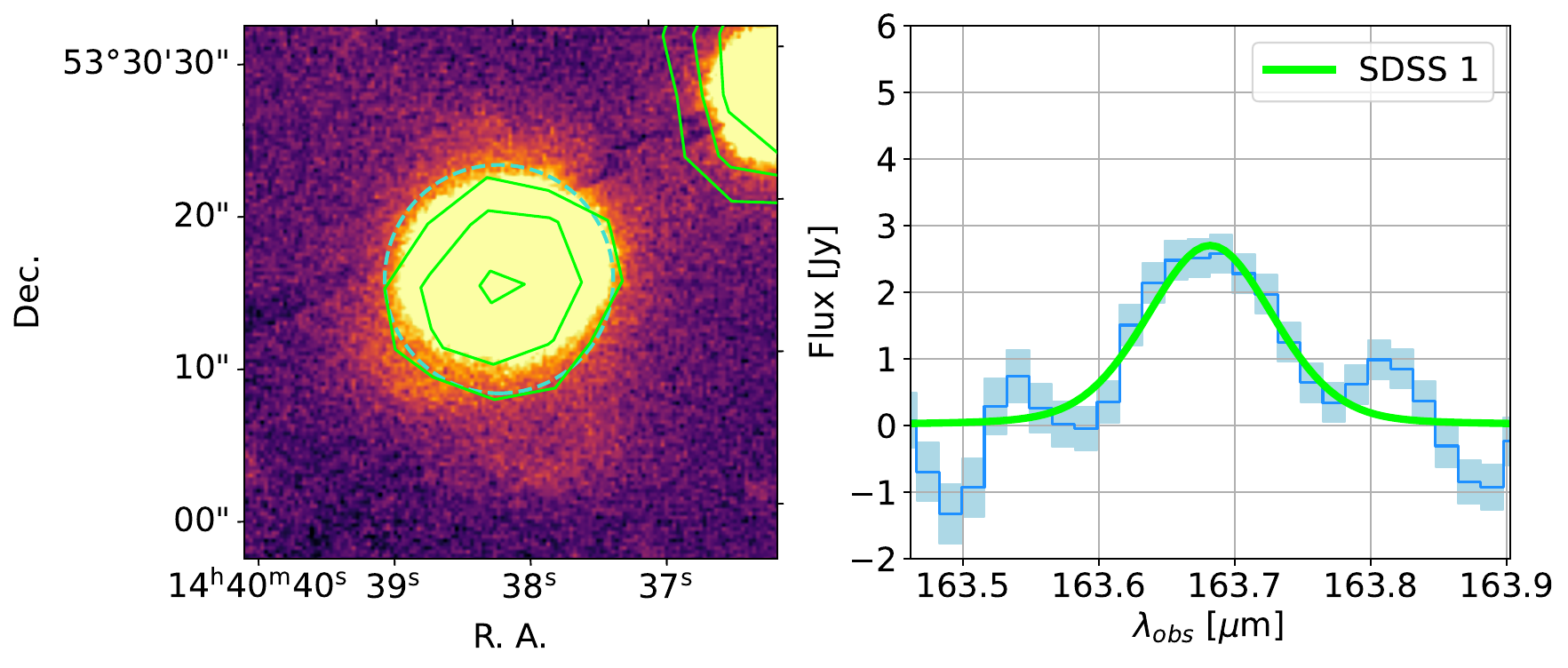}
\includegraphics[width=0.46\textwidth]{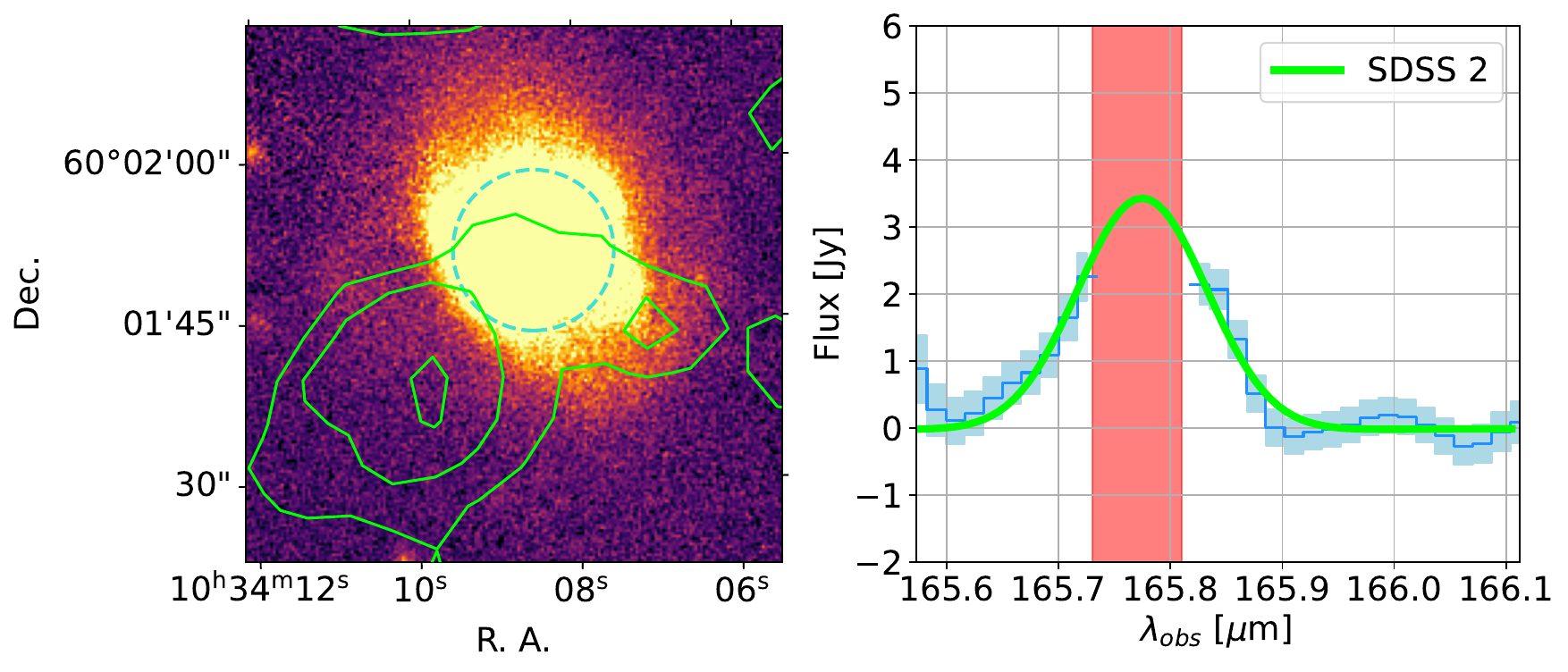}
\includegraphics[width=0.46\textwidth]{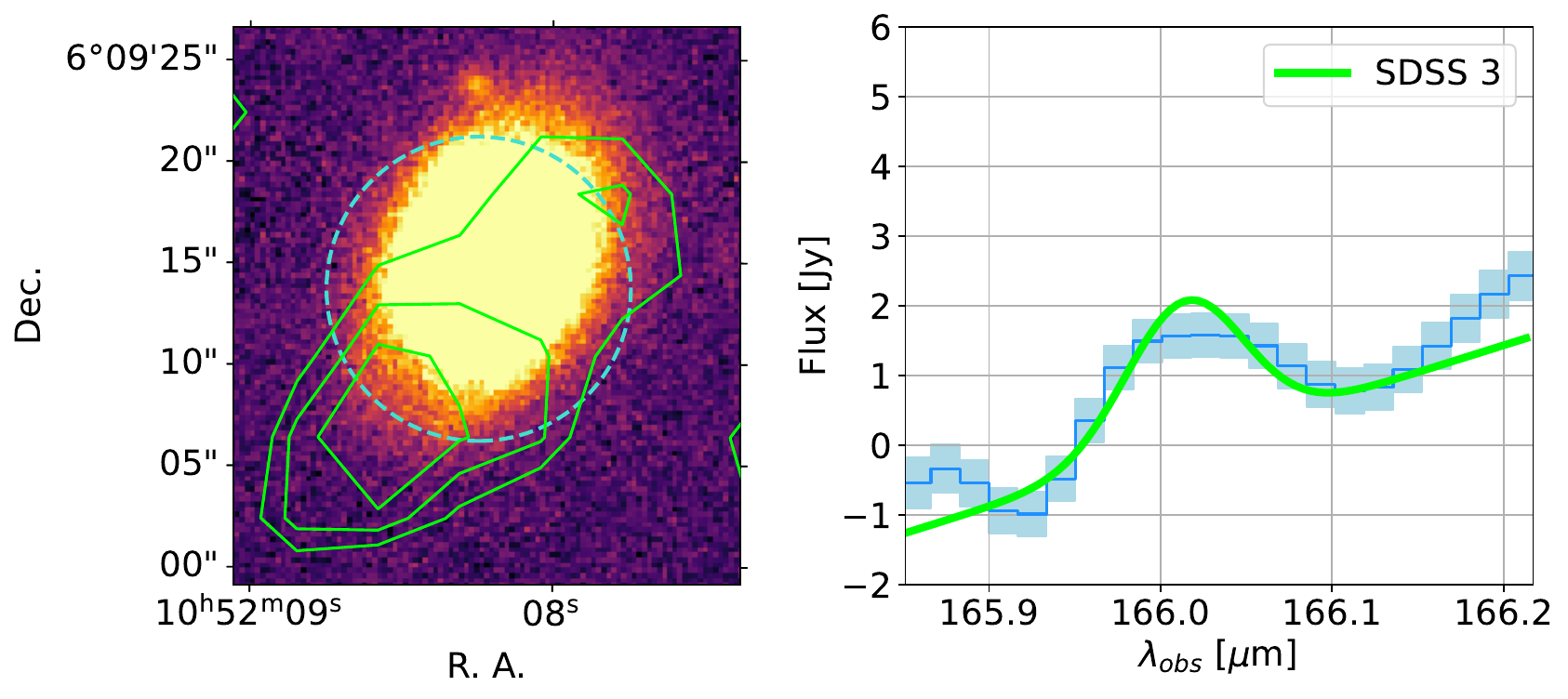}
\includegraphics[width=0.46\textwidth]{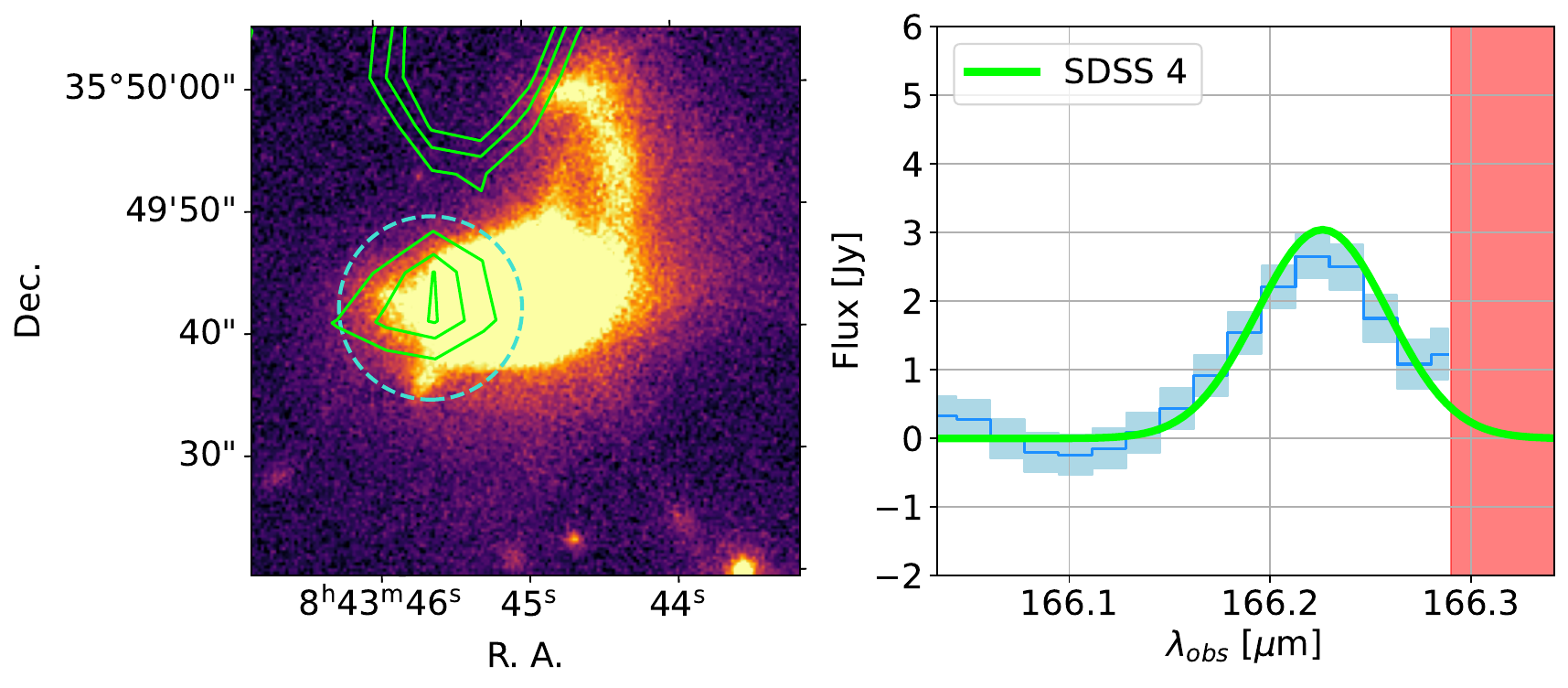}
\includegraphics[width=0.46\textwidth]{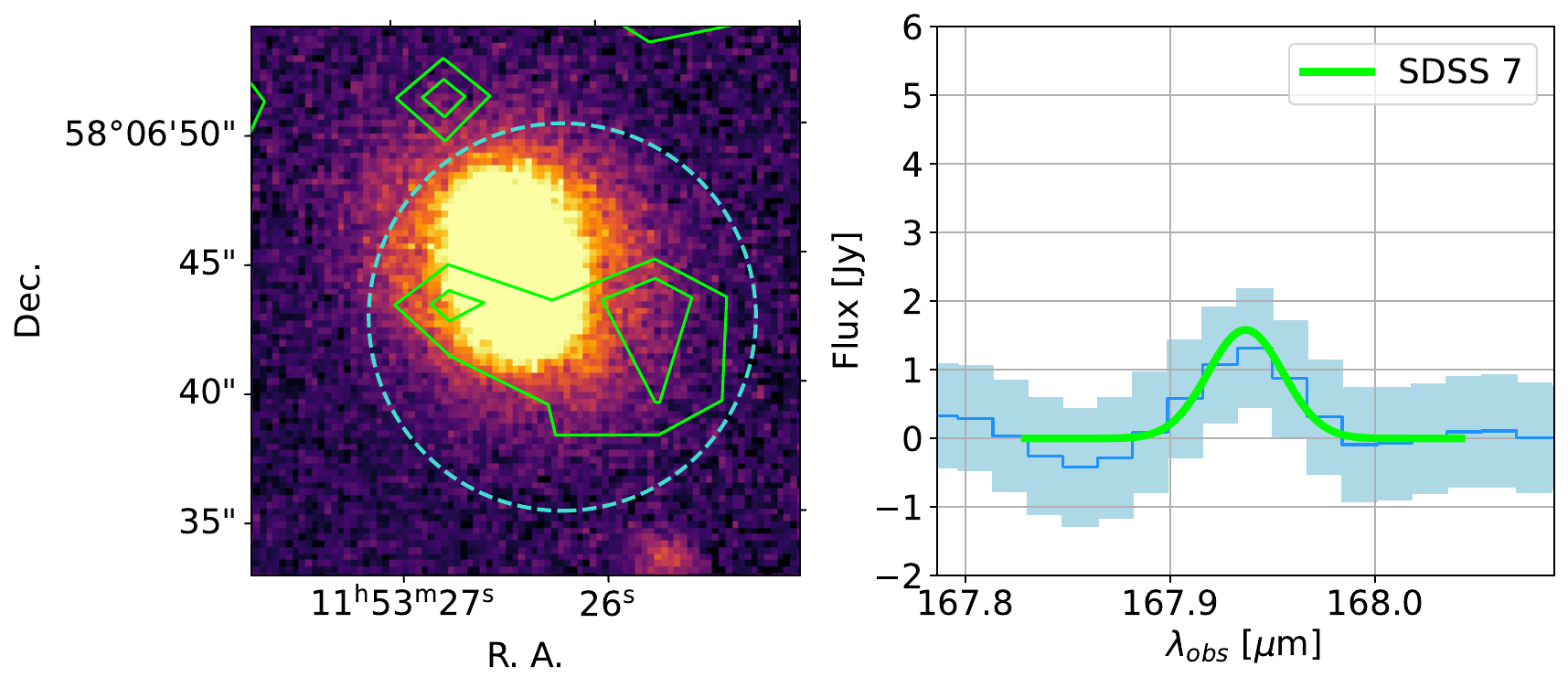}
\caption{[\ion{C}{2}] observations from SOFIA/FIFI-LS. Contours (solid lines) of the total [\ion{C}{2}] intensity are overlaid on PanSTARRS $i$-band images. The aperture (dashed circle) used to extract the spectrum has a diameter of 15.2\arcsec, corresponding to the FWHM of the FIFI-LS beam at this frequency. Spectra and their 3-$\sigma$ errors are shown as a blue line and a light blue band. The red vertical band in the spectra indicates the masked region due to strong atmospheric absorption. The green curves correspond to the fits of the [\ion{C}{2}] lines. The continuum baseline was modeled by a line with a slope for SDSS03.
\label{fig:SOFIA_detections}}
\end{figure*}

\subsection{SOFIA [\ion{C}{2}] Observations and Data Reduction}
The SOFIA Far Infrared Field-Imaging Line Spectrometer \citep[FIFI-LS;][]{Fischer_2018,Colditz_2018} observations presented here are observed during multiple flights scheduled across multiple campaigns between August 2020 and February 2023 with one-hour on-source integration except for the FIR luminous QSO1 UGC 05025, for which the total on-source integration is only 30 minutes. We use symmetric chop nodding centering the red channel of FIFI-LS on the [\ion{C}{2}] 158$\mu$m line. We center the blue channel on the [\ion{O}{3}] 88$\mu$m line when it does not coincide with telluric features. Several sources are observed in multiple observing runs due to weather or technical issues. 

The raw data are downloaded and reduced using the FIFI-LS pipeline \citep{FifiPL} and the latest calibration described in \citet{Fadda2023}. The reduction includes flagging bad pixels, removing scans affected by bad pointings or bad atmospheric transmission, and correcting for the atmospheric absorption using precipitated water vapor values evaluated during the flights \citep[see ][]{Iserlohe2021}. In particular, for lines close to a broad telluric absorption line, we apply the atmospheric transmission correction at the center of the line to avoid overcorrecting the line wings.

We then use the SOFIA package \texttt{SOSPEX} \citep{Fadda2018} to fit the [\ion{C}{2}] 158$\mu$m and [\ion{O}{3}] 88$\mu$m lines within an aperture corresponding to the PSF FWHM at the line wavelengths (8.8\arcsec and 15.2\arcsec diameter for [\ion{O}{3}] 88 $\mu$m and [\ion{C}{2}] 158 $\mu$m, respectively). Following \citet{Fadda2023}, the spectra were produced by subtracting an off-source observation from the on-source observation, and the line intensities were estimated from fitting the [\ion{C}{2}] and [\ion{O}{3}] lines with pseudo-Voigt and Gaussian profiles, respectively. In the case of extended emission, we also measure [\ion{C}{2}] 158 $\mu$m emission across the host galaxy. Our flux measurements are summarized in Table \ref{tab:CIIobservations} and the [\ion{C}{2}] spectra are shown in Fig.\ref{fig:SOFIA_detections}.

\begin{figure}
\centering
\includegraphics[width=\columnwidth]{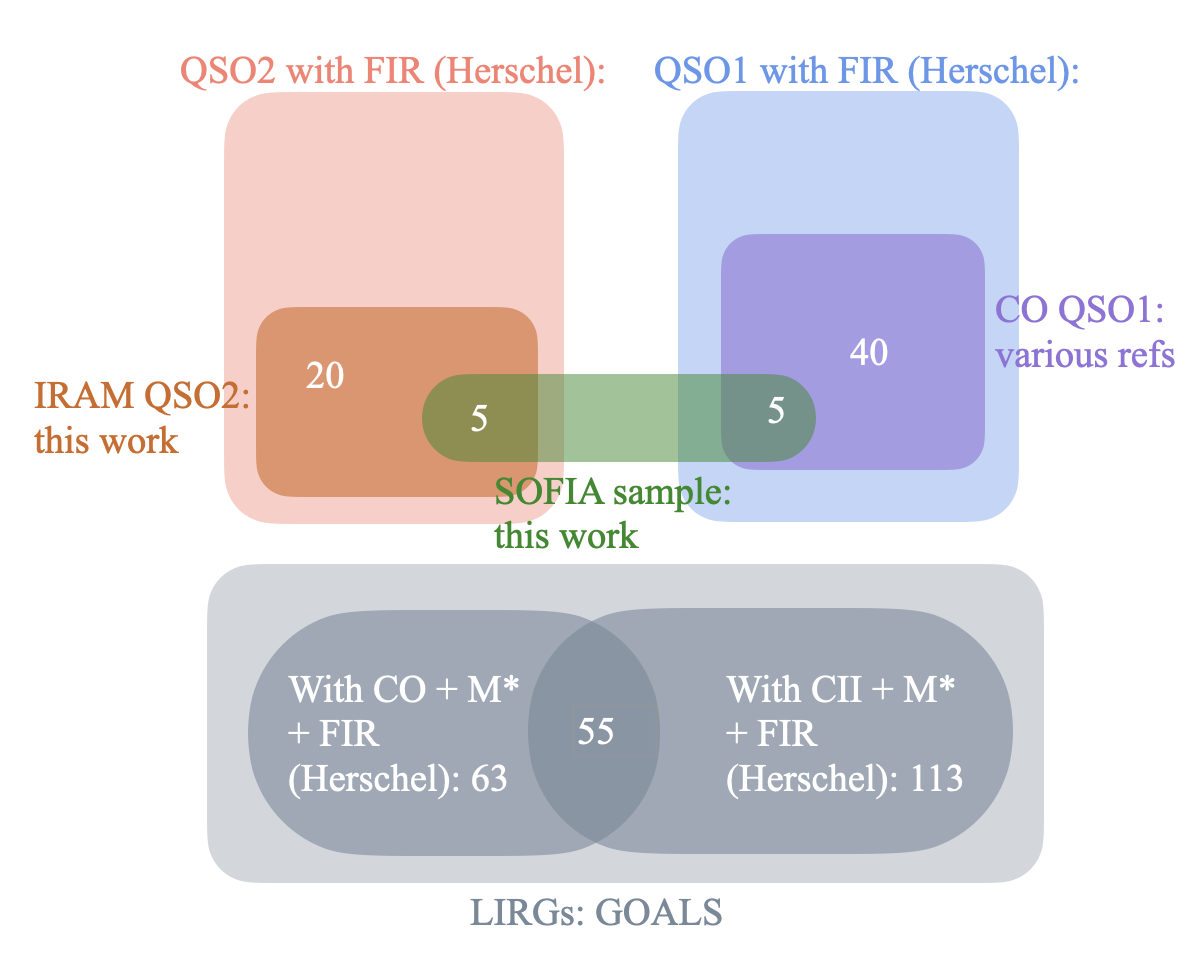}
\caption{The Venn diagram summarizing the different samples used in this work. \label{fig:comparison_samples}}
\end{figure}

\subsection{Reference Samples from the Literature} \label{sec:ref sample selection}
To explore the evolutionary link between LIRGs, QSO2s, and QSO1s, we select samples of local LIRGs and QSO1s with all CO or [\ion{C}{2}], FIR, and stellar mass (M$_*$) measurements, to compare with our CO and
[\ion{C}{2}] observations of local QSOs. We explain in the following paragraphs how we select our comparison samples and summarize the different data sets with a Venn diagram in  Fig.\ref{fig:comparison_samples}. For our QSO2 sample (SDSS02--SDSS20), we obtain M$_*$ and FIR data from \citet{Shangguan_2019}. The M$_*$ for SDSS01 is taken from \citet{Koss_2011} and the FIR data is from \citet{Melendez_2014}.

The LIRGs for comparison are from the Great Observatories All-sky LIRG Survey \citep[GOALS;][]{Armus_2009}. GOALS galaxies contain mergers which could be the trigger of obscured AGN activity as suggested by the evolutionary scenario, although the luminosities of GOALS galaxies are dominated by SF, with $\lesssim$10\% GOALS galaxies having AGN bolometric contribution to the infrared luminosities $\gtrsim$25\% \citep[e.g.,][]{petric2011,Alonso-Herrero_2012,Diaz-Santos_2017}. Our sample of QSOs is selected to be extremely bright (\S\ref{sec:data}) and thus despite possible small overlap with GOALS galaxies containing AGN, they should represent different populations of galaxies and provide informative comparison as possible different evolutionary stages. We construct the CO GOALS comparison sample by selecting galaxies that have 1) CO(1--0) measurements in \citet{Yamashita_2017} or \citet{Herrero-Illana_2019}\footnote{For GOALS galaxies in both \citet{Yamashita_2017} and \citet{Herrero-Illana_2019}, we use the measurements from \citet{Herrero-Illana_2019} because their observations were also from the IRAM 30m telescope.} 2) FIR measurements in \citet{Chu_2017} 3) M$_*$ estimates in \citet{Howell_2010}. For the [\ion{C}{2}] GOALS comparison sample, we use the same sources for FIR and M$_*$, but select those galaxies with [\ion{C}{2}] measurements presented in \citet{Diaz-Santos_2017}\footnote{We use their galaxy-integrated flux measured from the best aperture.}. We end up with 63 CO GOALS galaxies and 113 [\ion{C}{2}] GOALS galaxies.

For the CO QSO1 comparison sample, we select from the PG quasars that roughly match our IRAM QSO2s in redshift and [\ion{O}{3}] $\lambda$5007 luminosity (\S\ref{sec:data}). \citet{Shangguan_2020} summarized PG quasars with CO(1--0) data from the literature \citep{Casoli_2001,Evans_2001,Evans_2006,Bertram_2007,Evans_2009} in addition to presenting CO(2--1) observations of their selected PG quasars. To obtain a larger sample, we combine the CO(1--0) data converted from CO(2--1) measurements (Table 2 in \citealt{Shangguan_2020}) and those from the literature (Table 3 and 4 in \citealt{Shangguan_2020}). For objects with both CO(2-1)-converted and literature CO(1--0) luminosity, we adopt the more recent CO(2-1)-converted value if the literature value is an upper limit. CO(1--0) data for two additional PG quasars, PG 1534+580 and PG 0921+525, are taken from \citet{Wylezalek_2022} and \citet{Salome_2023}, respectively. For these QSO1s with CO data, we obtain FIR measurements from \citet{Petric_2015} and M$_*$ from \citet{Xie_2021}. M$_*$ values from \citet{Xie_2021} are scaled to the stellar initial mass function (IMF) of \citet{Salpeter_1955}. We divide these M$_*$ values by 1.5 to convert them to the \citet{Chabrier_2003} IMF to be consistent with the M$_*$ of GOALS galaxies and QSO2s.

The FIR observations for our samples of LIRGs, QSO1s, and QSO2s are from \textit{Herschel}. We use 160$\mu$m photometry because data at this wavelength, the longest observable wavelength of the Photodetector Array Camera and Spectrometer (PACS), are available for all samples. Emission at 160$\mu$m traces the coldest dust observable with PACS and they are not as impacted by confusion as longer wavelength observations with the Spectral and Photometric Imaging Receiver (SPIRE) because PACS data have higher spatial resolution than SPIRE. The 160$\mu$m fluxes obtained from the literature as described above have been corrected for aperture effects in corresponding works and represent the integrated flux for the entire galaxy. We convert observed fluxes at 160$\mu$m to rest-frame luminosities assuming a flat spectral shape around 160$\mu$m. This is a reasonable assumption based on existing spectral energy distribution (SED) analysis on our galaxies \citep[e.g.,][]{U_2012,Petric_2015,Shangguan_2019_sed,Shangguan_2019,Paspaliaris_2021}. We determine $F_{\rm{\lambda_{rest}}}$ from observed fluxes using $\lambda_{\rm rest} = \lambda_{\rm obs}/(1+z)$, and derive the spectral index ($\alpha: F_{\lambda} \propto \lambda^{\alpha}$) from observed fluxes at 100$\mu$m\footnote{When fluxes at 100$\mu$m are not available, we use fluxes at 70$\mu$m.} and 160$\mu$m. The final luminosity at rest-frame 160$\mu$m is estimated using the spectral index.

The data we obtain from the literature are summarized in \S\ref{sec:additional}. The M$_*$ distributions of our samples are similar (e.g., Fig.\ref{fig:co-relations} and Fig.\ref{fig:km}). Since our samples of LIRGs, QSO1s, and QSO2s are all nearby, our comparison of ISM properties between the samples should not be affected by redshift or M$_*$. We elaborate on the results of our comparison in \S\ref{sec:diss_comparison}.

\section{Results} \label{sec:result}

\subsection{Detection of CO in QSO2s}
As shown in Table \ref{tab:CO10observations} and \ref{tab:CO21observations}, we robustly detect CO molecular gas emission in the CO(1--0) line in 50\% of the 20 optically luminous QSO2s with a minimum detection limit of 0.4 $\pm$ 0.1 K~km~s$^{-1}$ and the brightest CO(1--0) emission at 1.2 $\pm$ 0.2 K~km~s$^{-1}$. A relatively faint source (SDSS06) is also detected with a 2.7$\sigma$ significance. Whether a source is detected is determined based on both the signal level above the continuum and the likelihood that a set of channels will have enhanced emission at the expected line frequency. The flux uncertainty estimation is described in \S2.1. The $I_{\mathrm{CO(1-0)}}$ upper limits range between 0.4 and 1.1 K~km~s$^{-1}$ and the median $I_{\mathrm{CO(1-0)}}$ of the detected sources is 0.8 K~km~s$^{-1}$.

We detect CO(2--1) in 40\% of the 15 QSO2s we target in this line. The faintest and brightest detections are 0.4 $\pm$ 0.2 and 2.5 $\pm$ 0.4 K~km~s$^{-1}$. The $I_{\mathrm{CO(2-1)}}$ upper limits range between 0.3 and 3.1 K~km~s$^{-1}$ and the median $I_{\mathrm{CO(2-1)}}$ of the detected sources is 1.3 K~km~s$^{-1}$.

Toward 4 sources we have robust detection of both the CO(1--0) and CO(2--1) emission while 2 sources have CO(2--1) measurements but not CO(1--0). We compute the line ratio R$_{21} = \frac{L^{\prime}_{\rm CO(2-1)}}{L^{\prime}_{\rm CO(1-0)}}$ for the 4 QSO2s with both lines detected and the values are recorded in Table \ref{tab:CO10observations} and \ref{tab:CO21observations}. The R$_{21}$ values for our QSO2s span a wide range of 0.45--1.35 which is roughly in agreement with the range of 0.49--0.90 found for PG QSO1s in \citet{Shangguan_2020}. Our median R$_{21}$ is 0.82 and we use this value to estimate the CO(1--0) luminosity of the 2 QSO2s that only have CO(2--1) measurements.

\subsection{Detection of [\ion{C}{2}] in QSO1s and QSO2s}
We robustly detect [\ion{C}{2}] 158$\mu$m emission in all 5 QSO2s and 4/5 QSO1s. The measured [\ion{C}{2}] fluxes are summarized in Table \ref{tab:CIIobservations} and we provide the upper limit on [\ion{C}{2}] flux for the QSO1 toward which we do not detect [\ion{C}{2}] emission. The [\ion{O}{3}] 88$\mu$m line is observed in parallel for 6 QSOs and 4 of them are detected in this line. We provide [\ion{O}{3}] 88$\mu$m fluxes for these detected sources. The uncertainties in the flux measurements for both lines are predominantly calibration errors, which we adopt to be 15\%\footnote{25\% for Sz II 10 due to higher background uncertainty.} of the reported fluxes.

3 QSO1s and 3 QSO2s have extended [\ion{C}{2}] 158$\mu$m emission. For those sources, the total flux which is larger than the flux enclosed in a single-resolution element is also reported in Table \ref{tab:CIIobservations}. 

\begin{figure*}
\centering
\includegraphics[width=\columnwidth]{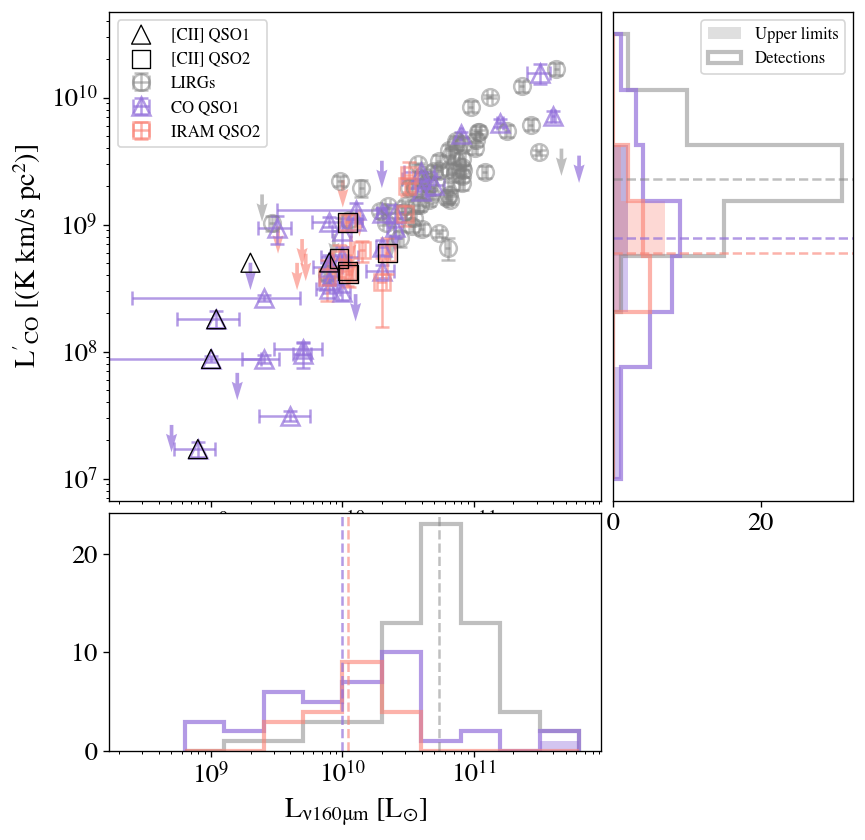}
\includegraphics[width=\columnwidth]{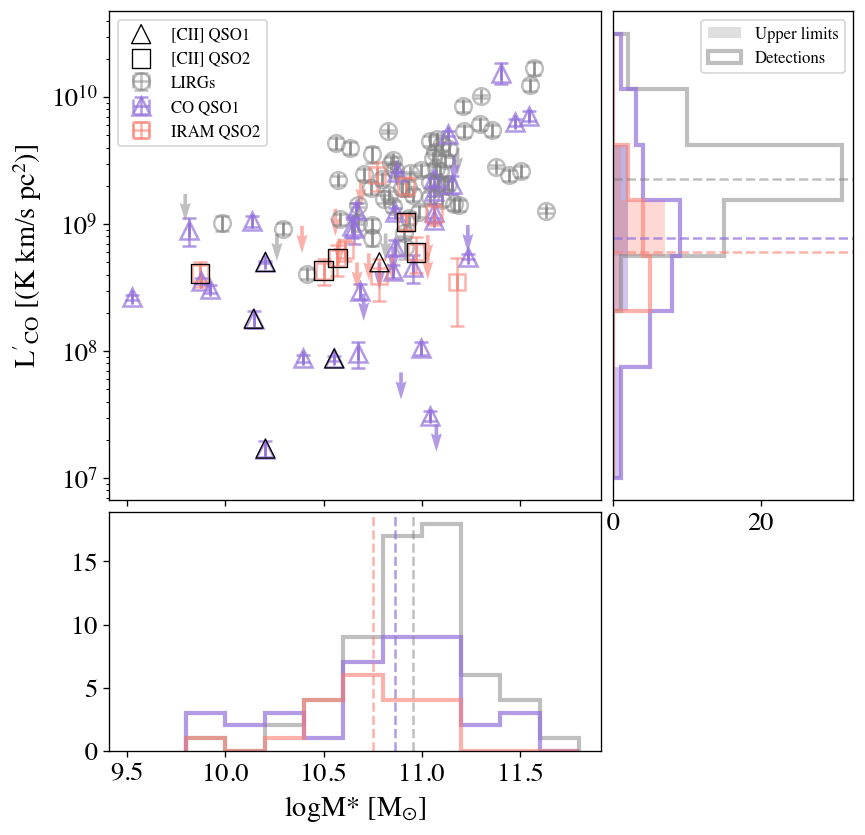}\\
\includegraphics[width=\columnwidth]{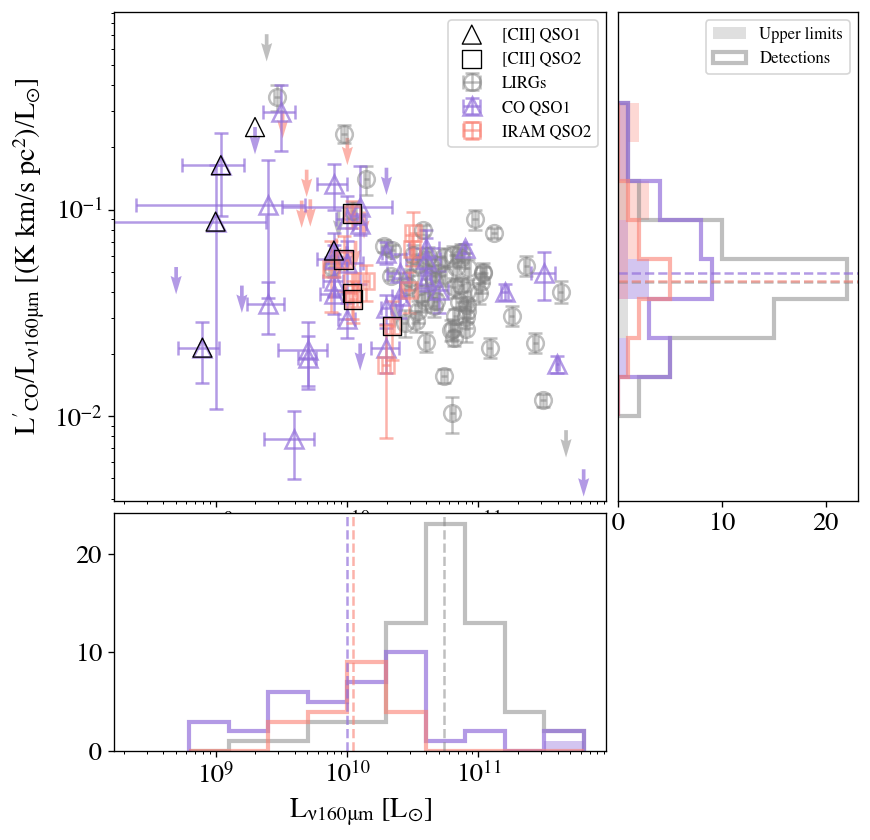}
\includegraphics[width=\columnwidth]{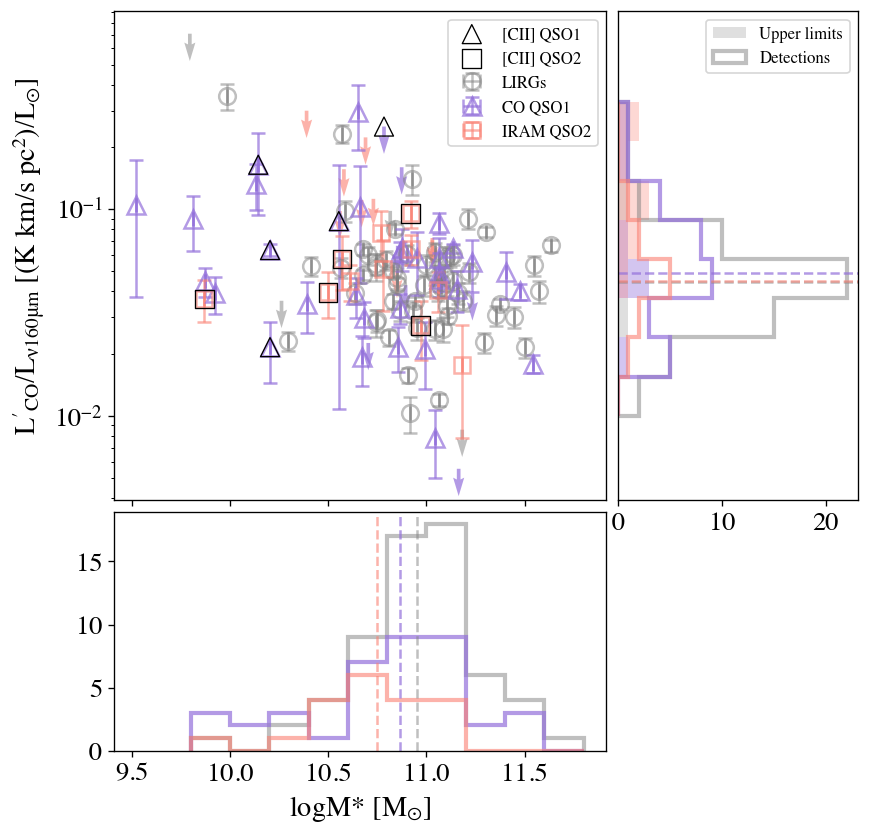}
\caption{\textbf{Upper left:} CO(1--0) luminosity versus FIR luminosity at 160$\mu$m. \textbf{Upper right:}  CO(1--0) luminosity versus galaxy stellar mass. \textbf{Lower left:} CO over FIR luminosity ratio versus FIR luminosity at 160$\mu$m.  \textbf{Lower right:} CO over FIR luminosity ratio versus galaxy stellar mass. Upper limits are indicated by downward arrows. Dashed lines at the histograms indicate the median of detected values. Values obtained from the literature are detailed in \S\ref{sec:ref sample selection} and summarized in Table \ref{tab:goals_sample}, \ref{tab:qso1_sample}, and \ref{tab:qso2_sample}. QSOs with [\ion{C}{2}] data available in this paper are indicated by additional black markers. \label{fig:co-relations}}
\end{figure*}

\section{Discussion} \label{sec:diss}
\subsection{Comparison between Nearby LIRGs, QSO2s, and QSO1s} \label{sec:diss_comparison}
\subsubsection{CO}\label{sec:co discussion}
We plot the CO, FIR, and M$_*$ data for our samples in Fig.\ref{fig:co-relations}. To compare the distributions of different physical quantities across samples, we derive cumulative distributions of different physical quantities and perform two-sample statistical tests using the Python package \texttt{lifelines}\footnote{\url{https://lifelines.readthedocs.io/en/latest/}}. Given that the data contains upper limits (left-censored data), the cumulative distributions are based on the Kaplan–Meier estimator \citep{km_1958}, and the two-sample tests we use here are the Logrank test, Peto–Peto test, Wilcoxon test, and Tarone–Ware test. \citep{Feigelson_1985,Feigelson_Babu_2012}. 

\begin{figure*}
\centering
\includegraphics[width=2\columnwidth]{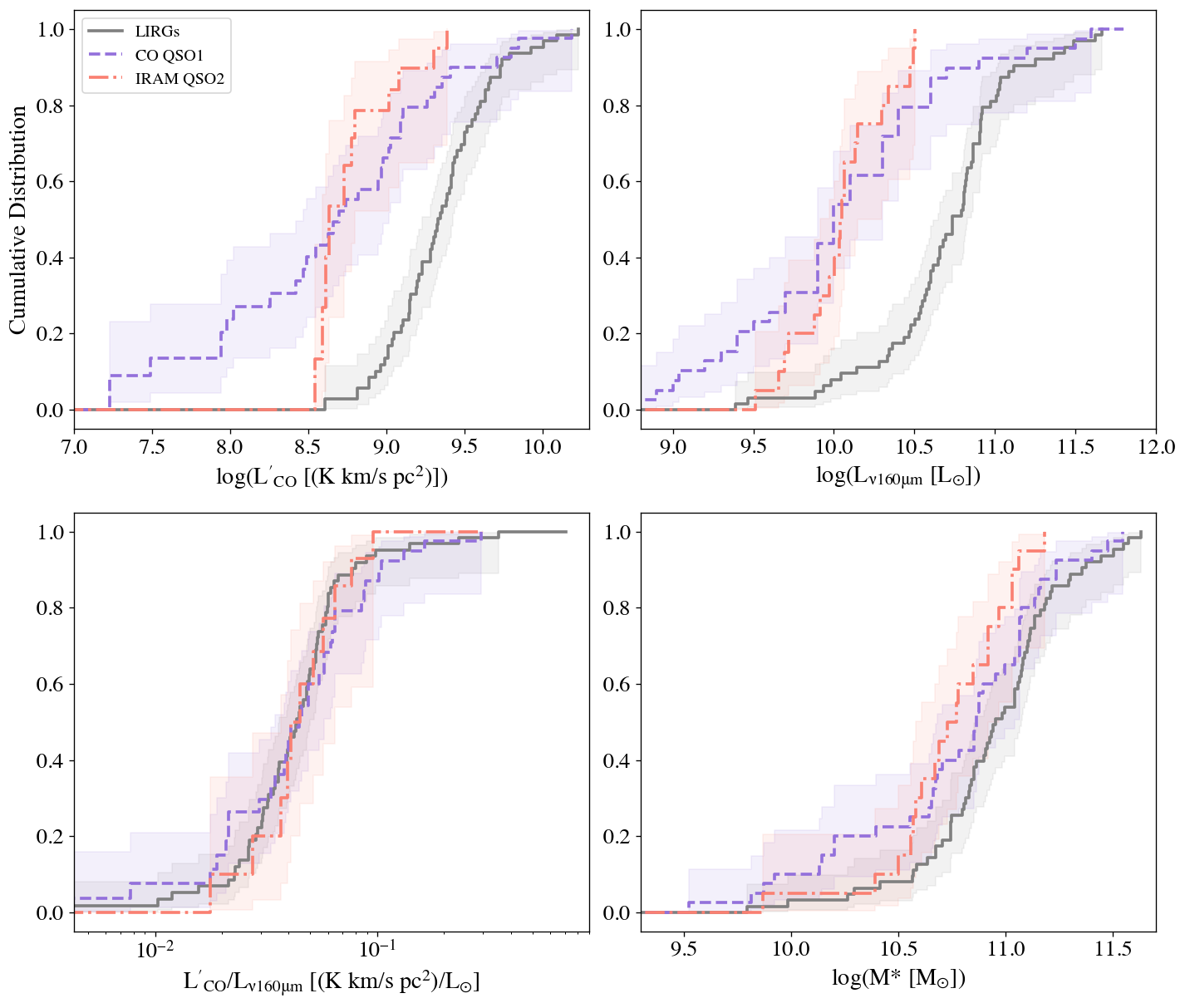}
\caption{Cumulative distributions of CO(1--0) luminosity, FIR luminosity at 160$\mu$m, the ratio of CO(1--0) over FIR luminosity, and galaxy stellar mass for our samples of LIRGs, QSO1s, and QSO2s. The distributions are derived from the Kaplan-Meier estimator taking into account undetected upper limits, and the shaded areas represent the 95\% confidence intervals. Values obtained from the literature are detailed in \S\ref{sec:ref sample selection} and summarized in Table \ref{tab:goals_sample}, \ref{tab:qso1_sample}, and \ref{tab:qso2_sample}. \label{fig:km}}
\end{figure*}

The cumulative distributions of $L^{\prime}_{\rm{CO}}$, $L_{\rm{160\mu m}}$, $L^{\prime}_{\rm{CO}}$/$L_{\rm{160\mu m}}$, and M$_{*}$ for QSO1s, QSO2s, and LIRGs are shown in Fig.\ref{fig:km} and the test results are summarized in Table \ref{tab:stats}. The test statistics represent how different the two samples are, and the $p$-values represent the likelihood that the two samples arise from the same underlying distribution (the smaller the $p$-value the less likely). We adopt here a conservative threshold that $p<10^{-3}$ indicates that the two samples are from distinctive distributions while $p>0.1$ indicates that the two samples are statistically indistinguishable. $p$-values in between imply marginal differences between the two samples and we interpret them as possible hints for future analysis.

QSO1s and QSO2s appear statistically indistinguishable in all four physical quantities. When compared to LIRGs which are hypothesized to be the precursors of both QSO1s and QSO2s, both types of QSOs show different distributions from LIRGs in CO and FIR luminosity, with LIRGs showing larger values of $L^{\prime}_{\rm{CO}}$ and $L_{\rm{160\mu m}}$. This result is not unexpected as LIRGs are IR-luminous by definition. Both higher $L^{\prime}_{\rm{CO}}$ and $L_{\rm{160\mu m}}$ suggest that there is more SF activity in LIRGs than in QSOs, which is consistent with the evolutionary scenario although larger samples probing a broader range of M$_{*}$ distributions would provide stronger support for this interpretation. Our three samples have statistically indistinguishable distributions for $L^{\prime}_{\rm{CO}}$/$L_{\rm{160\mu m}}$ and M$_{*}$. $L^{\prime}_{\rm{CO}}$/$L_{\rm{160\mu m}}$ is an indicator of the SFE since the numerator and denominator trace the SF fuel available and ongoing SF, respectively. The tight linear correlation between CO and FIR luminosities (upper left panel of Fig.\ref{fig:co-relations}) implies that FIR emission in our samples is mainly from dust heated by star formation rather than by AGN\footnote{We reiterate that our interpretations are from a global perspective based on galaxy-wide, unresolved data. Luminosity ratios such as $L_{\rm CII}$/$L_{\rm{160\mu m}}$ could exhibit significant differences at the nuclear regions of (U)LIRGs dominated by SF and AGN \citep[e.g.,][]{Herrera-Camus_2018}.} (see also \citealt{petric2011,Zakamska_2016,Shangguan_2020}). Therefore, despite showing more SF activity, the LIRGs in our sample are forming stars at the same efficiency as QSO1s and QSO2s. The similar distribution for M$_{*}$ ensures that the differences/similarities we observe in our samples are not affected by M$_{*}$.

To test for systematic differences between the ISM content of LIRGs, QSO2s, and QSO1s, we relate CO observations to ISM content by deriving the molecular gas mass (M$_{\rm{H_2}}$) and gas fraction (M$_{\rm{H_2}}$/M$_{*}$) for our samples. For consistency, we use M$_{\rm{H_2}}$ derived from the CO-to-H$_2$ conversion factor ($\alpha_{\rm{CO}}$): M$_{\rm{H_2,CO}}$ = $\alpha_{\rm{CO}}$$\times$$L^{\prime}_{\rm{CO}}$. We choose the most suitable value of $\alpha_{\rm{CO}}$ for each of our samples, based on previous works using samples that overlap significantly with our samples: $\alpha_{\rm{CO,QSO1}}$ = 3.1 (K km s$^{-1}$pc$^2$)$^{-1}$ for QSO1s (\citealt{Shangguan_2020}, based on total gas measurements from dust conetent), $\alpha_{\rm{CO,QSO2}}$ = 4.3 (K km s$^{-1}$pc$^2$)$^{-1}$ for QSO2s (\citealt{Ramos-Almeida_2022}, Milky Way $\alpha_{\rm{CO}}$, resulting gas mass in good agreement with other QSO2s in the litrature), and $\alpha_{\rm{CO,LIRG}}$ = 1.8 (K km s$^{-1}$pc$^2$)$^{-1}$ for LIRGs (\citealt{Herrero-Illana_2019,Montoya-Arroyave_2023}, consistent $\alpha_{\rm{CO,LIRG}}$ from [\ion{C}{1}]- and dust-derived gas mass). 

\begin{figure*}
\centering
\includegraphics[width=2\columnwidth]{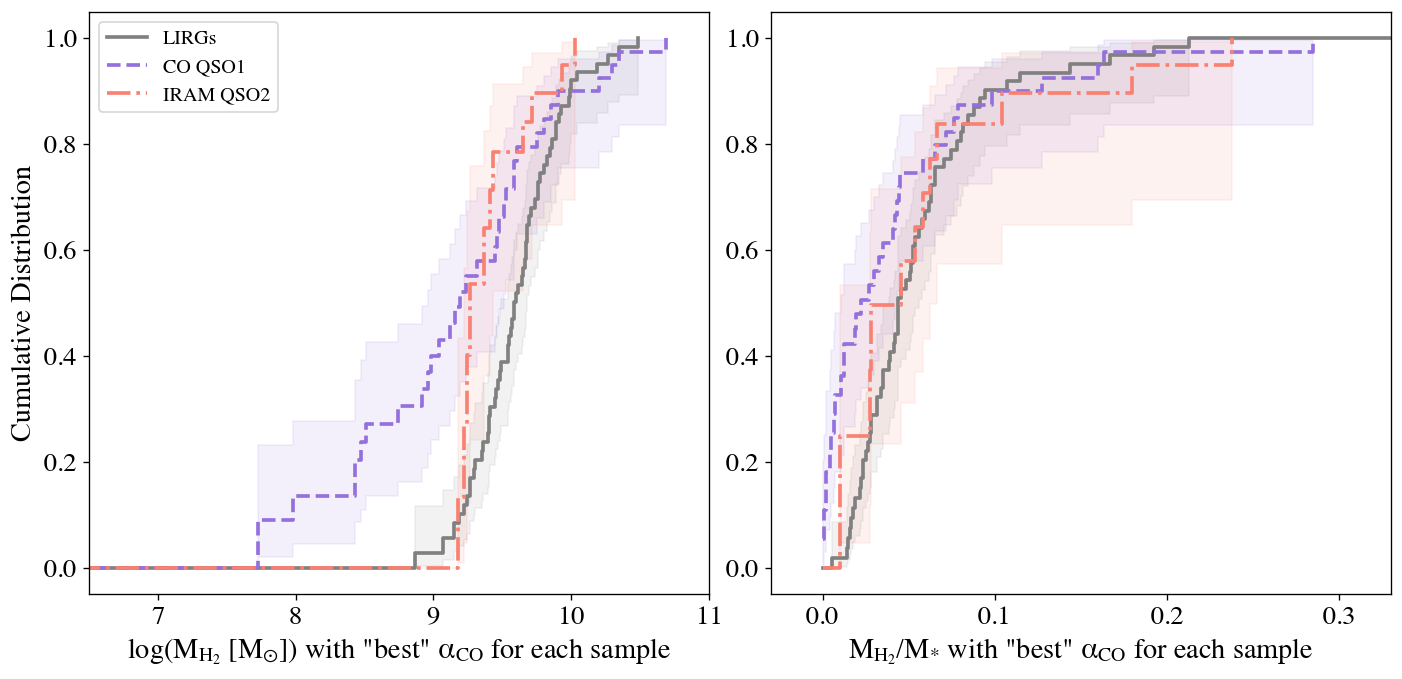}
\caption{Cumulative distributions of M$_{\rm{H_2,CO}}$ and M$_{\rm{H_2,CO}}$/M$_{*}$ derived using the CO-to-H$_2$ conversion factor as described in \S\ref{sec:co discussion}. The distributions are derived from the Kaplan-Meier estimator taking into account undetected upper limits, and the shaded areas represent the 95\% confidence intervals. \label{fig:km_mass}}
\end{figure*}

We plot the cumulative distributions of M$_{\rm{H_2,CO}}$ and  M$_{\rm{H_2,CO}}$/M$_{*}$ in Fig.\ref{fig:km_mass} and record the two-sample statistical test results in Table \ref{tab:stats}. QSO1s and QSO2s show statistically indistinguishable distributions for M$_{\rm{H_2,CO}}$ and  M$_{\rm{H_2,CO}}$/M$_{*}$. Based on the test statistics (Table \ref{tab:stats}), LIRGs appear to have similar M$_{\rm{H_2,CO}}$ distribution and identical M$_{\rm{H_2,CO}}$/M$_{*}$ distribution as QSO2s, but display relatively larger differences compared to QSO1s in both quantities (LIRGs show larger M$_{\rm{H_2,CO}}$ and  M$_{\rm{H_2,CO}}$/M$_{*}$). Although LIRGs tend to have brighter CO emission than QSO2s (Fig.\ref{fig:co-relations} and \ref{fig:km}), they have similar gas fractions and thus similar molecular gas reservoirs.
We note that $\alpha_{\rm{CO}}$ is known to be uncertain and can even vary within a single galaxy \citep[e.g.,][]{Bolatto_2013}. Although LIRGs show larger $L^{\prime}_{\rm{CO}}$ than QSOs, they are known to have smaller $\alpha_{\rm{CO}}$ due to complex gas dynamics from mergers (\citealt{Bolatto_2013} and references therein). The $\alpha_{\rm{CO,LIRG}}$ adopted here are tested against other molecular gas estimators \citep{Herrero-Illana_2019,Montoya-Arroyave_2023} and the similarity of molecular gas content in the two types of QSOs agrees with previous findings in the literature \citep[e.g.,][]{Krips_2012,VillarMartin_2013}.
To study the ISM content of our samples from a different perspective, we carry out a similar comparison using [\ion{C}{2}] as a proxy for ISM gas in the following section.

\subsubsection{[\ion{C}{2}]}
\begin{figure}
\centering
\includegraphics[width=\columnwidth]{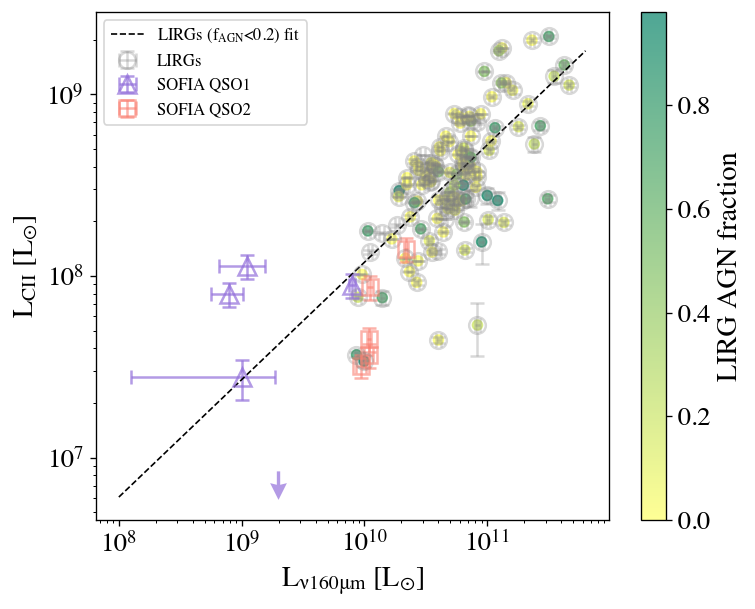}
\caption{[\ion{C}{2}] luminosity versus FIR luminosity at 160$\mu$m. The black dashed line is fit to LIRGs with AGN fractions less than 0.2. $L_{\rm{CII}}$ and $L_{\rm{160\mu m}}$ of LIRGs show the expected correlation. QSO2s lie within the scatter of LIRGs while QSO1s display more scatter.  \label{fig:cii-fir}}
\end{figure}

\begin{figure*}
\centering
\includegraphics[width=\columnwidth]{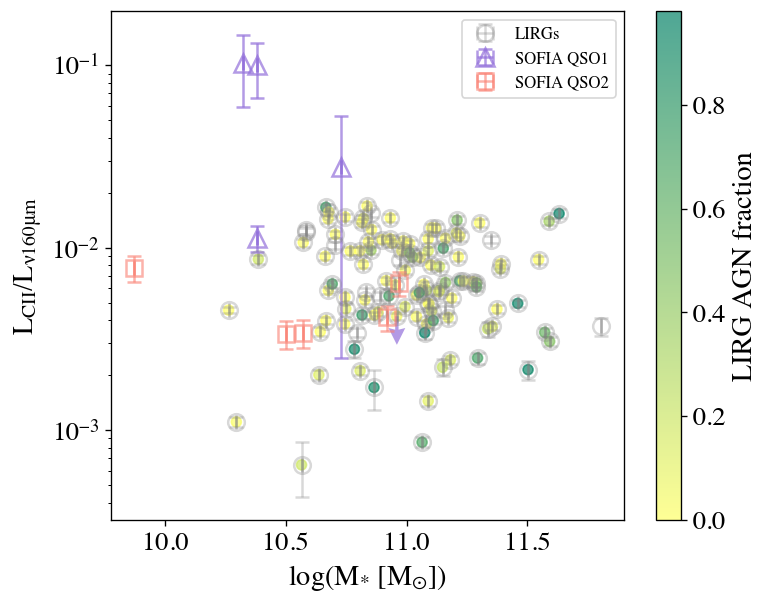}
\includegraphics[width=\columnwidth]{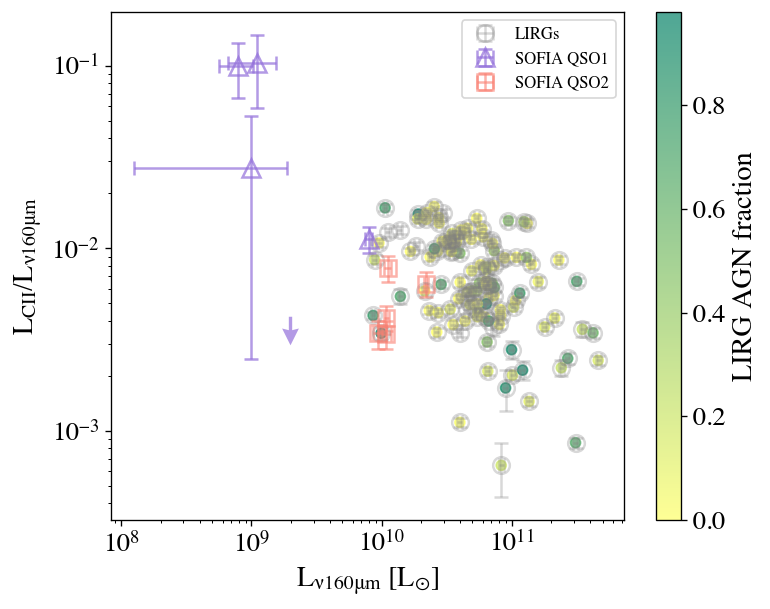}\\
\caption{\textbf{Left:} [\ion{C}{2}] over FIR luminosity ratio versus stellar mass. \textbf{Right:} [\ion{C}{2}] over FIR luminosity ratio versus FIR luminosity at 160$\mu$m. The LIRGs are color-coded by AGN fractions derived from multiple MIR diagnostics as described in \S4.1 of \citet{petric2011}. Values obtained from the literature are detailed in \S\ref{sec:ref sample selection} and summarized in Table \ref{tab:goals_sample}, \ref{tab:qso1_sample}, and \ref{tab:qso2_sample}. QSO2s lie within the scatter of LIRGs while QSO1s display more scatter. The two QSO1s exhibiting the largest $L_{\rm{CII}}$/$L_{\rm{160\mu m}}$ are Mrk 110 and Mrk 335, which have been observed to host outflows. \label{fig:cii/fir}}
\end{figure*}

The relation between [\ion{C}{2}] and FIR luminosity has been frequently used to probe the gas and dust content in high redshift galaxies \citep[e.g.,][]{Carilli_2013}. We plot $L_{\rm{CII}}$\footnote{We use $I_{\rm{CII}}$ for the entire galaxy, the extended emission column in Table \ref{tab:CIIobservations} when available.} versus $L_{\rm{160\mu m}}$ in Fig.\ref{fig:cii-fir}. The LIRGs are color-coded by the fractional contribution of AGN to MIR luminosity (AGN fractions) derived from multiple MIR diagnostics as described in \S4.1 of \citet{petric2011}. We observe the expected correlation between $L_{\rm{CII}}$ and $L_{\rm{160\mu m}}$ in our samples and fit a line to the LIRGs with AGN fractions less than 0.2. The QSOs also follow the correlation demonstrated by the line fit, though the QSO1s display more scatter ($\sigma_{\rm |log(L_{CII,obs})-log(L_{CII,line})|,QSO1}$ = 0.29, $\sigma_{\rm |log(L_{CII,obs})-log(L_{CII,line})|,QSO2}$ = 0.18). The slope of this line is less than 1, which indicates some ``deficit" in the [\ion{C}{2}] line emission compared to the underlying dust continuum emission. In Fig.\ref{fig:cii/fir} we plot the luminosity ratio between [\ion{C}{2}] and FIR versus M$_{*}$ and FIR luminosity. While we use the luminosity at 160$\mu$m instead of the total FIR luminosity, we observe in the right panel of Fig.\ref{fig:cii/fir} the negative trend between $L_{\rm{CII}}$/$L_{\rm{160\mu m}}$ and $L_{\rm{160\mu m}}$ with LIRG AGN spanning the whole y-axis range as reported in \citet{diaz-santos2013}. QSO1s tend to show larger [\ion{C}{2}]/FIR compared to LIRGs and QSO2s. Three (Sz II 10, Mrk 110, and Mrk 335) out of five QSO1s exhibit [\ion{C}{2}]/FIR ratios exceeding the highest values observed in LIRGs and QSO2s, with two QSO1s (Mrk 110 and Mrk 335) displaying [\ion{C}{2}]/FIR ratios an order of magnitude higher than those of LIRGs and QSO2s. In the left panel of Fig.\ref{fig:cii/fir}, the [\ion{C}{2}]/FIR ratio does not appear to correlate with M$_{*}$ for LIRGs and QSO2s, but seems to decrease with M$_{*}$ for QSO1s.

\begin{figure*}
\centering
\includegraphics[width=\columnwidth]{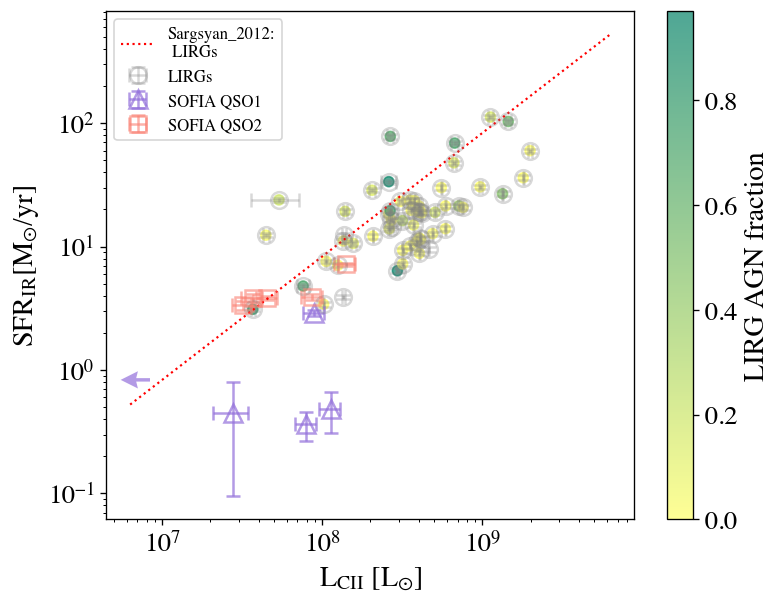}
\includegraphics[width=\columnwidth]{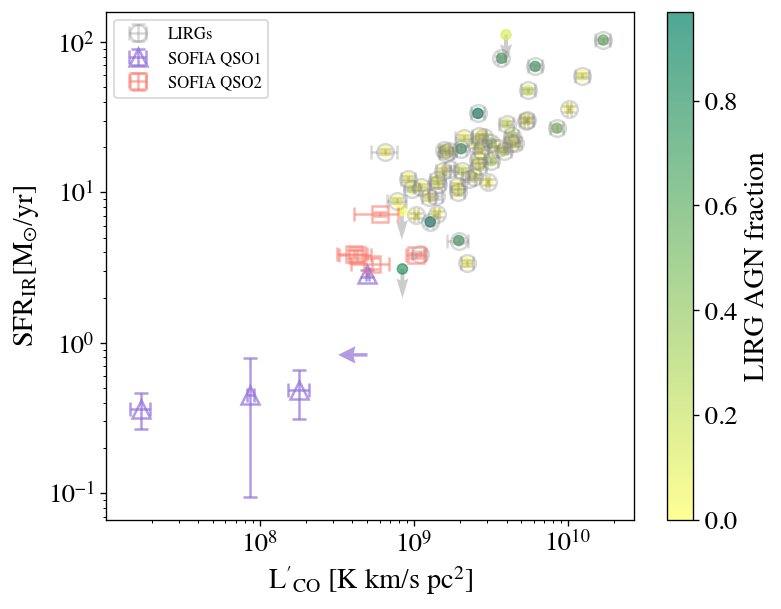}\\
\caption{\textbf{Left:} $L_{\rm{CII}}$ versus SFR$_{\rm{IR}}$ for our samples with both CO and [\ion{C}{2}] data. The correlation from \citet{sargsyan2012} based on starburst and AGN host galaxies is overplotted. LIRGs and QSO2s agree with this correlation while most of the QSO1s lie below the correlation. \textbf{Right:} $L^{\prime}_{\rm{CO}}$ versus SFR$_{\rm{IR}}$ for our samples with both CO and [\ion{C}{2}] data. $L^{\prime}_{\rm{CO}}$ in all three samples follows a tight correlation with SFR$_{\rm{IR}}$. \label{fig:cii-sfr}}
\end{figure*}

Given its close relation with PDRs where new stars form, the [\ion{C}{2}] luminosity has been used as a SF tracer in the literature \citep[e.g.,][]{De_Looze_2014, herrera2015, Narayanan2016, lagache2018}. Here we compare our measured $L_{\rm{CII}}$ with $L^{\prime}_{\rm{CO}}$ and SFRs derived from an independent indicator, the total infrared (IR) luminosity. We first convert $L_{\rm{160\mu m}}$ to the total infrared luminosity due to star formation using
\begin{equation}
    \rm{log}(L_{\rm{IR,SF}} [L_{\odot}]) = 1.49 + 0.90\ \rm{log}(L_{\rm{160\mu m}} [L_{\odot}]).
\end{equation}
This calibration was based on star-forming galaxies as presented in \citet{Symeonidis_2008}. The utility of this relation on similar samples of LIRGs and their validity for QSOs are tested in \citet{Zakamska_2016}. We then derive the SFR using the calibration from \citet{Kennicutt_1998}, adjusted to our \citet{Chabrier_2003} IMF by dividing by a factor of 1.5:
\begin{equation}
    \rm{SFR_{IR}} [M_{\odot}/yr] = 3 \times 10^{-44} L_{\rm{IR}} [erg/s].
\end{equation}
We plot SFR$_{\rm{IR}}$ versus $L_{\rm{CII}}$ and SFR$_{\rm{IR}}$ versus $L^{\prime}_{\rm{CO}}$ for our samples with both CO and [\ion{C}{2}] measurements (55 LIRGs, 5 QSO1s, 5 QSO2s) in Fig.\ref{fig:cii-sfr}. We observe a correlation between $L_{\rm{CII}}$ and SFR$_{\rm{IR}}$ for LIRGs and QSO2s, agreeing with the relation found in \citet{sargsyan2012} on a sample of starburst and AGN host galaxies. We find that QSO1s lie below this relation formed by LIRGs and QSO2s, but follow the $L^{\prime}_{\rm{CO}}$--SFR correlation formed by LIRGs and QSO2s, a trend also seen in the $L^{\prime}_{\rm{CO}}$ versus $L_{\rm{160\mu m}}$ plot with all samples that have CO data (upper left panel of Fig.\ref{fig:co-relations}).

\begin{figure}
\centering
\includegraphics[width=\columnwidth]{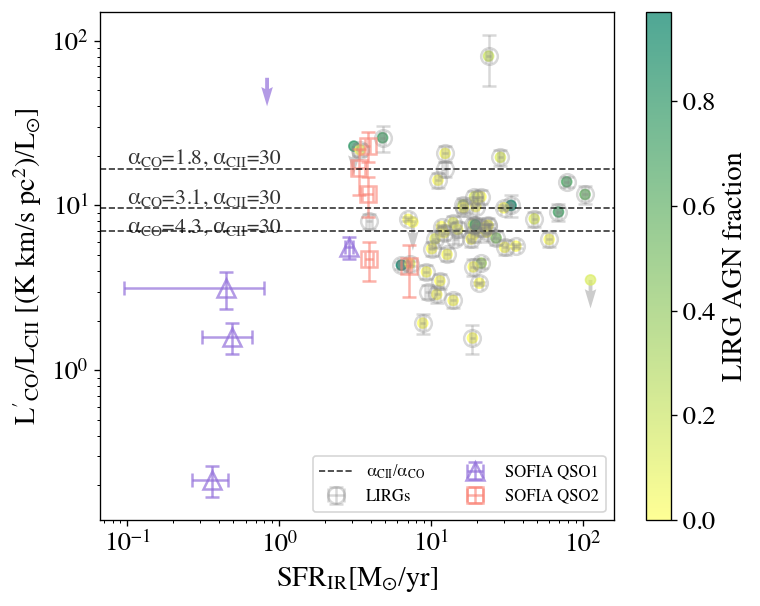}
\caption{$L^{\prime}_{\rm{CO}}$/$L_{\rm{CII}}$ versus SFR$_{\rm{IR}}$ with constant lines of $\alpha_{\rm{CII}}$/$\alpha_{\rm{CO}}$. The lines represent the case if [\ion{C}{2}] and CO are both tracing the same gas reservoir: $\alpha_{\rm{CII}}$$\times$$L_{\rm{CII}}$ = $\alpha_{\rm{CO}}$$\times$$L^{\prime}_{\rm{CO}}$. LIRGs and QSO2s show less scatter in $L^{\prime}_{\rm{CO}}$/$L_{\rm{CII}}$ and are more consistent with the constant lines. QSO1s show enhanced [\ion{C}{2}] emission not tracing SF compared to CO emission.\label{fig:sfr-co-cii}}
\end{figure}

We further compare $L^{\prime}_{\rm{CO}}$, $L_{\rm{CII}}$, and SFR in our samples by plotting $L^{\prime}_{\rm{CO}}$/$L_{\rm{CII}}$ versus SFR$_{\rm{IR}}$ in Fig.\ref{fig:sfr-co-cii}. The H$_2$ mass could also be derived from $L_{\rm{CII}}$: M$_{\rm{H_2,CII}}$ = $\alpha_{\rm{CII}}$$\times$$L_{\rm{CII}}$ \citep[e.g.,][]{Zanella_2018,Madden_2020}. If CO and [\ion{C}{2}] are tracing the same gas in galaxies, i.e., M$_{\rm{H_2,CO}}$ = M$_{\rm{H_2, CII}}$, we would expect $L^{\prime}_{\rm{CO}}$/$L_{\rm{CII}}$ to be the constant $\alpha_{\rm{CII}}$/$\alpha_{\rm{CO}}$. We plot in Fig.\ref{fig:sfr-co-cii} lines of constant $\alpha_{\rm{CII}}$/$\alpha_{\rm{CO}}$ using $\alpha_{\rm{CII}}$ = 30 M$_{\odot}$/$L_{\odot}$ from \citet{Zanella_2018} and the $\alpha_{\rm{CO}}$ values we used in \S\ref{sec:co discussion}. We observe that QSO1s show systematically smaller $L^{\prime}_{\rm{CO}}$/$L_{\rm{CII}}$ than QSO2s and LIRGs despite a large scatter, requiring $\alpha_{\rm{CO}}$ several times greater than the Milky Way value of 4.3 (K km s$^{-1}$pc$^2$)$^{-1}$ if M$_{\rm{H_2,CO}}$ = M$_{\rm{H_2,CII}}$. 

\citet{Zanella_2018} has tested with a heterogeneous sample including dwarf, starburst, and main-sequence galaxies that their $\alpha_{\rm{CII}}$ is invariant with redshift, metallicity, and gas depletion time. While there are uncertainties in the $\alpha$ conversion factors, they are unlikely large enough to explain the low values of $L^{\prime}_{\rm{CO}}$/$L_{\rm{CII}}$ shown by our sample of QSO1s. LIRGs and QSO2s show less scatter in $L^{\prime}_{\rm{CO}}$/$L_{\rm{CII}}$ and are consistent with typical ranges of $\alpha_{\rm{CII}}$/$\alpha_{\rm{CO}}$ as plotted in Fig.\ref{fig:sfr-co-cii}. This comparison of $L^{\prime}_{\rm{CO}}$/$L_{\rm{CII}}$ agrees with what we observe in Fig.\ref{fig:cii-sfr}: the CO emission in all three samples and the [\ion{C}{2}] emission in QSO2s and LIRGs trace gas related to SF activity, but there seems to be an excess of [\ion{C}{2}] emission that is unrelated to SF in some QSO1s.

In addition to molecular gas, [\ion{C}{2}] emission also traces warm and cold neutral gas, and can be affected by the presence of AGN. While some spatially-resolved observations reveal a drop in $L_{\rm{CII}}$/$L_{\rm{FIR}}$ due to a greater fraction of C+ ions transitioning to higher ionization states \citep[e.g.,][]{Langer_2015,Herrera-Camus_2018}, AGN-enhanced [\ion{C}{2}] emission has also been observed \citep[e.g.,][]{Appleton_2018,SmirnovaPinchukova_2019}. Shocks or turbulent heating (e.g., due to mergers or ram pressure) could also lead to excess [\ion{C}{2}] emission compared to what is expected from SF, increasing global [\ion{C}{2}]/FIR up to a factor of $\sim$5 \citep[e.g.,][]{Appleton_2013,Alatalo_2014,Minchin_2022,Fadda_2021,Fadda_2023}. LIRG, QSO1, and QSO2 samples in this work all contain galaxies showing disturbed morphology, although detailed studies on the relation between morphology and [\ion{C}{2}] emission require higher resolution imaging data and are outside the scope of this paper. In the sample of LIRGs studied in this work, those with high AGN fractions do not have more [\ion{C}{2}] emission than those without AGN (LIRGs with F$_{\rm AGN}$ $\geqslant$ 0.2 have median $L_{\rm{CII}}$ = 3.2 $\times$ 10$^8$ $L_{\odot}$ and those with F$_{\rm AGN}$ $<$ 0.2 have median $L_{\rm{CII}}$ = 3.7 $\times$ 10$^8$ $L_{\odot}$). While it is challenging to separate the effects of shocks on the observed [\ion{C}{2}] properties in LIRGs since they are complex, gravitationally interacting, star-forming systems, the differences in [\ion{C}{2}] emission between QSO1s and QSO2s do suggest that the multiphase ISM of these two populations differs either because of their formation history or geometric conditions. \citet{Decarli_2014} found that [\ion{C}{2}] emission in the QSO of an interacting system arises primarily in the neutral gas and an excess of neutral gas could lead to the observed excess of [\ion{C}{2}] emission in our SOFIA QSO1s. Furthermore, outflows can dissociate molecular gas, leading to a relatively larger fraction of ISM gas in the atomic or partially ionized phase. Black holes in the blow-out phase in QSO2s may destroy the dust and molecular gas, but the atomic gas might rain back and increase the neutral fraction of the total gas reservoirs in the QSO1 phase. \citet{SmirnovaPinchukova_2019} proposed that a significant [\ion{C}{2}] excess in luminous AGN can be used as an inference for a multiphase AGN-driven outflow with a high mass-loading factor, where [\ion{C}{2}] emission likely traces the interface between the warm and cold gas phase and is enhanced by the dissipation of the outflow kinetic energy (see also, e.g., \citealt{Lesaffre_2013}). Four out of five SOFIA QSO1s (except IISZ 010, including Mrk 110 and Mrk 335 which exhibit the largest $L_{\rm{CII}}$/$L_{\rm{160\mu m}}$) contain outflows studied in previous studies \citep[e.g.,][]{Tombesi_2010,Longinotti_2019,Molina_2022,Salome_2023} and these outflows might have altered their ISM content and contributed to their [\ion{C}{2}] emission.

\subsection{Implications for Origins and Evolution of QSOs}
The evolutionary picture where LIRGs evolve to QSO2s and then to QSO1s as ISM gas is removed during the black hole ``blow-out" phase has been challenged when QSO1s were found to contain gas and vigorously form stars \citep[e.g.,][]{Petric_2015,Shangguan_2020,Xie_2021}. Our comparison of the ISM content of nearby QSO1s, QSO2s, and LIRGs via observations of CO and [\ion{C}{2}] confirms that nearby QSO2s are unlikely the precursors of nearby QSO1s but could evolve from nearby LIRGs. 

LIRGs and QSO2s show similar distributions in $L^{\prime}_{\rm{CO}}$/$L_{\rm{160\mu m}}$, M$_{\rm{H_2, CO}}$, and M$_{\rm{H_2, CO}}$/M$_{*}$ (Fig.\ref{fig:km}, Fig.\ref{fig:km_mass}, Table \ref{tab:stats}), implying similar SFE and molecular gas reservoir between these two types of galaxies. Although the traditional evolutionary scenario predicts a decrease in the molecular gas fraction from mergers to QSO2s, the observed similarity from our CO data does not contradict the prediction as our sample of LIRGs contains mergers at different stages and a small fraction ($\sim$10\%) of objects with $\gtrsim$70\% AGN fraction which could be the immediate predecessors of our QSO2s. The absence of a decrease in SFE might indicate that AGN do not suppress SF on the galaxy-wide scale in which our CO observations probe. In terms of [\ion{C}{2}], QSO2s also show similarity to the LIRGs in $L_{\rm{CII}}$/$L_{\rm{160\mu m}}$ (Fig.\ref{fig:cii/fir}) and $L^{\prime}_{\rm{CO}}$/$L_{\rm{CII}}$ (Fig.\ref{fig:sfr-co-cii}). In Fig.\ref{fig:sfr-co-cii}, we notice that LIRGs with high AGN fractions show less scatter in $L^{\prime}_{\rm{CO}}$/$L_{\rm{CII}}$\footnote{$\sigma_{\rm L^{\prime}_{\rm{CO}}/L_{\rm{CII}}}$ for LIRGs with F$_{\rm AGN}$ $\geqslant$ 0.2, LIRGs with F$_{\rm AGN}$ $<$ 0.2, SOFIA QSO1s, SOFIA QSO2s are 3.5 $\times$ 10$^{-4}$, 6.3 $\times$ 10$^{-4}$, 1.1 $\times$ 10$^{-3}$, 3.5 $\times$ 10$^{-4}$.}, being more consistent with the simple model of a constant ratio between CO and [\ion{C}{2}] luminosity. This fact could suggest that gas of different phases and metallicity are better mixed in these LIRGs so the observed CO and [\ion{C}{2}] emission may predominantly trace the same gas reservoirs on global, galaxy-wide scales. Substantial gas mixing may occur at later stages of galaxy mergers, giving rise to luminous obscured quasars, akin to those observed in our QSO2 sample.

QSO1s, despite having similar SFE to LIRGs ($L^{\prime}_{\rm{CO}}$/$L_{\rm{160\mu m}}$, lower left panels in Fig.\ref{fig:co-relations} and \ref{fig:km}), show less SF traced by $L_{\rm{160\mu m}}$ and M$_{\rm{H_2,CO}}$ (Fig.\ref{fig:km} and \ref{fig:km_mass}). Our results also agree with previous studies that found QSO1s and QSO2s have the same availability of molecular gas traced by CO \citep[e.g.,][]{Krips_2012, VillarMartin_2013}. Without the [\ion{C}{2}] data, one might be under the impression that nearby QSO1s contain less gas than nearby LIRGs with similar M$_{*}$. Spatially resolved studies on the Milky Way and nearby galaxies have shown that either the ionized or atomic gas could contribute to up to 50\% of the total [\ion{C}{2}] emission \citep[e.g.,][]{Croxall_2017,Tarantino_2021}. Our finding of [\ion{C}{2}] emission in QSO1s that does not trace SF (Fig.\ref{fig:cii-sfr} and \ref{fig:sfr-co-cii}) suggests that QSO1s might as well contain a comparable amount of gas to star-forming LIRGs but with lower molecular fractions, e.g., more ionized or atomic gas. Although studies using FIR SED fitting and gas-to-dust ratios\footnote{Estimated from galaxy stellar masses using the stellar mass–metallicity relation and gas-to-dust ratio-metallicity relation from the literature. The median of the gas-to-dust ratio for their sample is 124 $\pm$ 6.} found that QSO1s and QSO2s are also indistinguishable in the total gas mass including molecular and neutral gas \citep[e.g.,][]{Shangguan_2018,Shangguan_2019}, future work with \ion{H}{1} data could better constrain the total gas mass and help reveal whether QSO1s contain more total gas than QSO2s and LIRGs. 

\citet{Shangguan_2019} found similar interstellar radiation fields in samples of QSO1s and QSO2s that contain our [\ion{C}{2}] targets and suggested that QSO1s and QSO2s have a similar spatial distribution of ISM. The difference in $L_{\rm{CII}}$ and $L_{\rm{CII}}$/$L_{\rm{160\mu m}}$ between QSO1s and QSO2s we observe in this work might suggest otherwise. Analysis of larger samples combined with theoretical simulations is needed to confirm these findings (Patil et al. in prep). Spatially resolved data is another key to understanding the ISM in different galaxy populations as they allow a separation between SF-, shock-, and AGN-dominated regions. In our context of LIRGs and QSOs, such data could offer valuable constraints on the specific effects of different physical mechanisms (e.g., AGN activity and shocks) on the ISM and SFR correlations. Comparing the morphology of gas and [\ion{C}{2}] emission can help determine whether [\ion{C}{2}] emission primarily arises from neutral or shock-heated gas. Moreover, examining the spatial correlation between CO, [\ion{C}{2}], and FIR emission could shed light on the dynamical history of these galaxies, such as merger events mix gas and outflows selectively destroy or displace gas in specific phases, thereby shedding light on their evolutionary connections.


The similarity of SFE, CO-, and [\ion{C}{2}]-traced gas in our LIRGs and QSO2s, as explained above, is consistent with the hypothesis that QSO2s could emerge from gas-rich mergers like local LIRGs. The similarity of SFE and CO-traced gas between QSO2s and QSO1s, coupled with evidence of potential additional gas in QSO1s from [\ion{C}{2}] observations, does not support the evolutionary scenario that require a ``blow-out" phase to remove gas to turn QSO2s to QSO1s. Our [\ion{C}{2}] observations of QSOs also disfavor the orientation-based unification theory. Our results are consistent with previous studies \citep[e.g.,][]{Veilleux_2009,Alonso-Herrero_2016} which highlighted the evolutionary complexity between LIRGs and QSOs. We note that this study focuses on the low redshift universe. To better characterize the ISM gas properties and understand the time evolution of QSOs, spatially resolved observations as well as samples covering different physical properties (e.g., redshifts, masses, and merger stages) are necessary.

\section{Conclusions} \label{sec:conclu} 
We present CO observations from the IRAM 30m telescope for 20 nearby, optically luminous QSO2s as well as [\ion{C}{2}] 158$\mu$m observations from SOFIA FIFI-LS for 5 QSO2s from the CO sample and 5 PG QSO1s. Combining those data with published measurements, we compare the ISM content (M$_{\rm{H_2}}$, M$_{\rm{H_2}}$/M$_{*}$) and SFE (indicated by $L^{\prime}_{\rm{CO}}$/$L_{\rm{160\mu m}}$) between M$_{*}$-matched QSO1s, QSO2s, and LIRGs using CO and [\ion{C}{2}] emission as probes. Our study using direct observations of gas tracers provides valuable insights into the evolutionary link between LIRGs, QSO2s, and QSO1s in the nearby universe. Our main findings are summarized as follows:
\begin{enumerate}
    \item QSO1s and QSO2s show statistically indistinguishable cumulative distributions in $L^{\prime}_{\rm{CO}}$, $L_{\rm{160\mu m}}$, $L^{\prime}_{\rm{CO}}$/$L_{\rm{160\mu m}}$, M$_{\rm{H_2,CO}}$, and M$_{\rm{H_2,CO}}$/M$_{*}$. The two types of QSOs appear to have similar SFE and molecular gas reservoirs traced by CO.
    
    \item LIRGs show larger $L^{\prime}_{\rm{CO}}$ (tracing gas available for SF) and $L_{\rm{160\mu m}}$ (tracing ongoing SF) than the QSOs, indicating more SF as expected. However, LIRGs show statistically indistinguishable cumulative distributions in $L^{\prime}_{\rm{CO}}$/$L_{\rm{160\mu m}}$ from the QSOs, implying that QSOs are forming stars as efficiently as LIRGs. The cumulative distributions of M$_{\rm{H_2,CO}}$ and M$_{\rm{H_2,CO}}$/M$_{*}$ for LIRGs resemble more closely those for QSO2s while showing statistical differences to those for QSO1s.
    
    \item LIRGs and QSO2s show close resemblance in [\ion{C}{2}] related quantities: $L_{\rm{CII}}$/$L_{\rm{160\mu m}}$, $L^{\prime}_{\rm{CO}}$/$L_{\rm{CII}}$. QSO1s show larger scatters in these quantities and tend to lie off the relations formed by LIRGs and QSO2s. Comparisons between $L_{\rm{CII}}$ and SFR derived from total IR luminosity imply that [\ion{C}{2}] emission in LIRGs and QSO2s trace SF. The [\ion{C}{2}] observations of 5 nearby QSO1s presented here provide tantalizing hints that [\ion{C}{2}] emission in QSO1s may have multiple origins, although more data are needed to corroborate this possibility. $L^{\prime}_{\rm{CO}}$ in all three samples follows a tight correlation with SFR$_{\rm{IR}}$.
    
    
    \item While [\ion{C}{2}] emission could be enhanced by AGN and shocks, we do not observe this enhancement in our samples of QSO2s and LIRGs with high AGN fractions. The extra [\ion{C}{2}] emission observed in QSO1s may have significant contributions from other ISM components that are more diffuse and warm, in neutral or ionized phases. The difference in observed $L_{\rm{CII}}$ hints at different interstellar radiation fields, indicating dissimilar ISM spatial distributions between QSO1s and QSO2s. Tighter constraints on the origins of the [\ion{C}{2}] emission in QSOs require a larger sample than what is available in this study as well as new FIR observational facilities. 
    
    \item The evolutionary scenario where LIRGs evolve into QSO2s and then to QSO1s predicts a decrease in gas fraction along the sequence observable on a galaxy-wide scale. The CO and [\ion{C}{2}] data we present here suggest that this scenario might not hold in the nearby universe. Nearby QSO2s could emerge from LIRGs based on their similar gas content, but they may not be the precursors of nearby QSO1s. The differences in ISM between QSO1s and QSO2s suggested by the CO and [\ion{C}{2}] data presented in this work also disagree with the simple orientation-based unified model. Future data of spatially resolved observations at different redshifts are required to further investigate the evolution of QSOs through cosmic time.
\end{enumerate}

\newpage

\figsetstart
\figsetnum{1}
\figsettitle{IRAM CO spectra with SDSS cutouts}

\figsetgrpstart
\figsetgrpnum{1.1}
\figsetgrptitle{SDSS01}
\figsetplot{./SDSS01IRAMlines-eps-converted-to.pdf}
\figsetgrpnote{\textbf{Left}: $60\arcsec \times 60\arcsec$ SDSS ($gri$) cutouts of the QSO2s. The solid and dotted yellow circles show the extent of the $3\arcsec$ SDSS spectroscopy fiber and the $22\arcsec$ IRAM beam at 3 mm, respectively. \textbf{Middle}: CO(1--0) spectra in units of mJy. \textbf{Right}: CO(2--1) spectra in units of mJy. For both spectra, the x-axis indicates the velocity offset of the line from the systemic velocity of the galaxy as determined by the optical spectrum. Spectra are shown at 25 MHz resolution and were taken from the WILMA backend. The dark grey shaded area indicates the CO line region used for the determination of $I_{\mathrm{CO}}$ and the red line is the Gaussian fit. Horizontal grey lines indicate the baseline (solid) and the $3\sigma$ scatter around the baseline (dashed).}
\figsetgrpend

\figsetgrpstart
\figsetgrpnum{1.2}
\figsetgrptitle{SDSS02}
\figsetplot{./SDSS02IRAMlines-eps-converted-to.pdf}
\figsetgrpnote{\textbf{Left}: $60\arcsec \times 60\arcsec$ SDSS ($gri$) cutouts of the QSO2s. The solid and dotted yellow circles show the extent of the $3\arcsec$ SDSS spectroscopy fiber and the $22\arcsec$ IRAM beam at 3 mm, respectively. \textbf{Middle}: CO(1--0) spectra in units of mJy. \textbf{Right}: CO(2--1) spectra in units of mJy. For both spectra, the x-axis indicates the velocity offset of the line from the systemic velocity of the galaxy as determined by the optical spectrum. Spectra are shown at 25 MHz resolution and were taken from the WILMA backend. The dark grey shaded area indicates the CO line region used for the determination of $I_{\mathrm{CO}}$ and the red line is the Gaussian fit. Horizontal grey lines indicate the baseline (solid) and the $3\sigma$ scatter around the baseline (dashed).}
\figsetgrpend

\figsetgrpstart
\figsetgrpnum{1.3}
\figsetgrptitle{SDSS03}
\figsetplot{./SDSS03IRAMlines-eps-converted-to.pdf}
\figsetgrpnote{\textbf{Left}: $60\arcsec \times 60\arcsec$ SDSS ($gri$) cutouts of the QSO2s. The solid and dotted yellow circles show the extent of the $3\arcsec$ SDSS spectroscopy fiber and the $22\arcsec$ IRAM beam at 3 mm, respectively. \textbf{Middle}: CO(1--0) spectra in units of mJy. \textbf{Right}: CO(2--1) spectra in units of mJy. For both spectra, the x-axis indicates the velocity offset of the line from the systemic velocity of the galaxy as determined by the optical spectrum. Spectra are shown at 25 MHz resolution and were taken from the WILMA backend. The dark grey shaded area indicates the CO line region used for the determination of $I_{\mathrm{CO}}$ and the red line is the Gaussian fit. Horizontal grey lines indicate the baseline (solid) and the $3\sigma$ scatter around the baseline (dashed).}
\figsetgrpend

\figsetgrpstart
\figsetgrpnum{1.4}
\figsetgrptitle{SDSS04}
\figsetplot{./SDSS04IRAMlines-eps-converted-to.pdf}
\figsetgrpnote{\textbf{Left}: $60\arcsec \times 60\arcsec$ SDSS ($gri$) cutouts of the QSO2s. The solid and dotted yellow circles show the extent of the $3\arcsec$ SDSS spectroscopy fiber and the $22\arcsec$ IRAM beam at 3 mm, respectively. \textbf{Middle}: CO(1--0) spectra in units of mJy. \textbf{Right}: CO(2--1) spectra in units of mJy. For both spectra, the x-axis indicates the velocity offset of the line from the systemic velocity of the galaxy as determined by the optical spectrum. Spectra are shown at 25 MHz resolution and were taken from the WILMA backend. The dark grey shaded area indicates the CO line region used for the determination of $I_{\mathrm{CO}}$ and the red line is the Gaussian fit. Horizontal grey lines indicate the baseline (solid) and the $3\sigma$ scatter around the baseline (dashed).}
\figsetgrpend

\figsetgrpstart
\figsetgrpnum{1.5}
\figsetgrptitle{SDSS05}
\figsetplot{./SDSS05IRAMlines-eps-converted-to.pdf}
\figsetgrpnote{\textbf{Left}: $60\arcsec \times 60\arcsec$ SDSS ($gri$) cutouts of the QSO2s. The solid and dotted yellow circles show the extent of the $3\arcsec$ SDSS spectroscopy fiber and the $22\arcsec$ IRAM beam at 3 mm, respectively. \textbf{Middle}: CO(1--0) spectra in units of mJy. \textbf{Right}: CO(2--1) spectra in units of mJy. For both spectra, the x-axis indicates the velocity offset of the line from the systemic velocity of the galaxy as determined by the optical spectrum. Spectra are shown at 25 MHz resolution and were taken from the WILMA backend. The dark grey shaded area indicates the CO line region used for the determination of $I_{\mathrm{CO}}$ and the red line is the Gaussian fit. Horizontal grey lines indicate the baseline (solid) and the $3\sigma$ scatter around the baseline (dashed).}
\figsetgrpend

\figsetgrpstart
\figsetgrpnum{1.6}
\figsetgrptitle{SDSS06}
\figsetplot{./SDSS06IRAMlines-eps-converted-to.pdf}
\figsetgrpnote{\textbf{Left}: $60\arcsec \times 60\arcsec$ SDSS ($gri$) cutouts of the QSO2s. The solid and dotted yellow circles show the extent of the $3\arcsec$ SDSS spectroscopy fiber and the $22\arcsec$ IRAM beam at 3 mm, respectively. \textbf{Middle}: CO(1--0) spectra in units of mJy. \textbf{Right}: CO(2--1) spectra in units of mJy. For both spectra, the x-axis indicates the velocity offset of the line from the systemic velocity of the galaxy as determined by the optical spectrum. Spectra are shown at 25 MHz resolution and were taken from the WILMA backend. The dark grey shaded area indicates the CO line region used for the determination of $I_{\mathrm{CO}}$ and the red line is the Gaussian fit. Horizontal grey lines indicate the baseline (solid) and the $3\sigma$ scatter around the baseline (dashed).}
\figsetgrpend

\figsetgrpstart
\figsetgrpnum{1.7}
\figsetgrptitle{SDSS07}
\figsetplot{./SDSS07IRAMlines-eps-converted-to.pdf}
\figsetgrpnote{\textbf{Left}: $60\arcsec \times 60\arcsec$ SDSS ($gri$) cutouts of the QSO2s. The solid and dotted yellow circles show the extent of the $3\arcsec$ SDSS spectroscopy fiber and the $22\arcsec$ IRAM beam at 3 mm, respectively. \textbf{Middle}: CO(1--0) spectra in units of mJy. \textbf{Right}: CO(2--1) spectra in units of mJy. For both spectra, the x-axis indicates the velocity offset of the line from the systemic velocity of the galaxy as determined by the optical spectrum. Spectra are shown at 25 MHz resolution and were taken from the WILMA backend. The dark grey shaded area indicates the CO line region used for the determination of $I_{\mathrm{CO}}$ and the red line is the Gaussian fit. Horizontal grey lines indicate the baseline (solid) and the $3\sigma$ scatter around the baseline (dashed).}
\figsetgrpend

\figsetgrpstart
\figsetgrpnum{1.8}
\figsetgrptitle{SDSS08}
\figsetplot{./SDSS08IRAMlines-eps-converted-to.pdf}
\figsetgrpnote{\textbf{Left}: $60\arcsec \times 60\arcsec$ SDSS ($gri$) cutouts of the QSO2s. The solid and dotted yellow circles show the extent of the $3\arcsec$ SDSS spectroscopy fiber and the $22\arcsec$ IRAM beam at 3 mm, respectively. \textbf{Middle}: CO(1--0) spectra in units of mJy. \textbf{Right}: CO(2--1) spectra in units of mJy. For both spectra, the x-axis indicates the velocity offset of the line from the systemic velocity of the galaxy as determined by the optical spectrum. Spectra are shown at 25 MHz resolution and were taken from the WILMA backend. The dark grey shaded area indicates the CO line region used for the determination of $I_{\mathrm{CO}}$ and the red line is the Gaussian fit. Horizontal grey lines indicate the baseline (solid) and the $3\sigma$ scatter around the baseline (dashed).}
\figsetgrpend

\figsetgrpstart
\figsetgrpnum{1.9}
\figsetgrptitle{SDSS09}
\figsetplot{./SDSS09IRAMlines-eps-converted-to.pdf}
\figsetgrpnote{\textbf{Left}: $60\arcsec \times 60\arcsec$ SDSS ($gri$) cutouts of the QSO2s. The solid and dotted yellow circles show the extent of the $3\arcsec$ SDSS spectroscopy fiber and the $22\arcsec$ IRAM beam at 3 mm, respectively. \textbf{Middle}: CO(1--0) spectra in units of mJy. \textbf{Right}: CO(2--1) spectra in units of mJy. For both spectra, the x-axis indicates the velocity offset of the line from the systemic velocity of the galaxy as determined by the optical spectrum. Spectra are shown at 25 MHz resolution and were taken from the WILMA backend. The dark grey shaded area indicates the CO line region used for the determination of $I_{\mathrm{CO}}$ and the red line is the Gaussian fit. Horizontal grey lines indicate the baseline (solid) and the $3\sigma$ scatter around the baseline (dashed).}
\figsetgrpend

\figsetgrpstart
\figsetgrpnum{1.10}
\figsetgrptitle{SDSS10}
\figsetplot{./SDSS10IRAMlines-eps-converted-to.pdf}
\figsetgrpnote{\textbf{Left}: $60\arcsec \times 60\arcsec$ SDSS ($gri$) cutouts of the QSO2s. The solid and dotted yellow circles show the extent of the $3\arcsec$ SDSS spectroscopy fiber and the $22\arcsec$ IRAM beam at 3 mm, respectively. \textbf{Middle}: CO(1--0) spectra in units of mJy. \textbf{Right}: CO(2--1) spectra in units of mJy. For both spectra, the x-axis indicates the velocity offset of the line from the systemic velocity of the galaxy as determined by the optical spectrum. Spectra are shown at 25 MHz resolution and were taken from the WILMA backend. The dark grey shaded area indicates the CO line region used for the determination of $I_{\mathrm{CO}}$ and the red line is the Gaussian fit. Horizontal grey lines indicate the baseline (solid) and the $3\sigma$ scatter around the baseline (dashed).}
\figsetgrpend

\figsetgrpstart
\figsetgrpnum{1.11}
\figsetgrptitle{SDSS11}
\figsetplot{./SDSS11IRAMlines-eps-converted-to.pdf}
\figsetgrpnote{\textbf{Left}: $60\arcsec \times 60\arcsec$ SDSS ($gri$) cutouts of the QSO2s. The solid and dotted yellow circles show the extent of the $3\arcsec$ SDSS spectroscopy fiber and the $22\arcsec$ IRAM beam at 3 mm, respectively. \textbf{Middle}: CO(1--0) spectra in units of mJy. \textbf{Right}: CO(2--1) spectra in units of mJy. For both spectra, the x-axis indicates the velocity offset of the line from the systemic velocity of the galaxy as determined by the optical spectrum. Spectra are shown at 25 MHz resolution and were taken from the WILMA backend. The dark grey shaded area indicates the CO line region used for the determination of $I_{\mathrm{CO}}$ and the red line is the Gaussian fit. Horizontal grey lines indicate the baseline (solid) and the $3\sigma$ scatter around the baseline (dashed).}
\figsetgrpend

\figsetgrpstart
\figsetgrpnum{1.12}
\figsetgrptitle{SDSS12}
\figsetplot{./SDSS12IRAMlines-eps-converted-to.pdf}
\figsetgrpnote{\textbf{Left}: $60\arcsec \times 60\arcsec$ SDSS ($gri$) cutouts of the QSO2s. The solid and dotted yellow circles show the extent of the $3\arcsec$ SDSS spectroscopy fiber and the $22\arcsec$ IRAM beam at 3 mm, respectively. \textbf{Middle}: CO(1--0) spectra in units of mJy. \textbf{Right}: CO(2--1) spectra in units of mJy. For both spectra, the x-axis indicates the velocity offset of the line from the systemic velocity of the galaxy as determined by the optical spectrum. Spectra are shown at 25 MHz resolution and were taken from the WILMA backend. The dark grey shaded area indicates the CO line region used for the determination of $I_{\mathrm{CO}}$ and the red line is the Gaussian fit. Horizontal grey lines indicate the baseline (solid) and the $3\sigma$ scatter around the baseline (dashed).}
\figsetgrpend

\figsetgrpstart
\figsetgrpnum{1.13}
\figsetgrptitle{SDSS13}
\figsetplot{./SDSS13IRAMlines-eps-converted-to.pdf}
\figsetgrpnote{\textbf{Left}: $60\arcsec \times 60\arcsec$ SDSS ($gri$) cutouts of the QSO2s. The solid and dotted yellow circles show the extent of the $3\arcsec$ SDSS spectroscopy fiber and the $22\arcsec$ IRAM beam at 3 mm, respectively. \textbf{Middle}: CO(1--0) spectra in units of mJy. \textbf{Right}: CO(2--1) spectra in units of mJy. For both spectra, the x-axis indicates the velocity offset of the line from the systemic velocity of the galaxy as determined by the optical spectrum. Spectra are shown at 25 MHz resolution and were taken from the WILMA backend. The dark grey shaded area indicates the CO line region used for the determination of $I_{\mathrm{CO}}$ and the red line is the Gaussian fit. Horizontal grey lines indicate the baseline (solid) and the $3\sigma$ scatter around the baseline (dashed).}
\figsetgrpend

\figsetgrpstart
\figsetgrpnum{1.14}
\figsetgrptitle{SDSS14}
\figsetplot{./SDSS14IRAMlines-eps-converted-to.pdf}
\figsetgrpnote{\textbf{Left}: $60\arcsec \times 60\arcsec$ SDSS ($gri$) cutouts of the QSO2s. The solid and dotted yellow circles show the extent of the $3\arcsec$ SDSS spectroscopy fiber and the $22\arcsec$ IRAM beam at 3 mm, respectively. \textbf{Middle}: CO(1--0) spectra in units of mJy. \textbf{Right}: CO(2--1) spectra in units of mJy. For both spectra, the x-axis indicates the velocity offset of the line from the systemic velocity of the galaxy as determined by the optical spectrum. Spectra are shown at 25 MHz resolution and were taken from the WILMA backend. The dark grey shaded area indicates the CO line region used for the determination of $I_{\mathrm{CO}}$ and the red line is the Gaussian fit. Horizontal grey lines indicate the baseline (solid) and the $3\sigma$ scatter around the baseline (dashed).}
\figsetgrpend

\figsetgrpstart
\figsetgrpnum{1.15}
\figsetgrptitle{SDSS15}
\figsetplot{./SDSS15IRAMlines-eps-converted-to.pdf}
\figsetgrpnote{\textbf{Left}: $60\arcsec \times 60\arcsec$ SDSS ($gri$) cutouts of the QSO2s. The solid and dotted yellow circles show the extent of the $3\arcsec$ SDSS spectroscopy fiber and the $22\arcsec$ IRAM beam at 3 mm, respectively. \textbf{Middle}: CO(1--0) spectra in units of mJy. \textbf{Right}: CO(2--1) spectra in units of mJy. For both spectra, the x-axis indicates the velocity offset of the line from the systemic velocity of the galaxy as determined by the optical spectrum. Spectra are shown at 25 MHz resolution and were taken from the WILMA backend. The dark grey shaded area indicates the CO line region used for the determination of $I_{\mathrm{CO}}$ and the red line is the Gaussian fit. Horizontal grey lines indicate the baseline (solid) and the $3\sigma$ scatter around the baseline (dashed).}
\figsetgrpend

\figsetgrpstart
\figsetgrpnum{1.16}
\figsetgrptitle{SDSS16}
\figsetplot{./SDSS16IRAMlines-eps-converted-to.pdf}
\figsetgrpnote{\textbf{Left}: $60\arcsec \times 60\arcsec$ SDSS ($gri$) cutouts of the QSO2s. The solid and dotted yellow circles show the extent of the $3\arcsec$ SDSS spectroscopy fiber and the $22\arcsec$ IRAM beam at 3 mm, respectively. \textbf{Middle}: CO(1--0) spectra in units of mJy. \textbf{Right}: CO(2--1) spectra in units of mJy. For both spectra, the x-axis indicates the velocity offset of the line from the systemic velocity of the galaxy as determined by the optical spectrum. Spectra are shown at 25 MHz resolution and were taken from the WILMA backend. The dark grey shaded area indicates the CO line region used for the determination of $I_{\mathrm{CO}}$ and the red line is the Gaussian fit. Horizontal grey lines indicate the baseline (solid) and the $3\sigma$ scatter around the baseline (dashed).}
\figsetgrpend

\figsetgrpstart
\figsetgrpnum{1.17}
\figsetgrptitle{SDSS17}
\figsetplot{./SDSS17IRAMlines-eps-converted-to.pdf}
\figsetgrpnote{\textbf{Left}: $60\arcsec \times 60\arcsec$ SDSS ($gri$) cutouts of the QSO2s. The solid and dotted yellow circles show the extent of the $3\arcsec$ SDSS spectroscopy fiber and the $22\arcsec$ IRAM beam at 3 mm, respectively. \textbf{Middle}: CO(1--0) spectra in units of mJy. \textbf{Right}: CO(2--1) spectra in units of mJy. For both spectra, the x-axis indicates the velocity offset of the line from the systemic velocity of the galaxy as determined by the optical spectrum. Spectra are shown at 25 MHz resolution and were taken from the WILMA backend. The dark grey shaded area indicates the CO line region used for the determination of $I_{\mathrm{CO}}$ and the red line is the Gaussian fit. Horizontal grey lines indicate the baseline (solid) and the $3\sigma$ scatter around the baseline (dashed).}
\figsetgrpend

\figsetgrpstart
\figsetgrpnum{1.18}
\figsetgrptitle{SDSS18}
\figsetplot{./SDSS18IRAMlines-eps-converted-to.pdf}
\figsetgrpnote{\textbf{Left}: $60\arcsec \times 60\arcsec$ SDSS ($gri$) cutouts of the QSO2s. The solid and dotted yellow circles show the extent of the $3\arcsec$ SDSS spectroscopy fiber and the $22\arcsec$ IRAM beam at 3 mm, respectively. \textbf{Middle}: CO(1--0) spectra in units of mJy. \textbf{Right}: CO(2--1) spectra in units of mJy. For both spectra, the x-axis indicates the velocity offset of the line from the systemic velocity of the galaxy as determined by the optical spectrum. Spectra are shown at 25 MHz resolution and were taken from the WILMA backend. The dark grey shaded area indicates the CO line region used for the determination of $I_{\mathrm{CO}}$ and the red line is the Gaussian fit. Horizontal grey lines indicate the baseline (solid) and the $3\sigma$ scatter around the baseline (dashed).}
\figsetgrpend

\figsetgrpstart
\figsetgrpnum{1.19}
\figsetgrptitle{SDSS19}
\figsetplot{./SDSS19IRAMlines-eps-converted-to.pdf}
\figsetgrpnote{\textbf{Left}: $60\arcsec \times 60\arcsec$ SDSS ($gri$) cutouts of the QSO2s. The solid and dotted yellow circles show the extent of the $3\arcsec$ SDSS spectroscopy fiber and the $22\arcsec$ IRAM beam at 3 mm, respectively. \textbf{Middle}: CO(1--0) spectra in units of mJy. \textbf{Right}: CO(2--1) spectra in units of mJy. For both spectra, the x-axis indicates the velocity offset of the line from the systemic velocity of the galaxy as determined by the optical spectrum. Spectra are shown at 25 MHz resolution and were taken from the WILMA backend. The dark grey shaded area indicates the CO line region used for the determination of $I_{\mathrm{CO}}$ and the red line is the Gaussian fit. Horizontal grey lines indicate the baseline (solid) and the $3\sigma$ scatter around the baseline (dashed).}
\figsetgrpend

\figsetgrpstart
\figsetgrpnum{1.20}
\figsetgrptitle{SDSS20}
\figsetplot{./SDSS20IRAMlines-eps-converted-to.pdf}
\figsetgrpnote{\textbf{Left}: $60\arcsec \times 60\arcsec$ SDSS ($gri$) cutouts of the QSO2s. The solid and dotted yellow circles show the extent of the $3\arcsec$ SDSS spectroscopy fiber and the $22\arcsec$ IRAM beam at 3 mm, respectively. \textbf{Middle}: CO(1--0) spectra in units of mJy. \textbf{Right}: CO(2--1) spectra in units of mJy. For both spectra, the x-axis indicates the velocity offset of the line from the systemic velocity of the galaxy as determined by the optical spectrum. Spectra are shown at 25 MHz resolution and were taken from the WILMA backend. The dark grey shaded area indicates the CO line region used for the determination of $I_{\mathrm{CO}}$ and the red line is the Gaussian fit. Horizontal grey lines indicate the baseline (solid) and the $3\sigma$ scatter around the baseline (dashed).}
\figsetgrpend

\figsetend

\begin{deluxetable*}{cccccccccccc}
\tablecaption{Summary of CO (1--0) observations for QSO2s with the IRAM 30m telescope.\label{tab:CO10observations}}
\tablehead{\colhead{Source} & \colhead{R.A}  & \colhead{Dec.} & \colhead{$z$} & \colhead{t$_{\rm{obs}}$\tablenotemark{a}} & \colhead{$\sigma(T_{A}^{*})$\tablenotemark{b}} & \colhead{$\nu_{\rm{CO(1-0)}}$\tablenotemark{c}} & \colhead{V$_{\rm offset}$\tablenotemark{d}} & \colhead{FWHM} & \colhead{$I_{\rm{CO(1-0)}}$\tablenotemark{e}} & \colhead{$L^{\prime}_{\rm{CO(1-0)}}$\tablenotemark{f}} & \colhead{R$_{21}$}\\ 
\colhead{} & \colhead{\textdegree} & \colhead{\textdegree} & \colhead{} & \colhead{hr} & \colhead{$\mu$K} & \colhead{GHz} & \colhead{km/s} & \colhead{km/s} & \colhead{K km/s} & \colhead{10$^8$ K km/s pc$^2$} & \colhead{} \\
\colhead{(1)} & \colhead{(2)} & \colhead{(3)} & \colhead{(4)} & \colhead{(5)} & \colhead{(6)} & \colhead{(7)} & \colhead{(8)} & \colhead{(9)} & \colhead{(10)} & \colhead{(11)} & \colhead{(12)}}
\startdata
SDSS01 & $ 220.159$ & $ 53.504$ & $ 0.0370$ & $ 2.0$ & $ 714$ & $ 111.158$ & $ 188\pm39$ & $ 370\pm100$ & $ 1.008\pm0.216$ & $ 4.1\pm0.9$ & \\
SDSS02 & $ 158.536$ & $ 60.031$ & $ 0.0511$ & $ 2.1$ & $ 788$ & $ 109.667$ & $ -91\pm43$ & $ 387\pm89$ & $ 0.790\pm0.247$ & $ 6.0\pm1.9$ & \\
SDSS03 & $ 163.034$ & $ 6.154$ & $ 0.0522$ & $ 2.5$ & $ 700$ & $ 109.552$ & $ -3\pm31$ & $ 304\pm60$ & $ 0.674\pm0.188$ & $ 5.4\pm1.5$ & \\
SDSS04 & $ 130.937$ & $ 35.828$ & $ 0.0539$ & $ 1.6$ & $ 602$ & $ 109.376$ & $ 17\pm23$ & $ 342\pm40$ & $ 1.232\pm0.174$ & $ 10.4\pm1.5$ & 0.80\\
SDSS05 & $ 119.921$ & $ 50.840$ & $ 0.0544$ & $ 1.2$ & $ 737$ & $ 109.324$ & $ -17\pm16$ & $ 176\pm30$ & $ 0.736\pm0.142$ & $ 6.3\pm1.2$ & \\
SDSS06 & $ 194.712$ & $ 52.654$ & $ 0.0552$ & $ 1.4$ & $ 522$ & $ 109.241$ & $ 56\pm51$ & $ 379\pm86$ & $ 0.441\pm0.161$ & $ 3.9\pm1.4$ & \\
SDSS07 & $ 178.360$ & $ 58.112$ & $ 0.0645$ & $ 3.5$ & $ 425$ & $ 108.287$ & $ 77\pm18$ & $ 187\pm31$ & $ 0.362\pm0.085$ & $ 4.3\pm1.0$ & 1.35\\
SDSS08 & $ 120.906$ & $ 39.443$ & $ 0.0656$ & $ 2.3$ & $ 542$ & $ 108.175$ &   &   & $ <0.488$ & $ <6.0$ & \\
SDSS09 & $ 178.190$ & $ 10.273$ & $ 0.0699$ & $ 2.5$ & $ 401$ & $ 107.740$ &   &   & $ <0.358$ & $ <5.0$ & \\
SDSS10 & $ 200.942$ & $ 61.067$ & $ 0.0713$ & $ 1.9$ & $ 457$ & $ 107.599$ &   &   & $ <0.409$ & $ <5.9$ & \\
SDSS11 & $ 238.107$ & $ 27.895$ & $ 0.0745$ & $ 0.0$ &   & $ 107.279$ &   &   &   &  & \\
SDSS12 & $ 124.676$ & $ 36.069$ & $ 0.0758$ & $ 1.0$ & $ 668$ & $ 107.149$ &   &   & $ <0.598$ & $ <9.7$ & \\
SDSS13 & $ 120.578$ & $ 30.773$ & $ 0.0766$ & $ 1.5$ & $ 523$ & $ 107.069$ &   &   & $ <0.468$ & $ <7.7$ & \\
SDSS14 & $ 165.554$ & $ 64.990$ & $ 0.0776$ & $ 0.7$ & $ 680$ & $ 106.970$ & $ 70\pm25$ & $ 334\pm42$ & $ 1.187\pm0.195$ & $ 20.1\pm3.3$ & 0.45\\
SDSS15 & $ 211.422$ & $ 40.442$ & $ 0.0806$ & $ 1.0$ & $ 810$ & $ 106.673$ &   &   & $ <0.727$ & $ <13.2$ & \\
SDSS16 & $ 120.721$ & $ 25.882$ & $ 0.0811$ & $ 0.0$ &   & $ 106.624$ &   &   &   &   & \\
SDSS17 & $ 189.681$ & $ 9.460$ & $ 0.0829$ & $ 2.5$ & $ 473$ & $ 106.447$ &   &   & $ <0.425$ & $ <8.2$ & \\
SDSS18 & $ 136.975$ & $ 52.191$ & $ 0.0848$ & $ 0.5$ & $ 1245$ & $ 106.260$ &   &   & $ <1.119$ & $ <22.4$ & \\
SDSS19 & $ 217.625$ & $ 13.653$ & $ 0.0851$ & $ 2.4$ & $ 292$ & $ 106.231$ & $ 57\pm44$ & $ 536\pm123$ & $ 0.599\pm0.115$ & $ 12.1\pm2.3$ & 0.85\\
SDSS20 & $ 30.599$ & $ 12.788$ & $ 0.0859$ & $ 1.2$ & $ 779$ & $ 106.153$ & $ -23\pm74$ & $ 556\pm101$ & $ 1.200\pm0.314$ & $ 24.6\pm6.4$ & \\
\enddata
\tablecomments{The columns are: (1) Source name (2) R.A. (3) Dec. (4) Redshift (5) On-source observation time (6) RMS baseline fluctuations (7) Expected line frequency (8) Offset with respect to the expected frequency of the best fit Gaussian (9) Best fit Gaussian line width (FWHM) (10) CO intensity (11) CO luminosity (12) $\frac{L^{\prime}_{\rm CO(2-1)}}{L^{\prime}_{\rm CO(1-0)}}$ if both lines are robustly detected.}
\tablenotetext{a}{Effective integration time is $2t_{\rm{obs}}$ by combining both polarizations.}
\tablenotetext{b}{Using 64 resolution elements of 25 MHz.}
\tablenotetext{c}{Based on optical redshift.}
\tablenotetext{d}{Relative to optical redshift.}
\tablenotetext{e}{In units of $T_{A}^{*}$.}
\tablenotetext{f}{In units of $T_{mb}$.}
\end{deluxetable*}

\begin{deluxetable*}{cccccccccccc}
\tablecaption{Summary of CO (2--1) observations for QSO2s with the IRAM 30m telescope.\label{tab:CO21observations}}
\tablehead{\colhead{Source} & \colhead{R.A}  & \colhead{Dec.} & \colhead{$z$} & \colhead{t$_{\rm{obs}}$\tablenotemark{a}} & \colhead{$\sigma(T_{A}^{*})$\tablenotemark{b}} & \colhead{$\nu_{\rm{CO(2-1)}}$\tablenotemark{c}} & \colhead{V$_{\rm offset}$\tablenotemark{d}} & \colhead{FWHM} & \colhead{$I_{\rm{CO(2-1)}}$\tablenotemark{e}} & \colhead{$L^{\prime}_{\rm{CO(2-1)}}$\tablenotemark{f}} & \colhead{R$_{21}$}\\ 
\colhead{} & \colhead{\textdegree} & \colhead{\textdegree} & \colhead{} & \colhead{hr} & \colhead{$\mu$K} & \colhead{GHz} & \colhead{km/s} & \colhead{km/s} & \colhead{K km/s} & \colhead{10$^8$ K km/s pc$^2$} & \colhead{} \\
\colhead{(1)} & \colhead{(2)} & \colhead{(3)} & \colhead{(4)} & \colhead{(5)} & \colhead{(6)} & \colhead{(7)} & \colhead{(8)} & \colhead{(9)} & \colhead{(10)} & \colhead{(11)} & \colhead{(12)}}
\startdata
SDSS01 & $ 220.159$ & $ 53.504$ & $ 0.0370$ & $ 0.3$ & $ 4794$ & $ 222.312$ &   &   & $ <2.594$ & $ <4.2$ & \\
SDSS02 & $ 158.536$ & $ 60.031$ & $ 0.0511$ & $ 1.2$ & $ 2367$ & $ 219.330$ &   &   & $ <1.287$ & $ <3.9$ & \\
SDSS03 & $ 163.034$ & $ 6.154$ & $ 0.0522$ & $ 0.7$ & $ 5630$ & $ 219.101$ &   &   & $ <3.061$ & $ <9.6$ & \\
SDSS04 & $ 130.937$ & $ 35.828$ & $ 0.0539$ & $ 1.6$ & $ 2372$ & $ 218.748$ & $ 35\pm39$ & $ 240\pm67$ & $ 2.501\pm0.428$ & $ 8.3\pm1.4$ & 0.80\\
SDSS05 & $ 119.921$ & $ 50.840$ & $ 0.0544$ & $ 0.0$ &   & $ 218.644$ &   &   &   &   & \\
SDSS06 & $ 194.712$ & $ 52.654$ & $ 0.0552$ & $ 1.4$ & $ 904$ & $ 218.478$ &   &   & $ <0.492$ & $ <1.7$ & \\
SDSS07 & $ 178.360$ & $ 58.112$ & $ 0.0645$ & $ 2.1$ & $ 586$ & $ 216.569$ & $ 70\pm14$ & $ 244\pm27$ & $ 1.226\pm0.107$ & $ 5.8\pm0.5$ & 1.35\\
SDSS08 & $ 120.906$ & $ 39.443$ & $ 0.0656$ & $ 1.1$ & $ 1170$ & $ 216.346$ &   &   & $ <0.651$ & $ <3.2$ & \\
SDSS09 & $ 178.190$ & $ 10.273$ & $ 0.0699$ & $ 0.6$ & $ 923$ & $ 215.476$ &   &   & $ <0.504$ & $ <2.8$ & \\
SDSS10 & $ 200.942$ & $ 61.067$ & $ 0.0713$ & $ 2.0$ & $ 570$ & $ 215.195$ &   &   & $ <0.312$ & $ <1.8$ & \\
SDSS11 & $ 238.107$ & $ 27.895$ & $ 0.0745$ & $ 1.0$ & $ 1396$ & $ 214.554$ &   &   & $ <0.775$ & $ <4.8$ & \\
SDSS12 & $ 124.676$ & $ 36.069$ & $ 0.0758$ & $ 0.0$ &   & $ 214.294$ &   &   &   &   & \\
SDSS13 & $ 120.578$ & $ 30.773$ & $ 0.0766$ & $ 0.0$ &   & $ 214.135$ &   &   &   &   & \\
SDSS14 & $ 165.554$ & $ 64.990$ & $ 0.0776$ & $ 1.0$ & $ 838$ & $ 213.937$ & $ 67\pm13$ & $ 243\pm22$ & $ 1.352\pm0.154$ & $ 9.0\pm1.0$ & 0.45\\
SDSS15 & $ 211.422$ & $ 40.442$ & $ 0.0806$ & $ 1.8$ & $ 989$ & $ 213.343$ & $ 57\pm18$ & $ 126\pm38$ & $ 0.409\pm0.119$ & $ 2.9\pm0.9$ & \\
SDSS16 & $ 120.721$ & $ 25.882$ & $ 0.0811$ & $ 1.5$ & $ 1252$ & $ 213.244$ & $ 11\pm37$ & $ 220\pm54$ & $ 0.403\pm0.215$ & $ 2.9\pm1.6$ & \\
SDSS17 & $ 189.681$ & $ 9.460$ & $ 0.0829$ & $ 0.0$ &   & $ 212.889$ &   &   &   &   & \\
SDSS18 & $ 136.975$ & $ 52.191$ & $ 0.0848$ & $ 1.6$ & $ 1195$ & $ 212.517$ &   &   & $ <0.656$ & $ <5.2$ & \\
SDSS19 & $ 217.625$ & $ 13.653$ & $ 0.0851$ & $ 1.4$ & $ 1352$ & $ 212.458$ & $ 160\pm35$ & $ 330\pm63$ & $ 1.292\pm0.308$ & $ 10.3\pm2.4$ & 0.85\\
SDSS20 & $ 30.599$ & $ 12.788$ & $ 0.0859$ & $ 0.0$ &   & $ 212.301$ &   &   &   &   & \\
\enddata
\tablecomments{The columns are: (1) Source name (2) R.A. (3) Dec. (4) Redshift (5) On-source observation time (6) RMS baseline fluctuations (7) Expected line frequency (8) Offset with respect to the expected frequency of the best fit Gaussian (9) Best fit Gaussian line width (FWHM) (10) CO intensity (11) CO luminosity (12) $\frac{L^{\prime}_{\rm CO(2-1)}}{L^{\prime}_{\rm CO(1-0)}}$ if both lines are robustly detected.}
\tablenotetext{a}{Effective integration time is $2t_{\rm{obs}}$ by combining both polarizations.}
\tablenotetext{b}{Using 64 resolution elements of 25 MHz.}
\tablenotetext{c}{Based on optical redshift.}
\tablenotetext{d}{Relative to optical redshift.}
\tablenotetext{e}{In units of $T_{A}^{*}$.}
\tablenotetext{f}{In units of $T_{mb}$.}
\end{deluxetable*}

\begin{deluxetable*}{cccccccc}
\tablecaption{Summary of [\ion{C}{2}] observations with SOFIA FIFI-LS.\label{tab:CIIobservations}}
\tablehead{\colhead{Source} & \colhead{Type} & \colhead{R.A}  & \colhead{Dec.} & \colhead{$z$} & \colhead{$I_{\rm{[OIII]}88\mu m}$} & \colhead{$I_{\rm{[CII]}158\mu m}$} & \colhead{$I_{\rm{[CII]}158\mu m,ext}$} \\ 
\colhead{} & \colhead{} & \colhead{\textdegree} & \colhead{\textdegree} & \colhead{} & \colhead{$\times 10^{-17}$ W/m$^2$} & \colhead{$\times 10^{-17}$ W/m$^2$} & \colhead{$\times 10^{-17}$ W/m$^2$} \\
\colhead{(1)} & \colhead{(2)} & \colhead{(3)} & \colhead{(4)} & \colhead{(5)} & \colhead{(6)} & \colhead{(7)} & \colhead{(8)}}
\startdata
Mrk 290 (PG 1534+580) & QSO1 & 233.970 & 57.902 & 0.0296 & 2.9 $\pm$ 0.44 & $<$1.6 & \\
UGC 05025 (PG 0923+129) & QSO1 & 141.514 & 12.734 & 0.0288 & 9.5 $\pm$ 1.4 & 7.2 $\pm$ 1.1 & 18 $\pm$ 2.7\\
Mrk 110 (PG 0921+525) & QSO1 & 141.304 & 52.286 & 0.0353 & & 7.0 $\pm$ 1.1 & 15 $\pm$ 2.25\\
Sz II 10 (PG 1310-108) & QSO1 & 198.274 & -11.128 & 0.0343 & & 3.9 $\pm$ 0.98& \\
Mrk 335 (PG 0003+199) & QSO1 & 1.581 & 20.203 & 0.0258 & 5.2 $\pm$ 0.78 & 6.0 $\pm$ 0.90 & 20 $\pm$ 3.0 \\
\hline
SDSS01 & QSO2 & 220.159 & 53.504 & 0.0370 & & 5.7 $\pm$ 0.86& 10 $\pm$ 1.5 \\
SDSS02 & QSO2 & 158.536 & 60.031 & 0.0511 & 7.5 $\pm$ 1.1 & 4.0 $\pm$ 0.60 &8.7 $\pm$ 1.3\\
SDSS03 & QSO2 & 163.034 & 6.154 & 0.0522 & & 1.9 $\pm$ 0.29 & \\
SDSS04 & QSO2 & 130.937 & 35.828 & 0.0539 & & 2.5 $\pm$ 0.38 & \\
SDSS07 & QSO2 & 178.360 & 58.112 & 0.0645 & & 0.8 $\pm$ 0.12 & 1.4 $\pm$ 0.21\\
\enddata
\tablecomments{The columns are: (1) Source name (2) Source type (3) R.A. (4) Dec. (5) Redshift (6) [\ion{O}{3}] 88$\mu$m line intensity within the PSF aperture (diameter 8.8\arcsec) (7) [\ion{C}{2}] 158$\mu$m line intensity within the PSF aperture (diameter 15.2\arcsec) (8) [\ion{C}{2}] 158$\mu$m line intensity within an extended aperture (diameter 30\arcsec for UGC 05025, diameter 22\arcsec for others to best capture all extended flux). The uncertainty in the measured line intensities is dominated by the calibration error (15\%, except for Sz II 10 which has a 25\% error due to higher background uncertainty).}
\end{deluxetable*}

\begin{table*}
\centering
\caption{Two-sample statistical test results on $L^{\prime}_{\rm{CO}}$, $L_{\rm{160\mu m}}$,  $L^{\prime}_{\rm{CO}}$/$L_{\rm{160\mu m}}$, M$_{*}$, M$_{\rm{H_2,CO}}$, and M$_{\rm{H_2,CO}}$/M$_{*}$ of QSO1s, QSO2s, and LIRGs}
\label{tab:stats}
\begin{tabular}{ccc}
\hline \hline
\textbf{Comparison pair} & \textbf{Test statistics} & \textbf{Test p-values} \\
& $t_{\rm{logrank}}$, $t_{\rm{peto}}$, $t_{\rm{wilcoxon}}$, $t_{\rm{tarone}}$$^a$  & $p_{\rm{logrank}}$, $p_{\rm{peto}}$, $p_{\rm{wilcoxon}}$, $p_{\rm{tarone}}$$^b$\\
\hline
\textbf{\textit{LIRGs (63), QSO1s (40)$^c$}} &  & \\
log($L^{\prime}_{\rm{CO}}$) & 36.4, 26.1, 25.1, 30.8 & 1.6$\times$10$^{-9}$, 3.3$\times$10$^{-7}$, 5.3$\times$10$^{-7}$, 2.9$\times$10$^{-8}$ \\
log($L_{\rm{160\mu m}}$) & 34.6, 29.8, 29.1, 32.8 & 4.1$\times$10$^{-9}$, 4.8$\times$10$^{-8}$, 7.0$\times$10$^{-8}$, 1.0$\times$10$^{-8}$\\
$L^{\prime}_{\rm{CO}}$/$L_{\rm{160\mu m}}$ & 0.03, 0.4, 0.4, 0.07 & 0.9, 0.6, 0.5, 0.8 \\
M$_{*}$ & 5.0, 3.4, 3.2, 4.1 & 0.03, 0.07, 0.07, 0.04\\
M$_{\rm{H_2,CO}}$ & 20.0, 11.4, 10.8, 14.9 & 7.8$\times$10$^{-6}$, 7.4$\times$10$^{-4}$, 1.0$\times$10$^{-3}$, 1.1$\times$10$^{-4}$\\
M$_{\rm{H_2,CO}}$/M$_{*}$ & 17.0, 8.2, 7.8, 11.6 & 3.7$\times$10$^{-5}$, 4.1$\times$10$^{-3}$, 5.3$\times$10$^{-3}$, 6.5$\times$10$^{-4}$\\
\hline
\textbf{\textit{LIRGs (63), QSO2s (20)}} &  & \\
log($L^{\prime}_{\rm{CO}}$) & 30.7, 24.6, 23.9, 27.9 & 3.1$\times$10$^{-8}$, 6.9$\times$10$^{-7}$, 1.0$\times$10$^{-6}$, 1.3$\times$10$^{-7}$ \\
log($L_{\rm{160\mu m}}$) & 16.7, 24.9, 24.9, 24.0 & 4.3$\times$10$^{-5}$, 6.0$\times$10$^{-7}$, 6.0$\times$10$^{-7}$, 9.9$\times$10$^{-7}$ \\
$L^{\prime}_{\rm{CO}}$/$L_{\rm{160\mu m}}$ & 0.20, 0.06, 0.04, 0.10 & 0.7, 0.8, 0.8, 0.8\\
M$_{*}$ & 3.7, 7.0, 7.0, 6.0 & 0.05, 0.01, 0.01, 0.01\\
M$_{\rm{H_2,CO}}$ & 4.5, 6.4, 6.3, 6.3 & 0.03, 0.01, 0.01, 0.01\\
M$_{\rm{H_2,CO}}$/M$_{*}$ & 0.63, 0.18, 0.14, 0.32 & 0.43, 0.67, 0.71, 0.57\\
\hline
\textbf{\textit{QSO1s (40), QSO2s (20)}} &  & \\
log($L^{\prime}_{\rm{CO}}$) & 0.16, 0.15, 0.45, 0.05 & 0.7, 0.7, 0.5, 0.8 \\
log($L_{\rm{160\mu m}}$) & 2.2, 0.17, 0.07, 0.65 & 0.1, 0.7, 0.8, 0.4 \\
$L^{\prime}_{\rm{CO}}$/$L_{\rm{160\mu m}}$ & 0.002, 0.2, 0.3, 0.1 & 1.0, 0.7, 0.6, 0.8 \\
M$_{*}$ & 0.002, 0.8, 0.9, 0.3 & 1.0, 0.4, 0.4, 0.6 \\
M$_{\rm{H_2,CO}}$ & 1.3, 0.18, 0.02, 0.3 & 0.3, 0.7, 0.9, 0.6\\
M$_{\rm{H_2,CO}}$/M$_{*}$ & 2.3, 1.5, 1.4, 1.7 & 0.13, 0.22, 0.24, 0.19\\
\hline \hline
\multicolumn{3}{p{16cm}}{$^a$The calculated test statistics from the Logrank, Peto-Peto, Wilcoxon and Tarone–Ware two-sample tests.}\\
\multicolumn{3}{p{16cm}}{$^b$The derived $p$-values from the Logrank, Peto-Peto, Wilcoxon and Tarone–Ware two-sample tests.}\\
\multicolumn{3}{p{16cm}}{$^c$The sizes of the samples that go into the tests. For $L^{\prime}_{\rm{CO}}$/$L_{\rm{160\mu m}}$, we do not include in tests the two QSO1s for which both $L^{\prime}_{\rm{CO}}$ and $L_{\rm{160\mu m}}$ are upper limits.}\\ 
\end{tabular}
\end{table*}

\begin{acknowledgments}
YL, AP, and DF gratefully acknowledge the support of grant SOF-08-0226 from USRA. YL acknowledges support from the Space Telescope Science Institute Director's Discretionary Research Fund grant D0101.90281. RJ was in part supported by an appointment to the NASA Postdoctoral Program at the NASA Jet Propulsion Laboratory, administered by Universities Space Research Association under contract with NASA. A.~M.~J. gratefully acknowledge the support of grant SOF-08-0038 from USRA and the Max Planck Society. This work is based on observations made with the NASA/ DLR Stratospheric Observatory for Infrared Astronomy (SOFIA). SOFIA is jointly operated by the Universities Space Research Association, Inc. (USRA), under NASA contract NNA17BF53C, and the Deutsches SOFIA Institut (DSI) under DLR contract 50 OK 0901 to the University of Stuttgart. The Herschel spacecraft was designed, built, tested, and launched under a contract to ESA managed by the Herschel/Planck Project team by an industrial consortium under the overall responsibility of the prime contractor Thales Alenia Space (Cannes), and including Astrium (Friedrichshafen) responsible for the payload module and for system testing at spacecraft level, Thales Alenia Space (Turin) responsible for the service module, and Astrium (Toulouse) responsible for the telescope, with in excess of a hundred subcontractors.
\end{acknowledgments}

%

\vspace{5mm}
\facilities{IRAM, SOFIA(FIFI-LS)}


\software{astropy \citep{astropy}, lifelines \citep{lifelines}, SOSPEX \citep{Fadda2018}}


\section{Additional information} \label{sec:additional}
Here we provide tables that summarize the ancillary data for our IRAM QSO2 sample as well as for the LIRGs (GOALS) and QSO1s (PG QSO) comparison samples. These data include the CO, [\ion{C}{2}], M$_*$, and FIR data we obtain from the literature and some derived properties such as CO luminosity in $L_{\odot}$ unit.

\begin{longrotatetable}
\begin{deluxetable*}{cccccccccccccccccc}
\tabletypesize{\tiny} 
\tablecaption{LIRGs literature data. \label{tab:goals_sample}}
\tablehead{\colhead{Name} & \colhead{$z$} & \colhead{M$_*$} & \colhead{$L_{\rm{160\mu m}}$} & \colhead{$\Delta$$L_{\rm{160\mu m}}$} & \colhead{F$_{\rm{AGN}}$} & \colhead{I$_{\rm{[CII]}}$} & \colhead{$\Delta$I$_{\rm{[CII]}}$} & \colhead{$L_{\rm{[CII]}}$} & \colhead{$\Delta$$L_{\rm{[CII]}}$} & \colhead{CO ref} & \colhead{$L^{\prime}_{\rm{CO}}$} & \colhead{$\Delta$$L^{\prime}_{\rm{CO}}$} & \colhead{$L_{\rm{CO}}$} & \colhead{$\Delta$$L_{\rm{CO}}$} & \colhead{M$_{\rm{H_2,CO}}$} & \colhead{SFR$_{\rm{IR}}$} & \colhead{$\Delta$SFR$_{\rm{IR}}$}  \\ 
\colhead{} & \colhead{} & \colhead{log(M$_{\odot}$)} & \colhead{log($L_{\odot}$)} & \colhead{log($L_{\odot}$)} & \colhead{} & \colhead{$\times 10^{-17}$ W/m$^2$} & \colhead{$\times 10^{-17}$ W/m$^2$} & \colhead{log($L_{\odot}$)} & \colhead{log($L_{\odot}$)} & \colhead{} & \colhead{log(K km/s pc$^2$)} & \colhead{log(K km/s pc$^2$)} & \colhead{log($L_{\odot}$)} & \colhead{log($L_{\odot}$)} & \colhead{log(M$_{\odot}$)} & \colhead{M$_{\odot}$/yr} & \colhead{M$_{\odot}$/yr} \\
\colhead{(1)} & \colhead{(2)} & \colhead{(3)} & \colhead{(4)} & \colhead{(5)} & \colhead{(6)} & \colhead{(7)} & \colhead{(8)} & \colhead{(9)} & \colhead{(10)} & \colhead{(11)} & \colhead{(12)} & \colhead{(13)} & \colhead{(14)} & \colhead{(15)} & \colhead{(16)} & \colhead{(17)} & \colhead{(18)}} 
\startdata
NGC 0232                &  0.02217  &  11.04  & 10.86  & 0.019  & 0.01    &  138.0   &    2.32   &  8.6     &  0.007     &    1       &    9.66    &     0.04     &   5.38    &    0.04    &    9.91    &  21.28      & 0.835 \\
NGC 0235A               &  0.02217  &  10.93  & 10.14  & 0.019  & 0.71    &   26.06  &    2.22   &  7.88    &  0.037     &    1       &    9.29    &     0.067    &   5.01    &    0.067   &    9.55    &   4.79      & 0.189 \\
IC 1623                 &  0.02007  &  11.21  & 10.98  & 0.022  & 0.47    &  561.57  &    2.78   &  9.13    &  0.002     &    2       &    9.93    &     0.043    &   5.62    &    0.002   &   10.18    &  26.79      & 1.213 \\
MCG-03-04-014           &  0.03349  &  11.21  & 11.04  & 0.019  & 0.0     &  143.43  &    3.11   &  8.99    &  0.009     &    2       &    9.73    &     0.044    &   5.38    &    0.005   &    9.99    &  30.46      & 1.215 \\
CGCG 436-030            &  0.03123  &  10.85  & 10.87  & 0.019  & 0.35    &  121.38  &    1.62   &  8.85    &  0.006     &    1       &    9.5     &     0.026    &   5.24    &    0.026   &    9.76    &  21.35      & 0.851 \\
IRAS F01364-1042        &  0.04825  &  10.64  & 11.01  & 0.019  & 0.12    &   14.25  &    0.62   &  8.31    &  0.019     &    2       &    9.6     &     0.046    &   5.29    &    0.016   &    9.86    &  28.59      & 1.122 \\
NGC 0695                &  0.03247  &  11.3   & 11.12  & 0.022  & 0.0     &  283.75  &    1.35   &  9.26    &  0.002     &    1       &   10.01    &     0.013    &   5.75    &    0.013   &   10.26    &  36.12      & 1.626 \\
UGC 01385               &  0.01875  &  10.68  & 10.33  & 0.019  & 0.0     &   60.28  &    0.54   &  8.1     &  0.004     &    1       &    9.01    &     0.012    &   4.74    &    0.012   &    9.27    &   7.06      & 0.285 \\
UGC 02238               &  0.02188  &  10.94  & 10.85  & 0.022  & 0.0     &  266.56  &    1.92   &  8.88    &  0.003     &    1       &    9.41    &     0.008    &   5.14    &    0.008   &    9.66    &  20.78      & 0.929 \\
NGC 1275                &  0.01756  &  11.63  & 10.28  & 0.022  & 0.97    &  162.31  &    3.73   &  8.47    &  0.01      &    1       &    9.11    &     0.02     &   4.83    &    0.02    &    9.36    &   6.37      & 0.288 \\
IRAS F03359+1523        &  0.0354   &  10.75  & 10.83  & 0.019  & 0.0     &   47.38  &    2.02   &  8.56    &  0.019     &    1       &    9.55    &     0.061    &   5.29    &    0.061   &    9.8     &  19.68      & 0.773 \\
CGCG 465-012N           &  0.02222  &  10.67  & 10.34  & 0.022  & 0.0     &  108.49  &    1.47   &  8.5     &  0.006     &    1       &    9.15    &     0.016    &   4.88    &    0.016   &    9.4     &   7.24      & 0.326 \\
CGCG 465-012S           &  0.02222  &  10.84  & 10.58  & 0.022  & 0.0     &  138.49  &    1.44   &  8.61    &  0.005     &    1       &    9.48    &     0.008    &   5.21    &    0.008   &    9.74    &  11.71      & 0.528 \\
CGCG 468-002 NED01      &  0.01819  &  10.82  &  9.93  & 0.016  & 0.79    &   18.88  &    0.6    &  7.57    &  0.014     &    1       &    8.93    &     \nodata\tablenotemark{a}  &   4.64    &    \nodata &    9.18    &   3.1       & 0.102 \\
CGCG 468-002 NED02      &  0.01819  &  10.26  & 10.37  & 0.017  & 0.0     &   54.22  &    0.65   &  8.02    &  0.005     &    1       &    8.92    &     \nodata  &   4.64    &    \nodata &    9.18    &   7.58      & 0.265 \\
IRAS 05083+2441         &  0.02307  &  10.75  & 10.44  & 0.019  & 0.0     &  127.98  &    1.35   &  8.61    &  0.005     &    1       &    8.89    &     0.062    &   4.61    &    0.062   &    9.15    &   8.81      & 0.34  \\
IRAS 05129+5128         &  0.02743  &  10.71  & 10.62  & 0.019  & 0.01    &  108.86  &    0.93   &  8.69    &  0.004     &    1       &    9.39    &     0.04     &   5.11    &    0.04    &    9.65    &  12.73      & 0.499 \\
IRAS F05189-2524        &  0.04256  &  11.5   & 11.08  & 0.019  & 0.94    &   23.53  &    2.6    &  8.42    &  0.048     &    2       &    9.42    &     0.044    &   5.1     &    0.007   &    9.67    &  33.57      & 1.315 \\
UGC 03410               &  0.01308  &  10.93  & 10.54  & 0.022  & 0.0     &  388.21  &    2.35   &  8.59    &  0.003     &    1       &    9.05    &     0.047    &   4.71    &    0.047   &    9.3     &  10.89      & 0.492 \\
UGC 03405               &  0.01308  &  10.57  &  9.98  & 0.022  & 0.0     &  101.92  &    1.0    &  8.01    &  0.004     &    1       &    9.35    &     0.038    &   5.0     &    0.038   &    9.6     &   3.41      & 0.154 \\
NGC 2342                &  0.0176   &  11.1   & 10.66  & 0.022  & 0.0     &  322.59  &    1.65   &  8.77    &  0.002     &    2       &    9.19    &     0.044    &   4.88    &    0.007   &    9.45    &  13.96      & 0.626 \\
NGC 2341                &  0.0176   &  10.85  & 10.48  & 0.022  & \nodata &  254.15  &    1.5    &  8.67    &  0.003     &    2       &    9.14    &     0.044    &   4.86    &    0.01    &    9.4     &   9.51      & 0.428 \\
NGC 2623                &  0.01851  &  10.81  & 10.82  & 0.019  & 0.13    &   68.72  &    1.87   &  8.14    &  0.012     &    1       &    9.2     &     0.032    &   4.86    &    0.032   &    9.45    &  19.29      & 0.767 \\
UGC 05101               &  0.03937  &  11.29  & 11.43  & 0.018  & 0.71    &   71.25  &    3.91   &  8.83    &  0.024     &    2       &    9.79    &     0.044    &   5.48    &    0.008   &   10.04    &  68.94      & 2.638 \\
MCG+08-18-013           &  0.02594  &  10.83  & 10.68  & 0.019  & \nodata &   67.88  &    1.52   &  8.44    &  0.01      &    1       &    9.23    &     0.031    &   4.92    &    0.031   &    9.49    &  14.42      & 0.57  \\
IRAS F10173+0828        &  0.04909  &  10.56  & 10.92  & 0.019  & 0.19    &    3.61  &    1.18   &  7.73    &  0.142     &    1       &    9.64    &     0.044    &   5.35    &    0.044   &    9.89    &  23.89      & 0.956 \\
CGCG 011-076            &  0.0249   &  11.12  & 10.73  & 0.019  & 0.25    &   85.87  &    2.0    &  8.5     &  0.01      &    2       &    9.5     &     0.045    &   5.2     &    0.01    &    9.76    &  16.24      & 0.647 \\
IC 2810                 &  0.034    &  11.08  & 10.91  & 0.019  & 0.0     &   44.12  &    1.56   &  8.49    &  0.015     &    1       &    9.33    &     0.054    &   5.0     &    0.054   &    9.58    &  23.2       & 0.892 \\
IC 2810E                &  0.034    &  10.8   & 10.6   & 0.018  & \nodata &   19.62  &    0.61   &  8.14    &  0.014     &    1       &    9.36    &     0.042    &   5.03    &    0.042   &    9.61    &  12.34      & 0.466 \\
ESO 507-G070            &  0.0217   &  11.09  & 10.74  & 0.019  & 0.0     &   96.86  &    2.69   &  8.43    &  0.012     &    2       &    9.42    &     0.045    &   5.12    &    0.012   &    9.67    &  16.35      & 0.649 \\
UGC 08387               &  0.0233   &  10.83  & 11.03  & 0.019  & 0.0     &  172.03  &    2.01   &  8.74    &  0.005     &    1       &    9.73    &     0.015    &   5.37    &    0.015   &    9.99    &  30.05      & 1.212 \\
NGC 5104                &  0.01861  &  11.13  & 10.66  & 0.019  & 0.08    &  128.49  &    1.69   &  8.42    &  0.006     &    2       &    9.31    &     0.044    &   5.01    &    0.01    &    9.57    &  13.95      & 0.554 \\
IC 4280                 &  0.01631  &  11.16  & 10.56  & 0.022  & 0.0     &  263.55  &    1.11   &  8.62    &  0.002     &    2       &    9.16    &     0.045    &   4.85    &    0.013   &    9.41    &  11.42      & 0.517 \\
UGC 08739               &  0.01679  &  11.0   & 10.69  & 0.022  & 0.05    &  224.15  &    1.76   &  8.57    &  0.003     &    2       &    9.43    &     0.044    &   5.12    &    0.006   &    9.68    &  14.95      & 0.67  \\
CGCG 247-020            &  0.02574  &  10.74  & 10.55  & 0.022  & 0.0     &   34.27  &    0.65   &  8.13    &  0.008     &    2       &    9.29    &     0.045    &   4.97    &    0.013   &    9.54    &  11.09      & 0.501 \\
IRAS F14348-1447        &  0.08273  &  11.57  & 11.63  & 0.019  & 0.57    &   32.8   &    1.32   &  9.16    &  0.017     &    2       &   10.23    &     0.049    &   5.92    &    0.023   &   10.48    & 103.14      & 4.052 \\
CGCG 049-057            &  0.013    &  10.29  & 10.6   & 0.02   & 0.02    &   44.94  &    1.86   &  7.65    &  0.018     &    2       &    8.96    &     0.043    &   4.65    &    0.004   &    9.22    &  12.36      & 0.505 \\
NGC 5936                &  0.01336  &  10.99  & 10.47  & 0.022  & 0.0     &  304.26  &    1.34   &  8.5     &  0.002     &    2       &    9.1     &     0.044    &   4.79    &    0.005   &    9.35    &   9.32      & 0.42  \\
Arp 220                 &  0.01813  &  11.06  & 11.49  & 0.02   & 0.67    &  137.44  &    6.1    &  8.43    &  0.019     &    1       &    9.57    &     0.022    &   5.18    &    0.022   &    9.82    &  78.07      & 3.248 \\
IRAS F16164-0746        &  0.02715  &  10.87  & 10.92  & 0.019  & 0.0     &   79.94  &    0.94   &  8.55    &  0.005     &    2       &    9.43    &     0.046    &   5.24    &    0.013   &    9.68    &  23.69      & 0.94  \\
CGCG 052-037            &  0.02449  &  11.07  & 10.81  & 0.019  & 0.0     &  114.93  &    1.95   &  8.61    &  0.007     &    2       &    9.44    &     0.044    &   5.13    &    0.006   &    9.69    &  19.15      & 0.762 \\
NGC 6286                &  0.01835  &  11.1   & 10.87  & 0.022  & 0.0     &  296.82  &    1.32   &  8.77    &  0.002     &    2       &    9.63    &     0.044    &   5.31    &    0.005   &    9.89    &  21.34      & 0.964 \\
NGC 6285                &  0.01835  &  10.58  & 10.05  & 0.019  & \nodata &   68.33  &    1.88   &  8.13    &  0.012     &    2       &    9.04    &     0.045    &   4.69    &    0.013   &    9.29    &   3.91      & 0.154 \\
IRAS F17207-0014        &  0.04281  &  11.18  & 11.67  & 0.019  & 0.14    &  100.4   &    2.97   &  9.05    &  0.013     &    1       &    9.6     &     \nodata  &   5.29    &    \nodata &    9.85    & 112.05      & 4.459 \\
UGC 11041               &  0.01628  &  10.9   & 10.5   & 0.022  & 0.0     &  226.91  &    1.65   &  8.55    &  0.003     &    2       &    9.29    &     0.044    &   4.98    &    0.006   &    9.54    &  10.09      & 0.456 \\
CGCG 141-034            &  0.01983  &  10.75  & 10.53  & 0.019  & 0.09    &   67.24  &    2.45   &  8.19    &  0.016     &    2       &    8.99    &     0.048    &   4.67    &    0.021   &    9.25    &  10.61      & 0.416 \\
ESO 593-IG008           &  0.04873  &  11.55  & 11.36  & 0.019  & 0.0     &  135.7   &    3.24   &  9.3     &  0.01      &    2       &   10.09    &     0.045    &   5.78    &    0.01    &   10.35    &  59.86      & 2.381 \\
NGC 6907                &  0.01064  &  11.19  & 10.59  & 0.022  & 0.0     &  314.32  &    1.55   &  8.32    &  0.002     &    2       &    9.15    &     0.044    &   4.84    &    0.007   &    9.41    &  12.14      & 0.548 \\
ESO 602-G025            &  0.02504  &  11.14  & 10.81  & 0.019  & 0.16    &  135.48  &    2.4    &  8.7     &  0.008     &    2       &    9.59    &     0.044    &   5.27    &    0.006   &    9.84    &  18.94      & 0.754 \\
UGC 12150               &  0.02139  &  11.04  & 10.79  & 0.019  & 0.01    &   95.97  &    2.96   &  8.42    &  0.013     &    2       &    9.21    &     0.046    &   4.9     &    0.015   &    9.47    &  18.34      & 0.724 \\
IRAS F22491-1808        &  0.07776  &  11.36  & 11.25  & 0.019  & 0.11    &   17.09  &    0.43   &  8.82    &  0.011     &    2       &    9.74    &     0.045    &   5.43    &    0.013   &   10.0     &  47.71      & 1.835 \\
CGCG 453-062            &  0.0251   &  10.95  & 10.81  & 0.019  & 0.0     &  106.99  &    2.43   &  8.6     &  0.01      &    2       &    9.23    &     0.049    &   4.93    &    0.022   &    9.49    &  18.92      & 0.754 \\
IC 5298                 &  0.02742  &  11.11  & 10.82  & 0.019  & 0.75    &   59.06  &    1.2    &  8.42    &  0.009     &    1       &    9.31    &     0.032    &   5.03    &    0.032   &    9.56    &  19.54      & 0.778 \\
NGC 7771                &  0.01427  &  11.37  & 10.91  & 0.022  & 0.0     &  315.9   &    2.63   &  8.58    &  0.004     &    1       &    9.45    &     0.011    &   5.16    &    0.011   &    9.71    &  23.54      & 1.057 \\
Mrk 331                 &  0.01848  &  10.92  & 10.8   & 0.019  & 0.0     &  207.63  &    1.37   &  8.62    &  0.003     &    1       &    8.82    &     0.085    &   4.53    &    0.085   &    9.07    &  18.71      & 0.742 \\
MCG+12-02-001           &  0.0157   &  10.91  & 10.74  & 0.022  & 0.0     &  \nodata &   \nodata &  \nodata &  \nodata   &    1       &    8.93    &     0.028    &   4.62    &    0.028   &    9.19    &   \nodata   & \nodata   \\
MCG+08-18-012           &  0.02594  &   9.8   &  9.39  & 0.026  & \nodata &  \nodata &   \nodata &  \nodata &  \nodata   &    1       &    9.24    &     \nodata  &   4.92    &    \nodata &    9.49    &   \nodata   & \nodata   \\
2MASX J11210825-0259399 &  0.0249   &   9.98  &  9.47  & 0.021  & \nodata &  \nodata &   \nodata &  \nodata &  \nodata   &    1       &    9.01    &     0.059    &   4.66    &    0.058   &    9.27    &   \nodata   & \nodata   \\
NGC 7770                &  0.01427  &  10.41  &  9.88  & 0.022  & 0.28    &  \nodata &   \nodata &  \nodata &  \nodata   &    1       &    8.61    &     0.039    &   4.32    &    0.039   &    8.86    &   \nodata   & \nodata   \\
NGC 0034                &  0.01962  &  11.04  & 10.65  & 0.02   & 0.17    &  \nodata &   \nodata &  \nodata &  \nodata   &    2       &    9.33    &     0.045    &   5.03    &    0.012   &    9.59    &   \nodata   & \nodata   \\
NGC 0958                &  0.01914  &  11.45  & 10.91  & 0.022  & \nodata &  \nodata &   \nodata &  \nodata &  \nodata   &    2       &    9.39    &     0.045    &   5.07    &    0.013   &    9.64    &   \nodata   & \nodata   \\
ESO 550-IG025           &  0.03209  &  11.06  & 10.91  & 0.022  & \nodata &  \nodata &   \nodata &  \nodata &  \nodata   &    2       &    9.52    &     0.047    &   5.21    &    0.017   &    9.78    &   \nodata   & \nodata   \\
CGCG 043-099            &  0.03748  &  11.08  & 11.02  & 0.019  & 0.0     &  \nodata &   \nodata &  \nodata &  \nodata   &    2       &    9.67    &     0.044    &   5.36    &    0.008   &    9.92    &   \nodata   & \nodata   \\
ESO 350-IG038           &  0.0206   &  10.67  & 10.03  & 0.019  & 0.75    &   70.56  &    0.79   &  8.25    &  0.005     &  \nodata   &  \nodata   &   \nodata    & \nodata   &  \nodata   &  \nodata   &   \nodata   & \nodata   \\
NGC 0876                &  0.01305  &  10.39  &  9.95  & 0.022  & 0.18    &   76.74  &    1.89   &  7.88    &  0.011     &  \nodata   &  \nodata   &   \nodata    & \nodata   &  \nodata   &  \nodata   &   \nodata   & \nodata   \\
NGC 0877                &  0.01305  &  11.14  & 10.7   & 0.022  & 0.31    &  290.92  &    1.66   &  8.46    &  0.002     &  \nodata   &  \nodata   &   \nodata    & \nodata   &  \nodata   &  \nodata   &   \nodata   & \nodata   \\
MCG+05-06-035           &  0.03371  &  11.28  & 10.6   & 0.018  & 0.0     &   38.14  &    1.84   &  8.42    &  0.021     &  \nodata   &  \nodata   &   \nodata    & \nodata   &  \nodata   &  \nodata   &   \nodata   & \nodata   \\
MCG+05-06-036           &  0.03371  &  10.99  & 11.01  & 0.019  & 0.12    &   71.72  &    2.09   &  8.69    &  0.013     &  \nodata   &  \nodata   &   \nodata    & \nodata   &  \nodata   &  \nodata   &   \nodata   & \nodata   \\
NGC 1068                &  0.00379  &  11.16  & 10.65  & 0.022  & \nodata & 2489.28  &    6.25   &  8.31    &  0.001     &  \nodata   &  \nodata   &   \nodata    & \nodata   &  \nodata   &  \nodata   &   \nodata   & \nodata   \\
UGC02982                &  0.0177   &  10.93  & 10.73  & 0.022  & 0.0     &  424.92  &    1.87   &  8.89    &  0.002     &  \nodata   &  \nodata   &   \nodata    & \nodata   &  \nodata   &  \nodata   &   \nodata   & \nodata   \\
ESO 203-IG001           &  0.05291  &  10.87  & 10.95  & 0.018  & 0.9     &    8.87  &    2.19   &  8.19    &  0.107     &  \nodata   &  \nodata   &   \nodata    & \nodata   &  \nodata   &  \nodata   &   \nodata   & \nodata   \\
MCG-05-12-006           &  0.01875  &  10.64  & 10.42  & 0.019  & 0.0     &   44.46  &    1.57   &  7.96    &  0.015     &  \nodata   &  \nodata   &   \nodata    & \nodata   &  \nodata   &  \nodata   &   \nodata   & \nodata   \\
NGC 1961                &  0.01312  &  11.59  & 10.81  & 0.022  & 0.46    &  196.93  &    2.97   &  8.3     &  0.007     &  \nodata   &  \nodata   &   \nodata    & \nodata   &  \nodata   &  \nodata   &   \nodata   & \nodata   \\
NGC 2146                &  0.00298  &  10.81  & 10.22  & 0.022  & 0.0     & 3130.97  &   10.07   &  8.2     &  0.001     &  \nodata   &  \nodata   &   \nodata    & \nodata   &  \nodata   &  \nodata   &   \nodata   & \nodata   \\
NGC 2369                &  0.01081  &  11.09  & 10.68  & 0.022  & 0.08    &  344.82  &    1.7    &  8.37    &  0.002     &  \nodata   &  \nodata   &   \nodata    & \nodata   &  \nodata   &  \nodata   &   \nodata   & \nodata   \\
NGC 2388                &  0.01379  &  10.89  & 10.64  & 0.02   & \nodata &  214.53  &    1.86   &  8.38    &  0.004     &  \nodata   &  \nodata   &   \nodata    & \nodata   &  \nodata   &  \nodata   &   \nodata   & \nodata   \\
NGC 2389                &  0.01379  &  10.58  & 10.14  & 0.022  & \nodata &  155.23  &    1.97   &  8.24    &  0.006     &  \nodata   &  \nodata   &   \nodata    & \nodata   &  \nodata   &  \nodata   &   \nodata   & \nodata   \\
MCG+02-20-003           &  0.01625  &  10.69  & 10.46  & 0.019  & 0.7     &  117.11  &    1.35   &  8.26    &  0.005     &  \nodata   &  \nodata   &   \nodata    & \nodata   &  \nodata   &  \nodata   &   \nodata   & \nodata   \\
ESO 60-IG016            &  0.04632  &  10.97  & 10.92  & 0.022  & \nodata &   53.63  &    1.52   &  8.85    &  0.012     &  \nodata   &  \nodata   &   \nodata    & \nodata   &  \nodata   &  \nodata   &   \nodata   & \nodata   \\
IRAS F08572+3915        &  0.05835  &  11.8   & 10.86  & 0.018  & \nodata &   12.6   &    1.36   &  8.43    &  0.047     &  \nodata   &  \nodata   &   \nodata    & \nodata   &  \nodata   &  \nodata   &   \nodata   & \nodata   \\
IRAS 09022-3615         &  0.05964  &  11.22  & 11.5   & 0.019  & 0.73    &   93.49  &    1.68   &  9.32    &  0.008     &  \nodata   &  \nodata   &   \nodata    & \nodata   &  \nodata   &  \nodata   &   \nodata   & \nodata   \\
NGC 3110                &  0.01686  &  11.12  & 10.77  & 0.022  & 0.0     &  451.52  &    2.49   &  8.88    &  0.002     &  \nodata   &  \nodata   &   \nodata    & \nodata   &  \nodata   &  \nodata   &   \nodata   & \nodata   \\
IC 2545                 &  0.0341   &  11.46  & 10.81  & 0.019  & 0.95    &   45.03  &    1.1    &  8.5     &  0.011     &  \nodata   &  \nodata   &   \nodata    & \nodata   &  \nodata   &  \nodata   &   \nodata   & \nodata   \\
NGC 3256                &  0.00935  &  11.06  & 10.95  & 0.022  & 0.0     & 1537.24  &    2.26   &  8.89    &  0.001     &  \nodata   &  \nodata   &   \nodata    & \nodata   &  \nodata   &  \nodata   &   \nodata   & \nodata   \\
ESO 264-G036            &  0.02101  &  11.39  & 10.81  & 0.022  & 0.12    &  188.1   &    1.28   &  8.69    &  0.003     &  \nodata   &  \nodata   &   \nodata    & \nodata   &  \nodata   &  \nodata   &   \nodata   & \nodata   \\
IRAS F10565+2448        &  0.0431   &  11.17  & 11.33  & 0.019  & 0.06    &   78.13  &    2.48   &  8.95    &  0.014     &  \nodata   &  \nodata   &   \nodata    & \nodata   &  \nodata   &  \nodata   &   \nodata   & \nodata   \\
ESO 320-G030            &  0.01078  &  10.67  & 10.64  & 0.02   & 0.0     &  257.94  &    1.94   &  8.24    &  0.003     &  \nodata   &  \nodata   &   \nodata    & \nodata   &  \nodata   &  \nodata   &   \nodata   & \nodata   \\
IRAS F12112+0305        &  0.07332  &  11.34  & 11.54  & 0.019  & 0.21    &   36.63  &    3.07   &  9.1     &  0.036     &  \nodata   &  \nodata   &   \nodata    & \nodata   &  \nodata   &  \nodata   &   \nodata   & \nodata   \\
ESO 267-G029            &  0.01849  &  10.98  & 10.43  & 0.022  & 0.0     &  128.03  &    1.6    &  8.41    &  0.005     &  \nodata   &  \nodata   &   \nodata    & \nodata   &  \nodata   &  \nodata   &   \nodata   & \nodata   \\
ESO 267-G030            &  0.01849  &  11.21  & 10.54  & 0.022  & 0.09    &  208.38  &    1.82   &  8.62    &  0.004     &  \nodata   &  \nodata   &   \nodata    & \nodata   &  \nodata   &  \nodata   &   \nodata   & \nodata   \\
VV 250b                 &  0.03107  &  10.66  & 10.37  & 0.016  & 0.0     &   36.5   &    1.04   &  8.33    &  0.012     &  \nodata   &  \nodata   &   \nodata    & \nodata   &  \nodata   &  \nodata   &   \nodata   & \nodata   \\
VV 250a                 &  0.03107  &  10.82  & 10.76  & 0.018  & 0.0     &   80.6   &    0.67   &  8.67    &  0.004     &  \nodata   &  \nodata   &   \nodata    & \nodata   &  \nodata   &  \nodata   &   \nodata   & \nodata   \\
MCG-03-34-064           &  0.01654  &  11.08  & 10.0   & 0.019  & 0.98    &   21.1   &    1.45   &  7.53    &  0.03      &  \nodata   &  \nodata   &   \nodata    & \nodata   &  \nodata   &  \nodata   &   \nodata   & \nodata   \\
NGC 5135                &  0.01369  &  11.1   & 10.74  & 0.022  & 0.14    &  225.47  &    1.54   &  8.39    &  0.003     &  \nodata   &  \nodata   &   \nodata    & \nodata   &  \nodata   &  \nodata   &   \nodata   & \nodata   \\
NGC 5256                &  0.02782  &  11.22  & 10.78  & 0.022  & 0.0     &  150.17  &    2.21   &  8.84    &  0.006     &  \nodata   &  \nodata   &   \nodata    & \nodata   &  \nodata   &  \nodata   &   \nodata   & \nodata   \\
NGC 5653                &  0.01188  &  11.01  & 10.51  & 0.02   & 0.0     &  414.99  &    1.66   &  8.53    &  0.002     &  \nodata   &  \nodata   &   \nodata    & \nodata   &  \nodata   &  \nodata   &   \nodata   & \nodata   \\
NGC 5734                &  0.01375  &  11.09  & 10.54  & 0.022  & 0.12    &  302.18  &    2.05   &  8.52    &  0.003     &  \nodata   &  \nodata   &   \nodata    & \nodata   &  \nodata   &  \nodata   &   \nodata   & \nodata   \\
NGC 5743                &  0.01375  &  10.82  & 10.28  & 0.022  & 0.07    &  247.23  &    1.92   &  8.44    &  0.003     &  \nodata   &  \nodata   &   \nodata    & \nodata   &  \nodata   &  \nodata   &   \nodata   & \nodata   \\
VV 340b                 &  0.03367  &  10.84  & 10.4   & 0.022  & 0.0     &   62.81  &    1.52   &  8.63    &  0.011     &  \nodata   &  \nodata   &   \nodata    & \nodata   &  \nodata   &  \nodata   &   \nodata   & \nodata   \\
VV 340a                 &  0.03367  &  11.39  & 11.15  & 0.022  & 0.0     &  169.55  &    3.38   &  9.06    &  0.009     &  \nodata   &  \nodata   &   \nodata    & \nodata   &  \nodata   &  \nodata   &   \nodata   & \nodata   \\
IRAS 15250+3609         &  0.05516  &  10.78  & 11.0   & 0.019  & 0.93    &   14.65  &    1.49   &  8.44    &  0.044     &  \nodata   &  \nodata   &   \nodata    & \nodata   &  \nodata   &  \nodata   &   \nodata   & \nodata   \\
NGC 5990                &  0.01281  &  11.15  & 10.41  & 0.022  & 0.72    &  264.1   &    1.81   &  8.4     &  0.003     &  \nodata   &  \nodata   &   \nodata    & \nodata   &  \nodata   &  \nodata   &   \nodata   & \nodata   \\
NGC 6052                &  0.01581  &  10.68  & 10.35  & 0.022  & 0.0     &  236.78  &    0.92   &  8.54    &  0.002     &  \nodata   &  \nodata   &   \nodata    & \nodata   &  \nodata   &  \nodata   &   \nodata   & \nodata   \\
NGC 6090                &  0.02984  &  11.35  & 10.84  & 0.019  & \nodata &  143.31  &    1.61   &  8.88    &  0.005     &  \nodata   &  \nodata   &   \nodata    & \nodata   &  \nodata   &  \nodata   &   \nodata   & \nodata   \\
NGC 6240                &  0.02448  &  11.59  & 11.09  & 0.019  & 0.34    &  487.24  &    5.8    &  9.24    &  0.005     &  \nodata   &  \nodata   &   \nodata    & \nodata   &  \nodata   &  \nodata   &   \nodata   & \nodata   \\
IRAS F17132+5313        &  0.05094  &  11.24  & 11.21  & 0.019  & 0.05    &   65.58  &    2.93   &  9.02    &  0.019     &  \nodata   &  \nodata   &   \nodata    & \nodata   &  \nodata   &  \nodata   &   \nodata   & \nodata   \\
IC 4689                 &  0.01735  &  10.7   & 10.27  & 0.019  & \nodata &  107.1   &    1.59   &  8.28    &  0.006     &  \nodata   &  \nodata   &   \nodata    & \nodata   &  \nodata   &  \nodata   &   \nodata   & \nodata   \\
NGC 6670A               &  0.0286   &  10.81  & 10.44  & 0.022  & 0.0     &   77.78  &    0.7    &  8.58    &  0.004     &  \nodata   &  \nodata   &   \nodata    & \nodata   &  \nodata   &  \nodata   &   \nodata   & \nodata   \\
IC 4734                 &  0.01561  &  10.96  & 10.74  & 0.019  & 0.06    &  160.72  &    1.85   &  8.36    &  0.005     &  \nodata   &  \nodata   &   \nodata    & \nodata   &  \nodata   &  \nodata   &   \nodata   & \nodata   \\
ESO 339-G011            &  0.0192   &  11.01  & 10.6   & 0.019  & 0.48    &  172.52  &    1.73   &  8.57    &  0.004     &  \nodata   &  \nodata   &   \nodata    & \nodata   &  \nodata   &  \nodata   &   \nodata   & \nodata   \\
NGC 6926a               &  0.01961  &  11.29  & 10.85  & 0.022  & 0.43    &  200.09  &    1.55   &  8.66    &  0.003     &  \nodata   &  \nodata   &   \nodata    & \nodata   &  \nodata   &  \nodata   &   \nodata   & \nodata   \\
ESO 286-IG019           &  0.043    &  11.05  & 11.06  & 0.019  & 0.77    &   57.93  &    1.04   &  8.82    &  0.008     &  \nodata   &  \nodata   &   \nodata    & \nodata   &  \nodata   &  \nodata   &   \nodata   & \nodata   \\
ESO 286-G035            &  0.01736  &  10.77  & 10.54  & 0.02   & 0.0     &  185.68  &    1.84   &  8.52    &  0.004     &  \nodata   &  \nodata   &   \nodata    & \nodata   &  \nodata   &  \nodata   &   \nodata   & \nodata   \\
NGC 7130                &  0.01615  &  11.16  & 10.75  & 0.022  & 0.41    &  237.74  &    0.95   &  8.56    &  0.002     &  \nodata   &  \nodata   &   \nodata    & \nodata   &  \nodata   &  \nodata   &   \nodata   & \nodata   \\
IC 5179                 &  0.01141  &  11.09  & 10.69  & 0.022  & 0.0     &  722.85  &    1.39   &  8.74    &  0.001     &  \nodata   &  \nodata   &   \nodata    & \nodata   &  \nodata   &  \nodata   &   \nodata   & \nodata   \\
NGC 7469                &  0.01632  &  11.29  & 10.87  & 0.019  & 0.58    &  286.81  &    1.7    &  8.65    &  0.003     &  \nodata   &  \nodata   &   \nodata    & \nodata   &  \nodata   &  \nodata   &   \nodata   & \nodata   \\
ESO 148-IG002           &  0.0446   &  11.03  & 11.11  & 0.019  & 0.4     &   94.85  &    1.85   &  9.06    &  0.008     &  \nodata   &  \nodata   &   \nodata    & \nodata   &  \nodata   &  \nodata   &   \nodata   & \nodata   \\
NGC 7552                &  0.00536  &  10.88  & 10.44  & 0.022  & 0.02    &  729.03  &    2.04   &  8.08    &  0.001     &  \nodata   &  \nodata   &   \nodata    & \nodata   &  \nodata   &  \nodata   &   \nodata   & \nodata   \\
ESO 077-IG014 NED01     &  0.04156  &  11.09  & 11.13  & 0.022  & 0.1     &   18.7   &    0.58   &  8.29    &  0.013     &  \nodata   &  \nodata   &   \nodata    & \nodata   &  \nodata   &  \nodata   &   \nodata   & \nodata   \\
IRAS F23365+3604        &  0.06448  &  11.15  & 11.38  & 0.019  & 0.22    &   20.34  &    1.92   &  8.73    &  0.041     &  \nodata   &  \nodata   &   \nodata    & \nodata   &  \nodata   &  \nodata   &   \nodata   & \nodata   \\
MCG-01-60-022           &  0.02324  &  10.85  & 10.61  & 0.019  & 0.0     &  160.18  &    1.61   &  8.71    &  0.004     &  \nodata   &  \nodata   &   \nodata    & \nodata   &  \nodata   &  \nodata   &   \nodata   & \nodata   \\   
\enddata
\tablecomments{The columns are: (1) Galaxy name (2) Redshift from \citet{Chu_2017} (3) Stellar mass from \citet{Howell_2010} (4)(5) Luminosity and its uncertainty at 160$\mu$m derived from fluxes in \citet{Chu_2017} as described in \S4.1. (6) The fractional contribution of AGN to MIR luminosity (AGN fractions) calculated following the method in \citet{petric2011} (7)(8) Galaxy-integrated flux and uncertainty of the [\ion{C}{2}] line measured from the best aperture in \citet{Diaz-Santos_2017} (9)(10) Columns (7)(8) converted to luminosity in solar unit (11) The reference for the CO luminosity (12)(13) CO luminosity and its uncertainty in brightness temperature unit (14)(15) Columns (12)(13) converted to solar unit. (16) The H$_2$ mass derived in \S4.1.1 using $\alpha_{\rm{CO}}$ = 1.8 (K km s$^{-1}$pc$^2$)$^{-1}$ \citep{Herrero-Illana_2019,Montoya-Arroyave_2023} (17)(18) Star formation rate and its uncertainty based on the total infrared luminosity derived from the luminosity at 160$\mu$m as described in \S4.1.2. }
\tablenotetext{a}{Unavailable uncertainty means the corresponding quantity is an upper limit.}
\tablerefs{(1)\citet{Yamashita_2017}, (2)\citet{Herrero-Illana_2019}}
\end{deluxetable*}
\end{longrotatetable}

\begin{longrotatetable}
\begin{deluxetable*}{ccccccccccccccc}
\tabletypesize{\scriptsize}
\tablecaption{QSO1s (PG QSO) literature and derived data.\label{tab:qso1_sample}}
\tablehead{\colhead{Name} & \colhead{$z$}  & \colhead{$L^{\prime}_{\rm{CO}}$} & \colhead{$\Delta$$L^{\prime}_{\rm{CO}}$} & \colhead{M$_*$} & \colhead{$L_{\rm{160\mu m}}$} & \colhead{$\Delta$$L_{\rm{160\mu m}}$} & \colhead{$L_{\rm{CO}}$} & \colhead{$\Delta$$L_{\rm{CO}}$}  & \colhead{CO ref} & \colhead{$L_{\rm{CII}}$} & \colhead{$\Delta$$L_{\rm{CII}}$} & \colhead{SFR$_{\rm{IR}}$} & \colhead{$\Delta$SFR$_{\rm{IR}}$} & \colhead{M$_{\rm{H_2,CO}}$} \\ 
\colhead{} & \colhead{} & \colhead{log(K km/s pc$^2$)} & \colhead{log(K km/s pc$^2$)} & \colhead{log(M$_{\odot}$)} & \colhead{log($L_{\odot}$)} & \colhead{log($L_{\odot}$)} & \colhead{log($L_{\odot}$)} & \colhead{log($L_{\odot}$)} & \colhead{}  & \colhead{log($L_{\odot}$)}  & \colhead{log($L_{\odot}$)}  & \colhead{M$_{\odot}$/yr}  & \colhead{M$_{\odot}$/yr}  & \colhead{log(M$_{\odot}$)} \\
\colhead{(1)} & \colhead{(2)} & \colhead{(3)} & \colhead{(4)} & \colhead{(5)} & \colhead{(6)} & \colhead{(7)} & \colhead{(8)} & \colhead{(9)} & \colhead{(10)} & \colhead{(11)} & \colhead{(12)} & \colhead{(13)} & \colhead{(14)} & \colhead{(15)}} 
\startdata
PG 0003+199 & 0.026 & 7.23 & 0.06 & 10.38 & 8.9 & 0.129 & 3.56 & 0.055 & 1 & 7.9 & 0.065 & 0.36 & 0.097 & 7.75 \\
PG 0007+106 & 0.089 & 8.63 & 0.05 & 11.03 & 10.3 & 0.095 & 4.92 & 0.047 & 1 & \nodata & \nodata & \nodata & \nodata & 9.15 \\
PG 0050+124 & 0.061 & 9.71 & 0.02 & 11.31 & 10.9 & 0.009 & 5.4 & 0.016 & 3 & \nodata & \nodata & \nodata & \nodata & 10.23 \\
PG 0934+013 & 0.05 & 8.49 & 0.03 & 10.1 & 9.9 & 0.083 & 4.79 & 0.025 & 1 & \nodata & \nodata & \nodata & \nodata & 9.01 \\
PG 1011-040 & 0.058 & 9.02 & 0.04 & 10.31 & 9.9 & 0.101 & 4.82 & 0.038 & 7 & \nodata & \nodata & \nodata & \nodata & 9.54 \\
PG 1119+120 & 0.05 & 8.47 & 0.06 & 10.86 & 10.0 & 0.059 & 4.18 & 0.064 & 3 & \nodata & \nodata & \nodata & \nodata & 8.99 \\
PG 1126-041 & 0.062 & 9.09 & 0.04 & 11.04 & 10.3 & 0.033 & 4.92 & 0.043 & 7 & \nodata & \nodata & \nodata & \nodata & 9.61 \\
PG 1211+143 & 0.081 & 7.94 & 0.03 & 10.57 & 9.4 & 0.117 & 4.19 & 0.034 & 1 & \nodata & \nodata & \nodata & \nodata & 8.46 \\
PG 1229+204 & 0.063 & 8.66 & 0.11 & 11.13 & 9.9 & 0.097 & 4.34 & 0.109 & 6 & \nodata & \nodata & \nodata & \nodata & 9.18 \\
PG 1310-108 & 0.034 & 7.94 & 0.02 & 10.73 & 9.0 & 0.38 & 4.21 & 0.017 & 1 & 7.44 & 0.109 & 0.45 & 0.352 & 8.46 \\
PG 1404+226 & 0.098 & 8.95 & 0.11 & 9.99 & 10.0 & 0.068 & 4.65 & 0.109 & 6 & \nodata & \nodata & \nodata & \nodata & 9.47 \\
PG 1426+015 & 0.086 & 9.09 & 0.07 & 11.24 & 10.4 & 0.077 & 4.79 & 0.072 & 6 & \nodata & \nodata & \nodata & \nodata & 9.61 \\
PG 1501+106 & 0.036 & 7.49 & 0.04 & 11.22 & 9.6 & 0.151 & 3.78 & 0.04 & 1 & \nodata & \nodata & \nodata & \nodata & 8.01 \\
PG 2130+099 & 0.063 & 8.82 & 0.06 & 11.04 & 10.3 & 0.038 & 4.55 & 0.056 & 3 & \nodata & \nodata & \nodata & \nodata & 9.34 \\
PG 2214+139 & 0.066 & 8.02 & 0.05 & 11.17 & 9.7 & 0.147 & 4.3 & 0.046 & 1 & \nodata & \nodata & \nodata & \nodata & 8.54 \\
PG 0052+251 & 0.155 & 9.36 & 0.0 & 11.24 & 10.6 & 0.111 & 5.05 & 0.0 & 5 & \nodata & \nodata & \nodata & \nodata & 9.88 \\
PG 0157+001 & 0.163 & 9.85 & 0.04 & 11.72 & 11.6 & 0.013 & 5.54 & 0.039 & 3 & \nodata & \nodata & \nodata & \nodata & 10.37 \\
PG 0804+761 & 0.1 & 8.97 & 0.11 & 10.83 & 9.5 & 0.105 & 4.66 & 0.109 & 6 & \nodata & \nodata & \nodata & \nodata & 9.49 \\
PG 0838+770 & 0.131 & 9.31 & 0.07 & 11.33 & 10.7 & 0.069 & 5.0 & 0.069 & 3 & \nodata & \nodata & \nodata & \nodata & 9.83 \\
PG 0844+349 & 0.064 & 8.45 & \nodata \tablenotemark{a} & 10.88 & 10.1 & 0.109 & 4.14 & \nodata & 6 & \nodata & \nodata & \nodata & \nodata & 8.97 \\
PG 1202+281 & 0.165 & 9.5 & \nodata & 11.05 & 10.3 & 0.183 & 5.19 & \nodata & 2 & \nodata & \nodata & \nodata & \nodata & 10.02 \\
PG 1309+355 & 0.183 & 8.99 & \nodata & 11.41 & 10.4 & 0.132 & 4.68 & \nodata & 5 & \nodata & \nodata & \nodata & \nodata & 9.51 \\
PG 1351+640 & 0.088 & 8.98 & 0.08 & 10.82 & 10.4 & 0.052 & 4.68 & 0.08 & 3 & \nodata & \nodata & \nodata & \nodata & 9.5 \\
PG 1402+261 & 0.164 & 9.41 & 0.0 & 11.05 & 10.6 & 0.1 & 5.11 & 0.0 & 5 & \nodata & \nodata & \nodata & \nodata & 9.93 \\
PG 1411+442 & 0.09 & 8.82 & \nodata & 11.03 & 10.0 & 0.191 & 4.53 & \nodata & 6 & \nodata & \nodata & \nodata & \nodata & 9.34 \\
PG 1415+451 & 0.114 & 9.11 & 0.06 & 10.84 & 10.1 & 0.242 & 4.8 & 0.062 & 3 & \nodata & \nodata & \nodata & \nodata & 9.63 \\
PG 1440+356 & 0.079 & 9.26 & 0.04 & 11.24 & 10.6 & 0.037 & 4.97 & 0.039 & 3 & \nodata & \nodata & \nodata & \nodata & 9.78 \\
PG 1545+210 & 0.264 & 9.54 & \nodata & 11.34 & 11.8 & \nodata & 5.23 & \nodata & 5 & \nodata & \nodata & \nodata & \nodata & 10.06 \\
PG 1613+658 & 0.129 & 9.8 & 0.03 & 11.65 & 11.2 & 0.025 & 5.49 & 0.033 & 3 & \nodata & \nodata & \nodata & \nodata & 10.32 \\
PG 1700+518 & 0.292 & 10.19 & 0.08 & 11.58 & 11.5 & 0.079 & 5.91 & 0.078 & 4 & \nodata & \nodata & \nodata & \nodata & 10.71 \\
PG 0049+171 & 0.064 & 7.83 & \nodata & 11.07 & 9.2 & 0.186 & 4.12 & \nodata & 1 & \nodata & \nodata & \nodata & \nodata & 8.35 \\
PG 0923+129 & 0.029 & 8.7 & 0.01 & 10.38 & 9.9 & 0.026 & 4.99 & 0.012 & 1 & 7.95 & 0.065 & 2.88 & 0.157 & 9.22 \\
PG 1244+026 & 0.048 & 8.42 & 0.02 & 9.7 & 9.4 & 0.279 & 4.71 & 0.018 & 1 & \nodata & \nodata & \nodata & \nodata & 8.94 \\
PG 1341+258 & 0.087 & 7.98 & 0.1 & 10.85 & 9.7 & 0.063 & 4.27 & 0.097 & 1 & \nodata & \nodata & \nodata & \nodata & 8.5 \\
PG 1351+236 & 0.055 & 9.03 & 0.02 & 11.24 & 10.1 & 0.047 & 5.33 & 0.015 & 1 & \nodata & \nodata & \nodata & \nodata & 9.55 \\
PG 1448+273 & 0.065 & 8.55 & 0.02 & 10.05 & 9.9 & 0.067 & 4.84 & 0.02 & 1 & \nodata & \nodata & \nodata & \nodata & 9.07 \\
PG 2209+184 & 0.07 & 8.74 & 0.02 & 11.41 & 10.0 & 0.119 & 5.04 & 0.019 & 1 & \nodata & \nodata & \nodata & \nodata & 9.26 \\
PG 2304+042 & 0.042 & 7.42 & \nodata & 11.25 & 8.7 & \nodata & 3.71 & \nodata & 1 & \nodata & \nodata & \nodata & \nodata & 7.94 \\
PG 1534+580 & 0.03 & 8.7 & \nodata & 10.96 & 9.3 & 0.221 & 4.39 & \nodata & 8 & 6.92 & \nodata & 0.83 & 0.382 & 9.19 \\
PG 0921+525 & 0.0353 & 8.26 & 0.07 & 10.32 & 9.04 & 0.174 & 3.93 & 0.081 & 9 & 8.05 & 0.065 & 0.49 & 0.175 & 8.75 \\
\enddata
\tablecomments{The columns are: (1) PG name (2) Redshift from \citet{Petric_2015} (3)(4) CO(1--0) luminosity and its uncertainty in brightness temperature unit (5) Stellar mass from \citet{Xie_2021} (6)(7) Luminosity and its uncertainty at 160$\mu$m from \citet{Petric_2015} (8)(9) Columns (3)(4) converted to solar unit (10) The reference for the CO luminosity (11)(12) [\ion{C}{2}] 158$\mu$m luminosity and its uncertainty used in plots. These values are converted from flux measurements listed in Table \ref{tab:CIIobservations}, using the measurements for extended emission when available (13)(14) Star formation rate and its uncertainty based on the total infrared luminosity derived from the luminosity at 160$\mu$m as described in \S4.1.2 (15) The H$_2$ mass derived in \S4.1.1 using $\alpha_{\rm{CO}}$ = 3.1 (K km s$^{-1}$pc$^2$)$^{-1}$ \citep{Shangguan_2020}. }
\tablenotetext{a}{Unavailable uncertainty means the corresponding quantity is an upper limit.}
\tablerefs{(1)\citet{Shangguan_2020} (2)\citet{Evans_2001} (3)\citet{Evans_2006} (4)\citet{Evans_2009} (5)\citet{Casoli_2001} (6)\citet{Scoville_2003} (7)\citet{Bertram_2007} (8)\citet{Wylezalek_2022} (9)\citet{Salome_2023}}
\end{deluxetable*}
\end{longrotatetable}

\begin{deluxetable*}{ccccccccccccc}[h]
\tablecaption{IRAM QSO2 literature and derived data.\label{tab:qso2_sample}}
\tablehead{\colhead{Name} & \colhead{M$_*$} & \colhead{$L_{\rm{160\mu m}}$} & \colhead{$\Delta$$L_{\rm{160\mu m}}$} & \colhead{$L_{\rm{CO}}$} & \colhead{$\Delta$$L_{\rm{CO}}$} & \colhead{M$_*$ ref} & \colhead{$L_{\rm{160\mu m}}$ ref} & \colhead{$L_{\rm{CII}}$} & \colhead{$\Delta$$L_{\rm{CII}}$} & \colhead{SFR$_{\rm{IR}}$} & \colhead{$\Delta$SFR$_{\rm{IR}}$} & \colhead{M$_{\rm{H_2,CO}}$} \\ 
\colhead{} & \colhead{log(M$_{\odot}$)} & \colhead{log($L_{\odot}$)} & \colhead{log($L_{\odot}$)} & \colhead{log($L_{\odot}$)} & \colhead{log($L_{\odot}$)} & \colhead{} & \colhead{} & \colhead{log($L_{\odot}$)}  & \colhead{log($L_{\odot}$)}  & \colhead{M$_{\odot}$/yr}  & \colhead{M$_{\odot}$/yr}  & \colhead{log(M$_{\odot}$)}\\
\colhead{(1)} & \colhead{(2)} & \colhead{(3)} & \colhead{(4)} & \colhead{(5)} & \colhead{(6)} & \colhead{(7)} & \colhead{(8)} & \colhead{(9)} & \colhead{(10)} & \colhead{(11)} & \colhead{(12)} & \colhead{(13)}} 
\startdata
SDSS01 & 9.87 & 10.05 & 0.028 & 4.29 & 0.093 & 1 & 2 & 7.94 & 0.065 & 3.91 & 0.218 & 9.25 \\
SDSS02 & 10.97 & 10.34 & 0.016 & 4.47 & 0.136 & 3 & 3 & 8.15 & 0.065 & 7.19 & 0.239 & 9.41 \\
SDSS03 & 10.57 & 9.97 & 0.043 & 4.42 & 0.121 & 3 & 3 & 7.51 & 0.065 & 3.36 & 0.286 & 9.37 \\
SDSS04 & 10.92 & 10.04 & 0.018 & 4.71 & 0.061 & 3 & 3 & 7.65 & 0.065 & 3.83 & 0.143 & 9.65 \\
SDSS05 & 10.61 & 10.15 & 0.022 & 4.49 & 0.084 & 3 & 3 & \nodata & \nodata & \nodata & \nodata & 9.43 \\
SDSS06 & 10.78 & 9.88 & 0.049 & 4.28 & 0.159 & 3 & 3 & \nodata & \nodata & \nodata & \nodata & 9.22 \\
SDSS07 & 10.5 & 10.04 & 0.039 & 4.33 & 0.102 & 3 & 3 & 7.57 & 0.065 & 3.83 & 0.299 & 9.27 \\
SDSS08 & 11.03 & 9.92 & 0.049 & 4.48 & \nodata\tablenotemark{a} & 3 & 3 & \nodata & \nodata & \nodata & \nodata & 9.41 \\
SDSS09 & 10.67 & 9.65 & 0.091 & 4.4 & \nodata & 3 & 3 & \nodata & \nodata & \nodata & \nodata & 9.33 \\
SDSS10 & 10.73 & 9.72 & 0.05 & 4.48 & \nodata & 3 & 3 & \nodata & \nodata & \nodata & \nodata & 9.4 \\
SDSS11 & 10.85 & 10.06 & 0.056 & 4.88 & \nodata & 3 & 3 & \nodata & \nodata & \nodata & \nodata & 9.4 \\
SDSS12 & 10.39 & 9.51 & 0.092 & 4.69 & \nodata & 3 & 3 & \nodata & \nodata & \nodata & \nodata & 9.62 \\
SDSS13 & 10.58 & 9.69 & 0.064 & 4.6 & \nodata & 3 & 3 & \nodata & \nodata & \nodata & \nodata & 9.52 \\
SDSS14 & 10.92 & 10.49 & 0.015 & 5.01 & 0.071 & 3 & 3 & \nodata & \nodata & \nodata & \nodata & 9.94 \\
SDSS15 & 10.56 & 10.13 & 0.051 & 4.83 & \nodata & 3 & 3 & \nodata & \nodata & \nodata & \nodata & 9.75 \\
SDSS16 & 11.18 & 10.3 & 0.032 & 4.67 & 0.232 & 3 & 3 & \nodata & \nodata & \nodata & \nodata & 9.18 \\
SDSS17 & 11.03 & 10.06 & 0.04 & 4.62 & \nodata & 3 & 3 & \nodata & \nodata & \nodata & \nodata & 9.55 \\
SDSS18 & 10.69 & 10.0 & 0.027 & 5.07 & \nodata & 3 & 3 & \nodata & \nodata & \nodata & \nodata & 9.98 \\
SDSS19 & 11.06 & 10.47 & 0.048 & 4.8 & 0.083 & 3 & 3 & \nodata & \nodata & \nodata & \nodata & 9.72 \\
SDSS20 & 10.77 & 10.51 & 0.033 & 5.11 & 0.114 & 3 & 3 & \nodata & \nodata & \nodata & \nodata & 10.02 \\
\enddata
\tablecomments{The columns are: (1) Object name (2) Stellar mass (3)(4) Luminosity and its uncertainty at 160$\mu$m from fluxes as described in \S4.1. (5)(6) CO(1--0) luminosity and its uncertainty in solar unit (7) The reference for the stellar mass (8) The reference for the FIR luminosity. (9)(10) [\ion{C}{2}] 158$\mu$m luminosity and its uncertainty used in plots. These values are converted from flux measurements listed in Table \ref{tab:CIIobservations}, using the measurements for extended emission when available (11)(12) Star formation rate and its uncertainty based on the total infrared luminosity derived from the luminosity at 160$\mu$m as described in \S4.1.2 (13) The H$_2$ mass derived in \S4.1.1 using $\alpha_{\rm{CO}}$ = 4.3 (K km s$^{-1}$pc$^2$)$^{-1}$ \citep{Ramos-Almeida_2022}.}
\tablenotetext{a}{Unavailable uncertainty means the corresponding quantity is an upper limit.}
\tablerefs{(1)\citet{Koss_2011} (2)\citet{Melendez_2014} (3)\citet{Shangguan_2019}}
\end{deluxetable*}

\clearpage
\bibliography{sample631,bibliography_p,galaxies,mseP}{}
\bibliographystyle{aasjournal}

\begin{figure*}[h!]
\includegraphics[width=0.95\columnwidth]{./SDSS01IRAMlines-eps-converted-to.pdf}\\
\includegraphics[width=0.95\columnwidth]{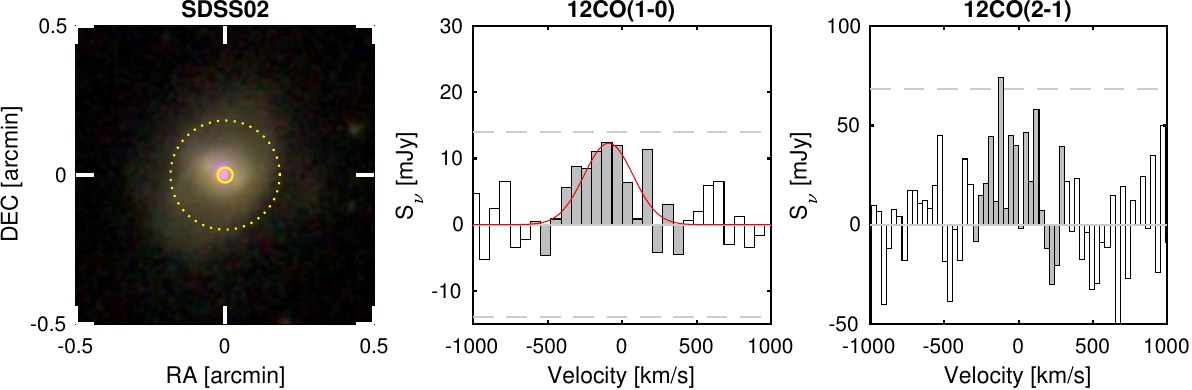}\\
\includegraphics[width=0.95\columnwidth]{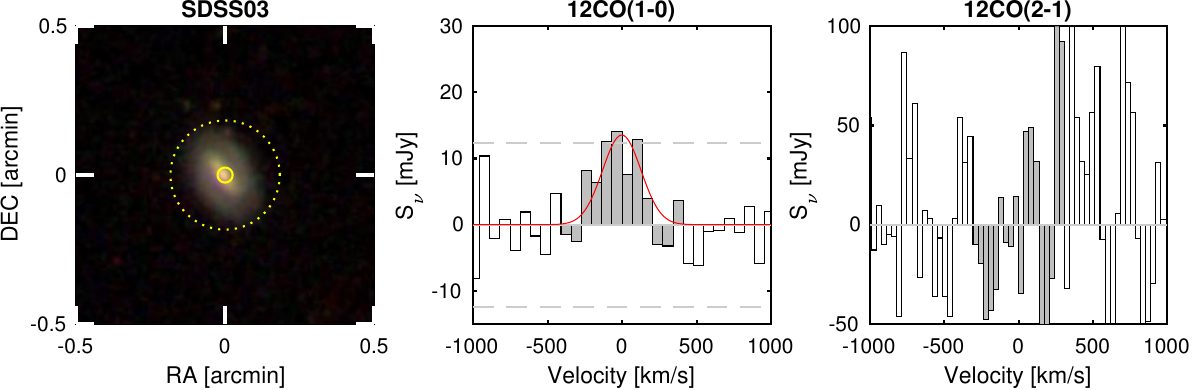}\\
\includegraphics[width=0.95\columnwidth]{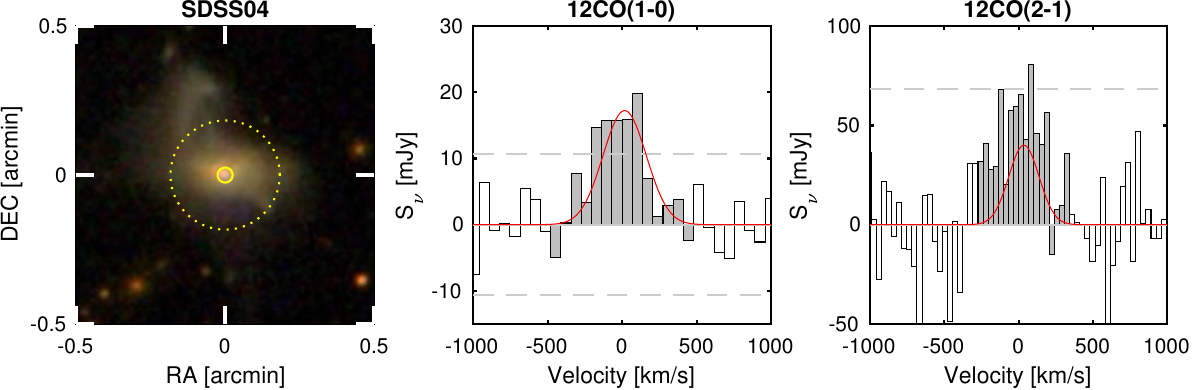}\\
\end{figure*}
\begin{figure*}[h!]
\includegraphics[width=0.95\columnwidth]{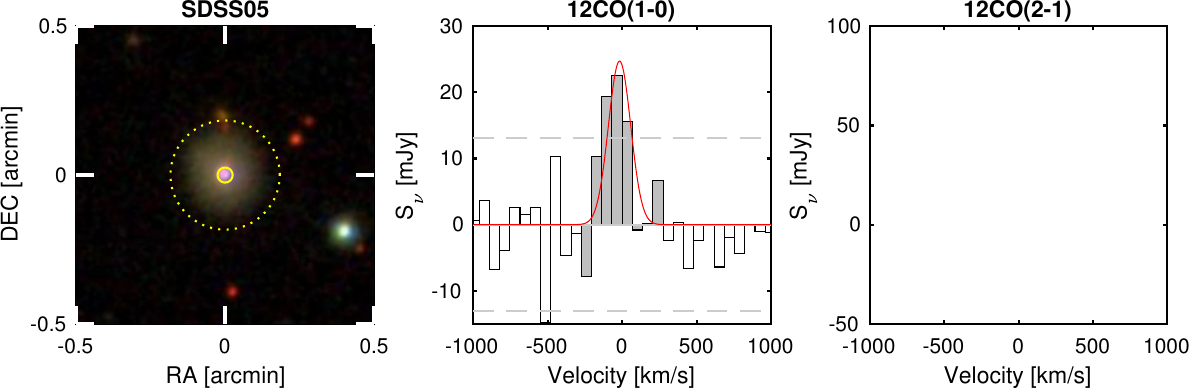}\\
\includegraphics[width=0.95\columnwidth]{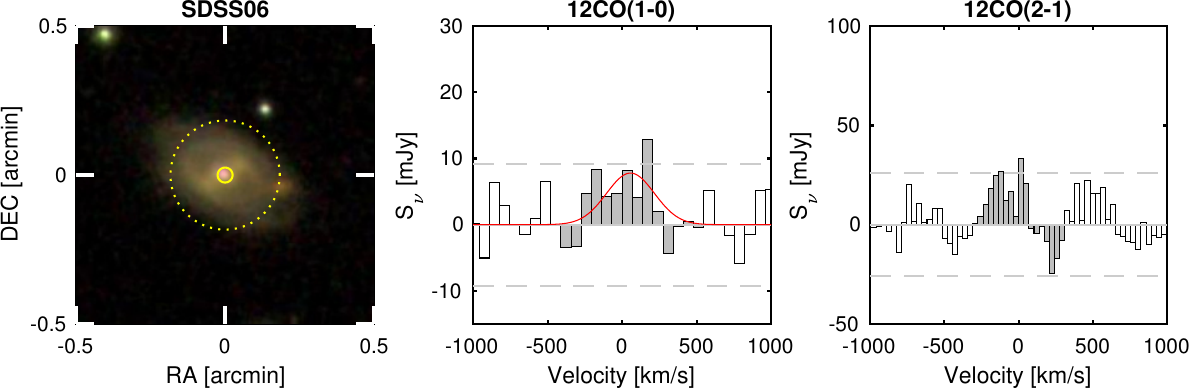}\\
\includegraphics[width=0.95\columnwidth]{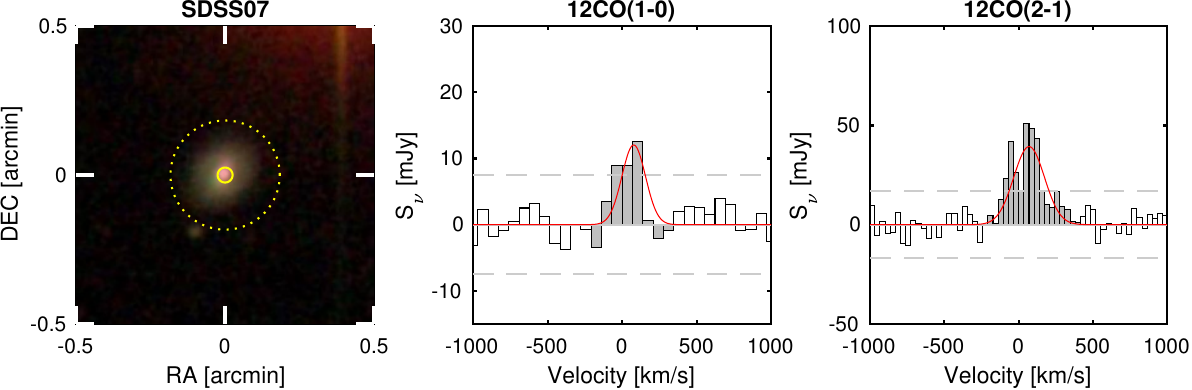}\\
\includegraphics[width=0.95\columnwidth]{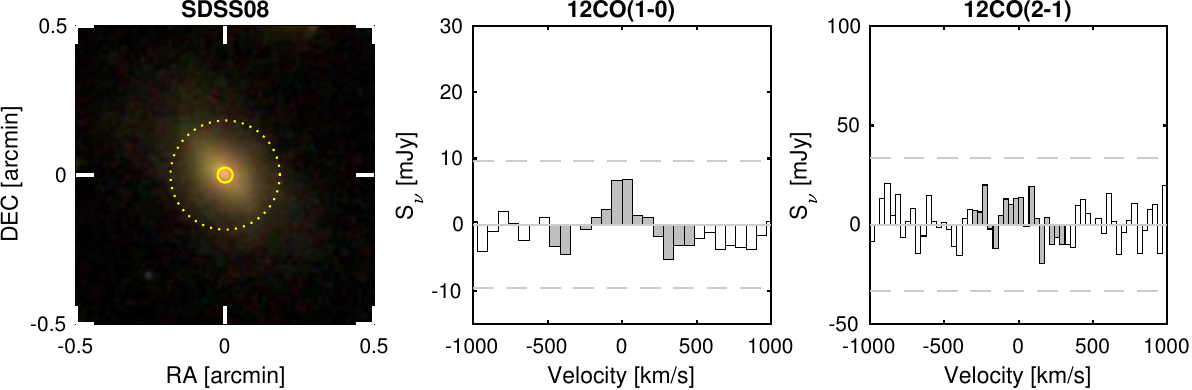}\\
\end{figure*}
\begin{figure*}[h!]
\includegraphics[width=0.95\columnwidth]{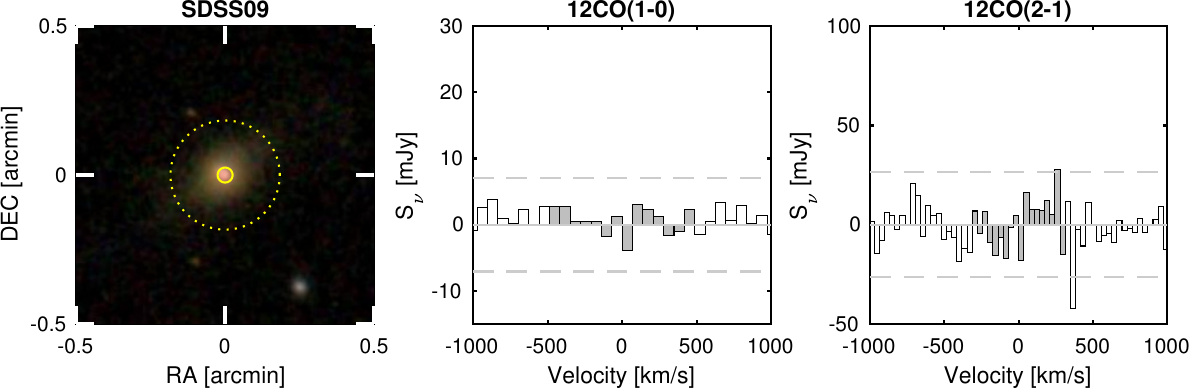}\\
\includegraphics[width=0.95\columnwidth]{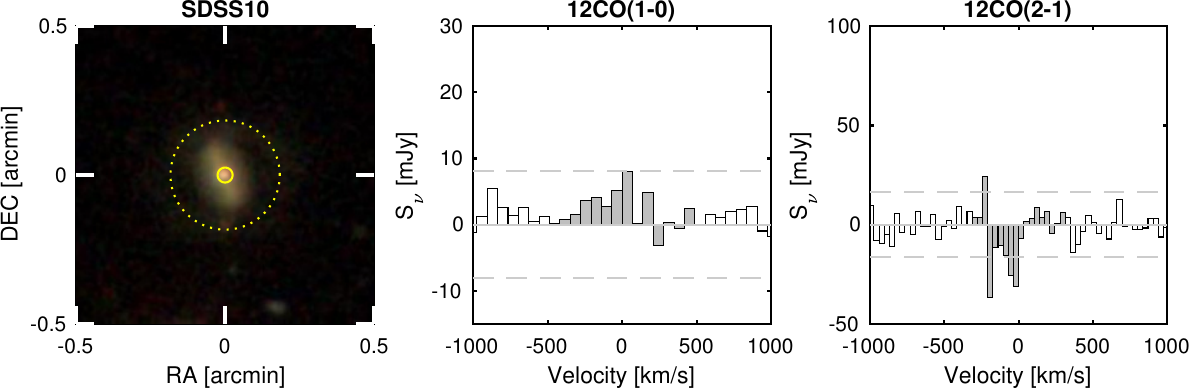}\\
\includegraphics[width=0.95\columnwidth]{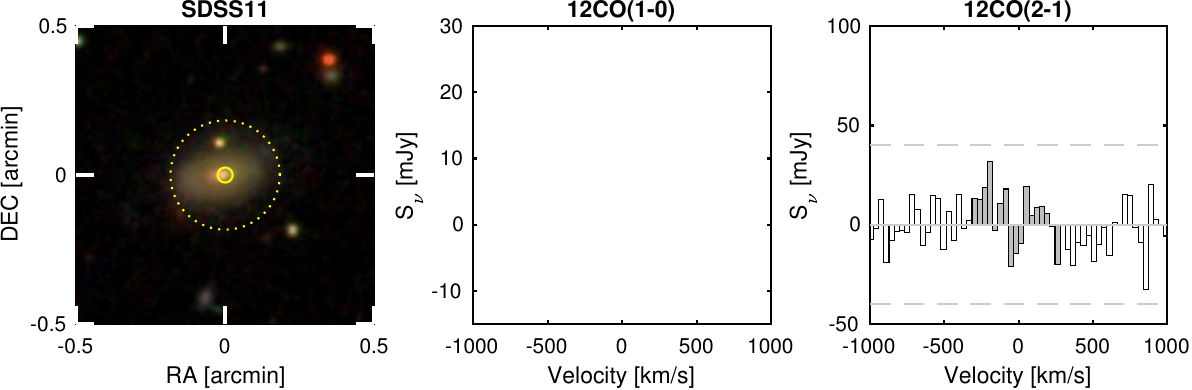}\\
\includegraphics[width=0.95\columnwidth]{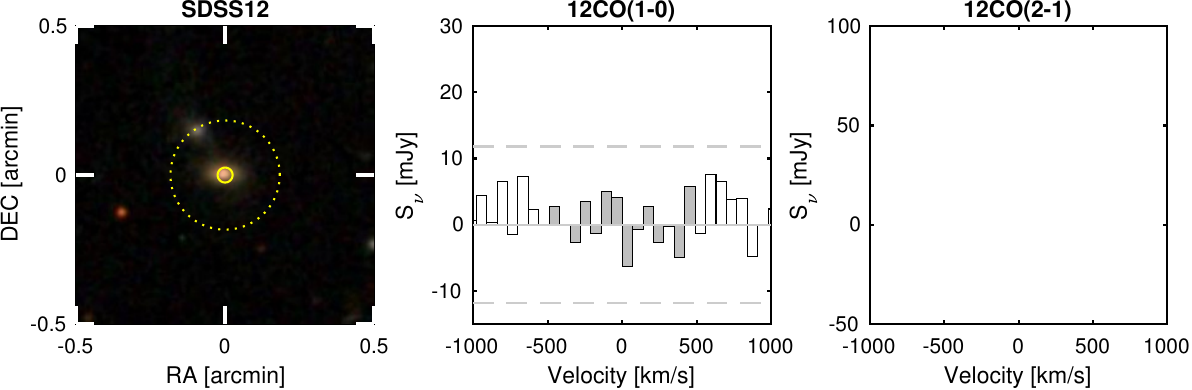}\\
\end{figure*}
\begin{figure*}[h!]
\includegraphics[width=0.95\columnwidth]{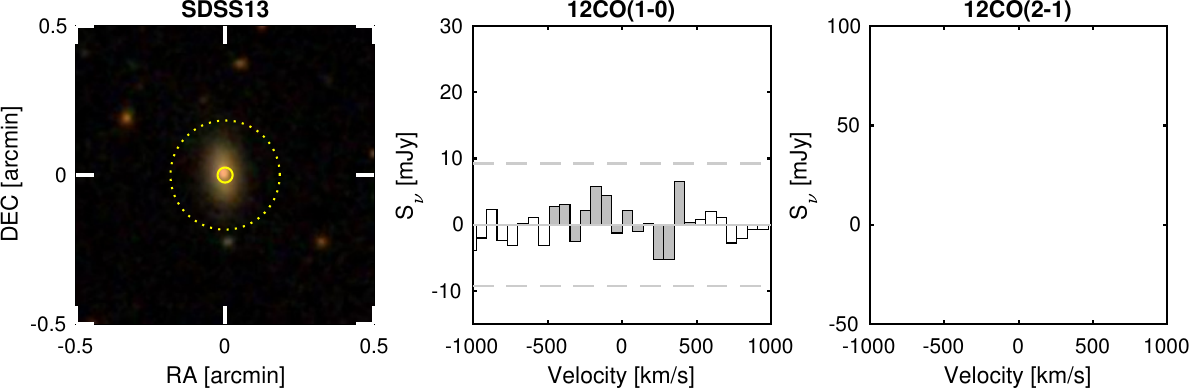}\\
\includegraphics[width=0.95\columnwidth]{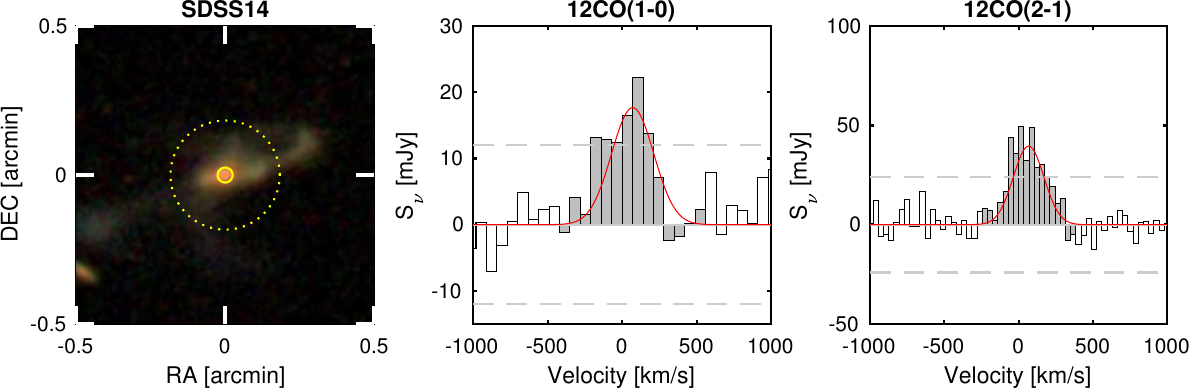}\\
\includegraphics[width=0.95\columnwidth]{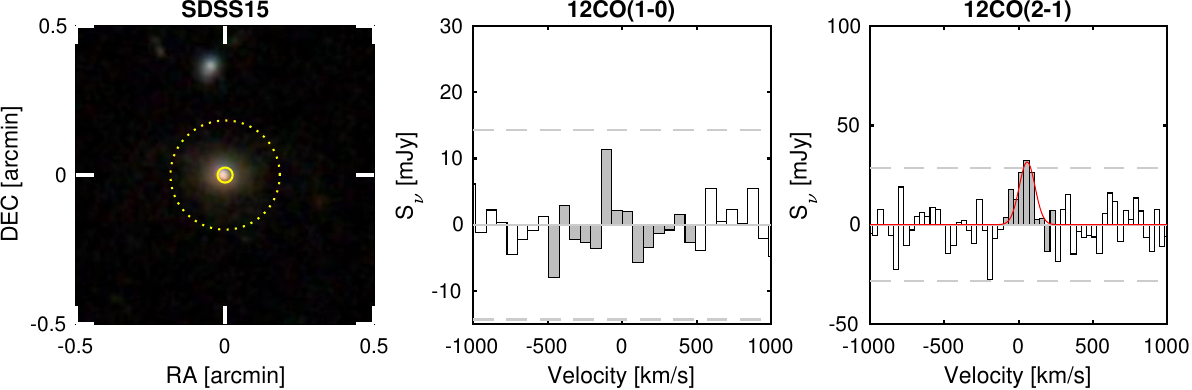}\\
\includegraphics[width=0.95\columnwidth]{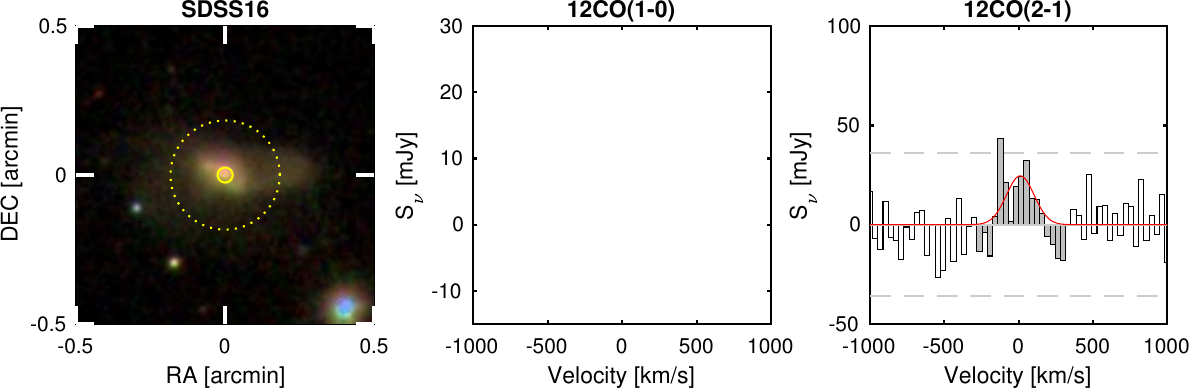}\\
\end{figure*}
\begin{figure*}[h!]
\includegraphics[width=0.95\columnwidth]{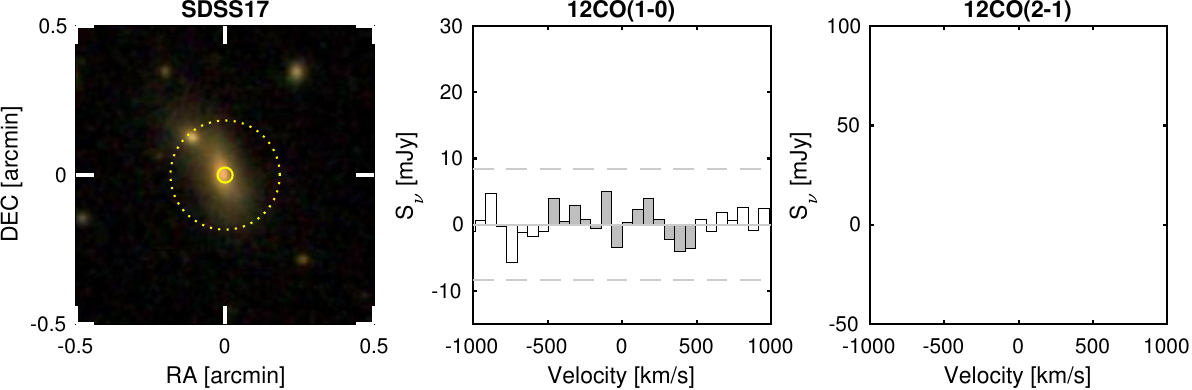}\\
\includegraphics[width=0.95\columnwidth]{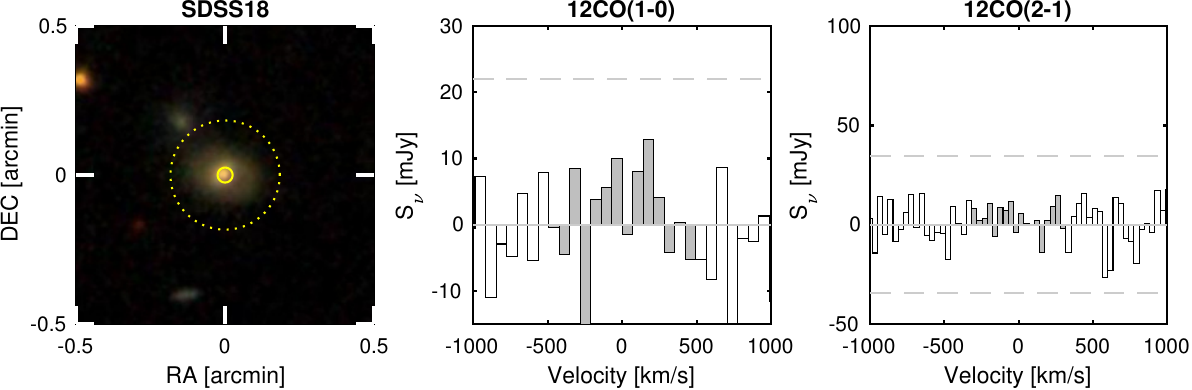}\\
\includegraphics[width=0.95\columnwidth]{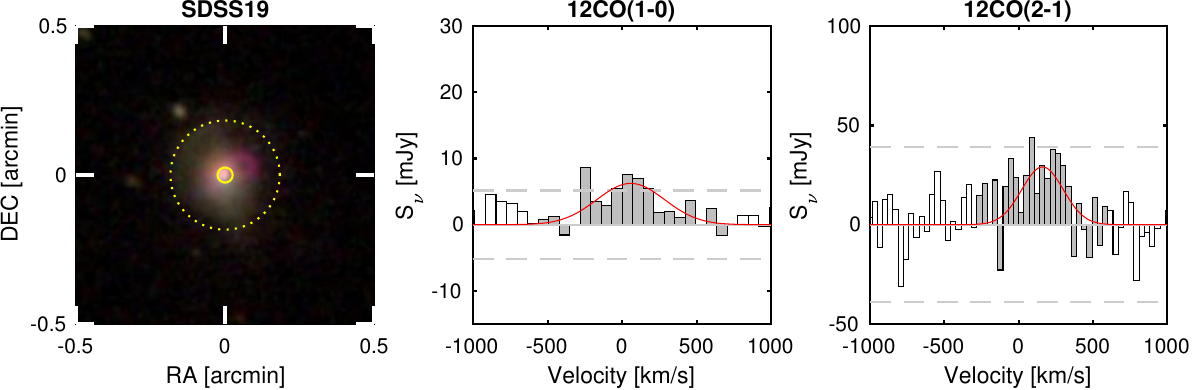}\\
\includegraphics[width=0.95\columnwidth]{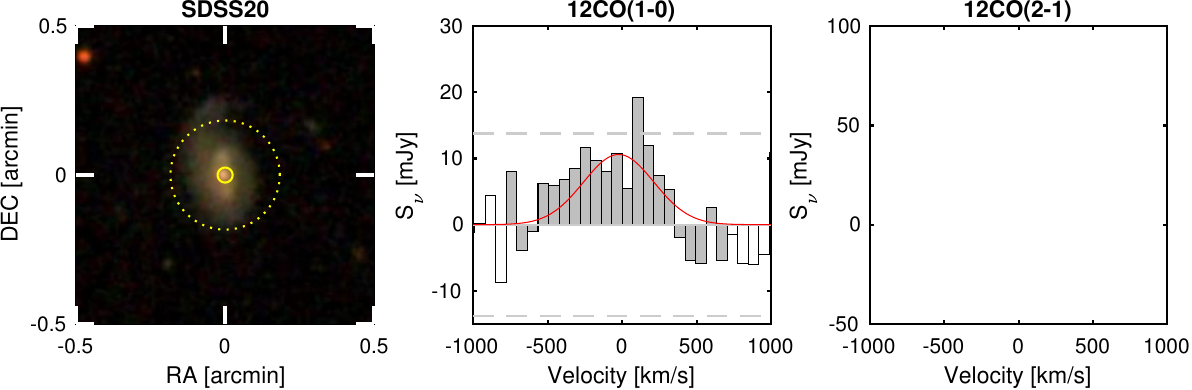}\\
\end{figure*}

\end{document}